\documentstyle[12pt]{article}

\input epsf.tex
\catcode`\@=11
\def\numberbysection{\@addtoreset{equation}{section}
        \def\theequation{\thesection.\arabic{equation}}}
\numberbysection

\catcode`\@=11
\font\tenmsa=msam10 at 12truept
\font\sevenmsa=msam7
\font\fivemsa=msam5
\font\tenmsb=msbm10 at 12truept
\font\sevenmsb=msbm7 at 9truept
\font\fivemsb=msbm5 at 7truept
\newfam\msafam
\newfam\msbfam
\textfont\msafam=\tenmsa  \scriptfont\msafam=\sevenmsa
  \scriptscriptfont\msafam=\fivemsa
\textfont\msbfam=\tenmsb  \scriptfont\msbfam=\sevenmsb
  \scriptscriptfont\msbfam=\fivemsb

\def\hexnumber@#1{\ifcase#1 0\or1\or2\or3\or4\or5\or6\or7\or8\or9\or
        A\or B\or C\or D\or E\or F\fi }

\def\Bbb{\ifmmode\let\next\Bbb@\else
 \def\next{\errmessage{Use \string\Bbb\space only in math mode}}\fi\next}
\def\Bbb@#1{{\Bbb@@{#1}}}
\def\Bbb@@#1{\fam\msbfam#1}
\catcode`\@=11
\font\teneuf=eufm10 at 12truept
\font\seveneuf=eufm7
\font\fiveeuf=eufm5
\newfam\euffam
\textfont\euffam=\teneuf  \scriptfont\euffam=\seveneuf
  \scriptscriptfont\euffam=\fiveeuf

\def\hexnumber@#1{\ifcase#1 0\or1\or2\or3\or4\or5\or6\or7\or8\or9\or
        A\or B\or C\or D\or E\or F\fi }

\def\goth{\ifmmode\let\next\goth@\else
 \def\next{\errmessage{Use \string\goth\space only in math mode}}\fi\next}
\def\goth@#1{{\goth@@{#1}}}
\def\goth@@#1{\fam\euffam#1}
\def\beq{\begin{equation}}
\def\eeq{\end{equation}}
\def\bea{\begin{eqnarray}}
\def\eea{\end{eqnarray}}

\def\N{\Bbb N}
\def\z{\Bbb Z}
\def\Q{\Bbb Q}
\def\R{\Bbb R}
\def\C{\Bbb C}
\def\PP{\Bbb P}
\def\\{\hfil\break}
\def\hrst{H_{r,s,t}}
\def\lrst{L_{r,s,t}}
\def\wrst{W_{r,s,t}}
\def\crst{C_{r,s,t}}
\def\ze{\zeta}
\def\om{\omega}
\def\ga{\gamma}
\def\Z#1{\Bbb Z_{#1}^*}
\def\rest#1{\langle #1 \rangle}
\def\jac{{\rm Jac}}
\def\gcd{{\rm gcd}}
\def\endo{{\rm End}}
\def\endoq{{\rm End}_{\Q}}
\def\e#1{\varepsilon_{\sigma }(#1)} 

\def\h{\goth H}
\def\G{\Gamma}
\def\g{\gamma}
\def\S{\Sigma}
\def\oS{\overline\S}
\def\sl{SL_2({\Bbb Z})}
\def\psl{PSL_2({\Bbb Z})}

\begin{document}

\pagenumbering{alph}
\setcounter{page}{0}
\renewcommand{\thefootnote}{\alph{footnote}}

\rightline{SPhT--96--031}
\rightline{CPT--96/P.3332}
\rightline{UCL--IPT--96--04}

\vskip 2cm
{\LARGE \centerline{Comments on the Links}
\centerline{between $su(3)$ Modular Invariants,}
\centerline{Simple Factors in the Jacobian of Fermat Curves,}
\centerline{and Rational Triangular Billiards}}

\vskip 1cm

\centerline{\large M. Bauer$^1$, A. Coste$^2$\footnote{\ \ Member of the CNRS}, 
C. Itzykson$^1$ and  P. Ruelle$^3$\footnote{\ \ Chercheur Qualifi\'e FNRS}}

\vskip 1truecm
\centerline{$^1$ Service de Physique Th\'eorique de Saclay}
\centerline{F--91191 \hskip 0.5truecm Gif-sur-Yvette, France}

\bigskip \smallskip
\centerline{$^2$ Centre de Physique Th\'eorique, UPR 7061}
\centerline{CNRS Luminy case 907}
\centerline{F--13288 \hskip 0.5truecm Marseille cedex 09, France}

\bigskip \smallskip
\centerline{$^3$ Institut de Physique Th\'eorique}
\centerline{Universit\'e Catholique de Louvain}
\centerline{B--1348 \hskip 0.5truecm Louvain-La-Neuve, Belgium}

\vskip 1.5truecm
\begin{abstract}
We examine the proposal made recently that the $su(3)$ modular invariant
partition functions could be related to the geometry of the complex Fermat
curves. Although a number of coincidences and similarities emerge between them
and certain algebraic curves related to triangular billiards, their meaning
remains obscure. In an attempt to go beyond the $su(3)$ case, we show
that any rational conformal field theory determines canonically a Riemann
surface.
\end{abstract}

\renewcommand{\thefootnote}{\arabic{footnote}}
\setcounter{footnote}{0}

\newpage
\pagenumbering{arabic}
\section{Introduction}

The partition function of a rational conformal field theory (RCFT) on a torus  
is subjected to modular invariance constraints. These constraints turn out to
be very strong, and have led to the classification of families of models. The
most celebrated achievement is the ADE classification of $su(2)$ 
Wess--Zumino--Novikov--Witten models \cite{ciz}. Its relationship with the
classification of simply--laced Lie algebras, a one--to--one correspondence, is
an {\it a posteriori} observation. Two very different problems lead to the same
classification pattern, but the proofs have very little in
common. This remarkable coincidence sustained the hope that RCFTs, combined
with the requirement of modular invariance, could perhaps be organized by known
mathematical structures, thereby bringing order in the so--called conformal zoo, 
and possibly much deeper connections with seemingly unrelated problems. At the
moment however, such general connections remain very uncertain. Even in the case
of the $su(2)$ models, a deep reason behind the ADE correspondence has remained
elusive. The few other families of theories classified up to now are either
closely related to ADE, or clearly related to arithmetical peculiarities, which
look like mere facts, and for that reason, are not understood. 

The classification of $su(3)$ modular invariants, due to Gannon \cite{gan1},
belongs to this second class. But even if a clear interpretation of the
result is lacking, the work on $su(3)$ has led to fundamental progress in our
understanding of general methods to address the problem of modular invariance.
In particular a very powerful (but weaker than the full modular invariance)
selection rule, called the parity rule, has emerged. First defined in a
restricted context \cite{gan2,rtw}, it has now been shown to hold in any
rational conformal field theory as an application of Galois theory
\cite{cg}. A few years ago, Thiran, Weyers and one of the authors \cite{rtw}
observed that the parity rule for $su(3)$ appears in a totally different
context, as an isomorphism criterion for Abelian varieties that build up the
decomposition of the Jacobian of Fermat curves in simple factors. In this
case, existing mathematical results about Fermat curves apply directly to the
problem of modular invariance, and the work of Koblitz and Rohrlich \cite{kr} 
was used to classify the modular invariants when the height (to be defined in
Section 2) is prime to 6 \cite{rtw}. Our first aim in this paper is to
explore this connection in more detail. In particular, we shall show that
$su(3)$ conformal field theories and Fermat curves have striking similarities
that might go beyond the above observation, but they have important differences
as well. Also, quite unexpectedly, the problem of rational triangular billiards
\cite{auritz} is naturally related to the parity rule and to modular invariants. 

We will present a certain number of ``strange coincidences'', relating in a
curious way the three topics we discuss in this article, namely the $su(3)$
models, the complex Fermat curves and the triangular billiards. These
observations take place at various levels, but as intriguing as they may be, 
they remain obscure. In fact the obvious observation that conformal field
theories are often organized in families (indexed by the level for
Wess--Zumino--Novikov--Witten models, or the degree of Fermat curves for
instance) is the starting point for other puzzling remarks. The paper is
organized as follows.

The second section is a general reminder of conformal theories with an
affine Lie algebra, which we take to be $su(3)$ for definiteness.  
The material is not new, but presented in such a way as to emphasize the links
with Fermat curves. This section contains a short review on the modular group
and the modular invariance problem, and recalls the parity selection rule. We
also prove some character identities related to lattice summations.

Section 3 is an introduction for non--experts to the geometry of complex
Fermat curves and their Jacobians. Complex multiplication in Abelian varieties
is briefly discussed. Again there is no claim to originality. We present the
criterion of Shimura--Taniyama to study the isogeny classes of Abelian
varieties and we show its equivalence with the parity rule for $su(3)$. 

In the fourth section, we start with a short introduction to the notion of
``dessins d'enfants'' (see for instance the collective contribution
\cite{dessins}). They give a convenient framework to discuss combinatorial and
analytic aspects of certain (special, but ubiquitous for the objects we
study) ramified coverings between Riemann surfaces. This general discussion puts
on the same footing Fermat curves and their holomorphic differentials, rational
triangular billiards and some aspects of their trajectories. We then attempt to
make a list of similarities between $su(3)$ affine characters and holomorphic
differentials on Fermat curves. In particular we show that the identity block of
the exceptional modular invariants for $su(3)$ is encoded in a sequence of
rational maps between the degree 24 Fermat curve and algebraic curves associated
with rational triangular billiards. We also show that holomorphic differentials
on Fermat curves can be reinterpreted, via uniformisation theory, as modular
forms that share some of the properties of the $su(3)$ characters, and for which
we solve the modular problem.

Finally we explore in Section 5 the algebraic consequences of the fact that the
genus one characters of an arbitrary rational conformal field theory are
automorphic functions for a finite index subgroup of the modular group. We prove
that the characters are all algebraic over $\Q(j)$, a property that allows
to associate a well--defined Riemann surface with any rational conformal field
theory (or with any chiral algebra). We study some general features of the
Riemann surfaces arising in this way, and show how they can be computed in
actual cases. This is illustrated by determining the surface associated to the
$su(3)$, level 1 ($su(3)$, level 2, is relegated to a separate appendix).

There are two appendices containing technicalities and computational details.

\medskip

Claude Itzykson's premature death is a tragedy for his friends and
collaborators. This article tries to address questions that were raised
more than three years ago and Claude participated very actively to the
early stages of this work. He not only did actual computations
(the link between billiards and blocks of modular invariants is only one
of those), but he also pointed out some possible hidden facets of the problem.
We tried to put his ideas in a form as close as possible to Claude's
standards. Anyway, it is fair to say that he should be credited for most
of the ideas while the other authors should be blamed for the inaccuracies.
We miss him very much. 

\section{Modular invariance for su(3) theories}

We review in this first section the basic features of affine Lie algebras and  
the problem of modular invariance. We will mainly consider the so--called 
untwisted $su(3)$ affine Lie algebra, but most of the material presented here
has straightforward generalizations to other algebras
\cite{kac,drouffeitz,kawa}.

\subsection{Affine representations}

The Wess--Zumino--Novikov--Witten (WZNW) models are rational conformal theories
which describe two--dimensional massless physical systems possessing an
affine Lie algebra as dynamical current symmetry \cite{witten,godol,drouffeitz}.
The building block for the chiral algebra pertaining to the $su(3)$--based
models is the  current algebra known as $\widehat{su(3)}_k$
\beq
[x_l,y_m] = [x,y]_{l+m} + kl\,\delta_{l+m,0}\,\langle x,y \rangle, 
\qquad x,y \in su(3), \;\; l,m \in \z.
\label{aff}
\eeq
$k$ is a central element, $[k,x_l]=0$, called the level, and
$\langle \cdot,\cdot \rangle$ is the Killing form  on the finite dimensional 
algebra $su(3)$. The full symmetry algebra $\cal A$ is built on the direct 
product $\widehat{su(3)_k} \otimes_I \widehat{su(3)_k}$, where $\otimes_I$ 
means that the central extensions are identified. An appropriate completion
of the envelopping algebra of (\ref{aff}) contains a central extension of the
conformal algebra with central charge $c_k={k\, {\rm dim}\,su(3) \over k+g}
\equiv 8-{24 \over n}$, where $g=3$ is the dual Coxeter number of $su(3)$ and
$n \equiv k+3$ is called the height. If we write $x(z) = \sum_m \, x_m 
z^{-1-m}$, the Virasoro algebra is generated by the density $L(z) = \sum_m \, 
L_m z^{-m-2} = \alpha \, :\langle x(z), x(z) \rangle:$ for a suitable choice of 
the constant $\alpha$. One traditionally denotes the generators of the symmetry 
algebra $\cal A$ by $x_l\otimes \overline x_m$, and those of the Virasoro
algebra by $L_l \otimes \overline L_m$.

The Hilbert space ${\cal H}$ of the theory is the direct sum of highest weight
$\cal A$--modules:
\beq
{\cal H} = \bigoplus_{p,p'} \; N_{p,p'}\,{\cal R}_p \otimes {\cal R}_{p'},
\label{hilbert}
\eeq
with $N_{p,p'} \in \N$ giving the multiplicities. If one requires that the
representations ${\cal R}_p$ be unitary, as appropriate in the case of WZNW 
models which are unitary field theories, the level hence the height must be a
positive integer (implying $c_k \geq 0$), and only a finite number of
representations are possible. They are labeled by strictly ({\it i.e.}
shifted) dominant $su(3)$ weights $p=(r,s)$ whose Dynkin labels satisfy $r+s<n$.
For what follows, it is convenient to introduce a third label $t=n-r-s$, which
can be interpreted as the zero--th label corresponding to the extra affine
fundamental weight \cite{kac}, and define the alc\^ove as the set of triplets
(or affine weights) 
\beq
B_n = \{p=(r,s,t) \;:\; r,s,t \geq 1 \hbox{ and } r+s+t = n \}.
\eeq
$B_n$ consists of the portion of the $su(3)$ weight lattice that lies in the
interior of the region delimited by the three lines (affine walls) $\alpha_1
\cdot p = \alpha_2 \cdot p = \psi \cdot p - n = 0$ ($\alpha_i$ are the two
simple roots and $\psi$ is the highest root). Equivalently, $B_n$ is a
fundamental domain for the action of the affine Weyl group $\widehat{W}_n$ on 
elements of the weight lattice with trivial little group. Its cardinality is 
$(n-1)(n-2) \over 2$.

In addition to being graded by a Cartan subalgebra of $su(3)$, which is the
reason why we could label the affine representations (and all their states) by
weights, all modules ${\cal R}_p$ are graded by $L_0$. On ${\cal R}_p$, the
spectrum of $L_0$ is equal to $h_p + \N$, with $h_p$ given by ($\rho={1 \over 2}
\sum_{\alpha > 0} \alpha$)
\beq
h_p = h_{(r,s)} = {p^2 \over 2n} - {\rho^2 \over 2n} = {r^2 + rs + s^2 \over 3n} 
- {1 \over n}.
\eeq
It follows that $h_p > 0$ for all $p$ except $p=(1,1)$ for which $h_{(1,1)}=0$.
The vacuum of the theory, which is annihilated by $L_0$ due to the
global conformal symmetry, necessarily belongs to the $\cal A$--module ${\cal
R}_{(1,1)} \otimes {\cal R}_{(1,1)}$. Its unicity then implies we should impose
$N_{(1,1),(1,1)} = 1$.

In complete analogy with the finite dimensional Lie algebras, one defines the 
character $\chi_p$ of the representation ${\cal R}_p$ as the function
\beq
\chi_p(q,M) = {\rm Tr}_{{\cal R}_p}(q^{L_0-c_k/24}M) = q^{h_p-c_k/24} \,
\sum_{m=0}^\infty \; {\rm Tr}_m(M) q^m, \qquad |q|<1.
\label{char}
\eeq
The notation ${\rm Tr}_m$ means that one traces over the subspace
of ${\cal R}_p$ where $L_0=h_p+m$. In (\ref{char}), $M$ is a function which
takes its values in the Cartan subalgebra. A traditional choice is
$M=\exp{(i\sum_j z_jH_j)}$, in which case one can show that, as functions of 
$q$ and $z_j$, the $\chi_p$ are linearly independent as $p$ runs over $B_n$.

We will exclusively use the specialized (or restricted) characters $\chi_p(q)
\equiv \chi_p(q,I)$. They can be very explicitly computed from the Weyl--Kac
formula \cite{kac}. If we denote the co--root lattice by $\tilde R$, the 
formula yields in the case of $su(3)$ 
\beq
\chi_{(r,s,t)}(q) = [\eta(q)]^{-8}\,\sum_{(a,b)=(r,s)+n \tilde R} 
{\textstyle {1 \over 2}}ab(a+b) \, q^{(a^2+ab+b^2)/3n},
\label{weylkac}
\eeq
where $\eta(q)=q^{1/24} \prod_{m \geq 1}(1-q^m)$ is the Dedekind function. Note
that the charge conjugation $C(r,s)=(s,r)$ stabilizes $B_n$, and also leaves
the specialized characters invariant: $\chi_p = \chi_{Cp}$. In the case of
$su(3)$ and for fixed $n$, there is no other linear relation among the
specialized characters, but, as we will show in Section 5, any two of them are
algebraically related (they satisfy a polynomial equation with coefficients in
$\Q$). However there are linear relations among characters corresponding to
different values of $n$, as we now show.

Let us define the functions ${\cal F}_{(r,s,t)}^{[n]}(q) \equiv [\eta(q)]^8
\chi_{(r,s,t)}$ as the numerators of the characters. We
added an extra superscript $n$ to stress the height dependence. Let $\widehat
W_n$ be the affine Weyl group corresponding to height $n$, that is, $\widehat
W_n$ is the semi--direct product of the finite Weyl group (the symmetric group
$S_3$ for $su(3)$) by the group $n \tilde R$ of translations by $n$--multiples 
of co--roots. Let also $\varepsilon(w)$ be the parity of a Weyl transformation.
Then for all integers $j$ in $\N^*$ and all $p$ in $B_n$, we claim that the
following relations hold 
\beq
{\cal F}_{p}^{[n]}(q) = \sum_{\scriptstyle w \in \widehat W_n \atop 
\scriptstyle w(p) \in B_{jn}} \varepsilon (w) \, {\cal F}_{w(p)}^{[jn]}(q^j).
\label{relation}
\eeq
The proof is easy. First of all, the formula
(\ref{weylkac}) allows to extend the functions $\chi_p$ to the whole weight
lattice, but one may check that $\chi_{p+n\tilde R}=\chi_p$ and
$\chi_{w(p)}=\varepsilon(w) \chi_p$ for any (finite) Weyl transformation, where
$\varepsilon(w)$ is the parity of the transformation. In particular, if $p$ lies
on a boundary of $B_n$, one has $\chi_p(q)=0$ identically. Obviously, the
functions ${\cal F}_p^{[n]}$ have the same properties. The sum over
$(a,b)-(r,s) \in n\tilde R$ can be split into a sum over the classes $\{n\tilde
r + nj\tilde R\}$ for $\tilde r \in \tilde R/j\tilde R$, from which it
follows that
\beq
{\cal F}_{p}^{[n]}(q) = \sum_{\tilde r \in \tilde R/j\tilde R} 
{\cal F}_{p+n\tilde r}^{[jn]}(q^j) = \sum_{\tilde r \in \tilde 
R/j\tilde R} \varepsilon(w_{\tilde r})\,{\cal F}_{w_{\tilde r}(p+n\tilde
r)}^{[jn]}(q^j).
\eeq
The last equality is proved by observing that, although $p+n\tilde r$ is not
in $B_{jn}$, there exists a Weyl transformation $w_{\tilde r}$ in $\widehat
W_{jn}$ such that $w_{\tilde r}(p+n \tilde r)$ is in $B_{jn}$. Since $\widehat
W_{jn} \subset \widehat W_n$, the $j^2$ weights $w_{\tilde r}(p+n\tilde r)$ are
images of $p$ under affine Weyl transformations of $\widehat W_n$. Conversely,
the intersection $\widehat W_n(p)
\cap B_{jn}$ is precisely equal to these weights, and the formula
(\ref{relation}) follows. 

Another straightforward consequence of (\ref{weylkac}) is the identity
\beq
{\cal F}_{jp}^{[jn]}(q) = j^3 \, {\cal F}_p^{[n]} (q^j).
\eeq
Combined with (\ref{relation}), it leads to an identity for the characters
\beq
\chi_p^{[n]}(q) = {1 \over j^3} \sum_{\scriptstyle w \in \widehat W_n \atop 
\scriptstyle w(p) \in B_{jn}} \varepsilon(w) \chi_{jw(p)}^{[j^2n]}(q),
\qquad \forall j \in \N^*.
\eeq
These formulas, written here for $su(3)$, have a strict analogue in more
general algebras, and constitute the generalization of relations that have
appeared in \cite{ciz} in the case of $su(2)$. 

As illustration, we write the relations for $j=2$ and $j=3$ in the case of
$su(3)$:
\bea
&& 8\chi_{(r,s)}^{[n]} = \chi_{(2r,2s)}^{[4n]} + \chi_{(2s+2n,2t)}^{[4n]} +
\chi_{(2t,2r+2n)}^{[4n]} - \chi_{(2n-2s,2n-2r)}^{[4n]}, \\
\noalign{\smallskip}
&& 27\chi_{(r,s)}^{[n]} = \chi_{(3r,3s)}^{[9n]} + \chi_{(3s+3n,3t)}^{[9n]} +
\chi_{(3t,3r+3n)}^{[9n]} + \chi_{(3r+3n,3s+3n)}^{[9n]} +
\chi_{(3s,3t+6n)}^{[9n]} \nonumber\\ 
&& \hskip 1.7truecm + \chi_{(3t+6n,3r)}^{[9n]} - \chi_{(3n-3s,3n-3r)}^{[9n]} -
\chi_{(3n-3r,6n-3t)}^{[9n]} - \chi_{(6n-3t,3n-3s)}^{[9n]}. \qquad \qquad
\eea
Setting $n=3$ and using $\chi_{(1,1)}^{[3]}=1$, one obtains linear relations
between affine characters and the constant function. It is amusing to note that
the above relation for $j=2$ and $n=3$ is precisely the one Moore and Seiberg
used to discover the exceptional $su(3)$ modular invariant at height $k+3=12$
\cite{moosei}, namely
\beq
\chi_{(2,2,8)} + \chi_{(2,8,2)} + \chi_{(8,2,2)} - \chi_{(4,4,4)} = 8.
\eeq

\subsection{Modular invariance}

Besides their group theoretical importance, the characters are intimately related
to the partition function of physical models on Riemann surfaces. The simplest
and by now classical case, namely tori, has been first considered in
\cite{cardy}. There it was shown that the partition function of a rational
conformal field theory put on a torus $\C/\z + \tau \z$ of modulus $\tau$, has
the general form
\beq 
Z(\tau) = {\rm Tr}_{\cal H}\big(q^{L_0-c/24} \otimes \overline 
q^{\overline L_0-c/24}\big),
\eeq
where the complex number $q$ is related to the modulus of the torus by
$q=e^{2i\pi \tau}$. The trace is taken over the Hilbert space of the model, and
$\overline q$ is the complex conjugate of $q$. In virtue of the decomposition
(\ref{hilbert}) ---it is completely general, just insert the representations of
whatever the symmetry algebra $\cal A$ is---, one obtains
\beq
Z(\tau) = \sum_{p,p'} N_{p,p'}\,\chi_p(q) \, \chi_{p'}(\overline q).
\label{partition}
\eeq
The problem of modular invariance stems from the fact that a same torus may be
given in terms of a whole class of moduli, namely all $\tau$ related by 
$\psl$ transformations, also known as modular transformations, in fact specify a
single torus. Which representative one chooses in this class should not affect
the physical partition function, and as a consequence, it must be modular
invariant, $Z(\tau)=Z({a\tau + b \over c\tau + d})$. It is actually sufficient 
to check the invariance of the partition function under $S\,:\,\tau \rightarrow
{-1 \over \tau}$ and $T\,:\,\tau \rightarrow \tau+1$, since together they
generate the whole of $\psl = \langle S,T \,|\, S^2,(ST)^3 \rangle$. This is what
the modular invariance (on the torus) requires: to check that the partition
function satisfies
\beq
Z(\tau) = Z(\tau + 1) = Z({-1 \over \tau}).
\eeq
But in fact this argument can be turned around. Since the partition function
must have the general form (\ref{partition}), the modular invariance constrains
the choice of the integers $N_{p,p'}$, and hence the model itself. This is how
the criterion of modular invariance led to the possibility of classifying the
consistent (candidates of) conformal theories. It turns out that the modular
invariance is a fantastically strong constraint, as very few choices 
of integers $N_{p,p'}$ lead to modular invariant partition functions. In the
first case for which the classification has been carried out, namely ${\cal A} =
\widehat{su(2)_k} \otimes_I \widehat{su(2)_k}$, unexpected connections emerged
with other mathematical areas. Indeed the results showed that the list of
$su(2)$ modular invariants is isomorphic to the list of simply--laced simple
complex Lie algebras ADE (or equivalently to the list of finite subgroups of
$SO(3)$) \cite{ciz}. This surprising correspondence has remained largely
mysterious (see \cite{nahm} however), but prompted further investigations. As 
far as affine Lie algebras are concerned, the next case is
$\widehat{su(3)}$. Here too the complete list is known for all levels
\cite{gan1}, but it shows no obvious pattern. An attempt to link the structure
of modular invariants (for $su(3)$ and more general cases) to graphs has been
made in \cite{dizu}. Based on technical similarities, another connection was
suggested in \cite{rtw}, which relates the affine $su(3)$ modular problem to 
the geometry of the complex Fermat curves. This connection is precisely the
problem we want to address in this article. 

It would probably be inspiring to see the solution to the modular problem for
higher rank affine simple Lie algebras, but no complete list is known beyond
rank 2. Partial results for affine algebras include: all simple algebras at
level 1 \cite{gan2}, all $su(N)$ algebras at level 2 and 3 \cite{gan3},
products of $su(2)$ factors with the restriction $\gcd(k_i+2,k_j+2) \leq 3$
(except for a product of two factors for which the classification is complete)
\cite{gan4}. Other approaches to the classification problem have produced
complete lists of modular invariants of specific types \cite{gan4,grw,gan5}.

One may check the modular invariance of $Z(\tau)$ by looking at the way the
affine characters transform \cite{kac}. From the general form (\ref{char})
of the characters, the transformation under $T$ is easy to compute, while that 
under $S$ can be obtained from the Poisson formula. The results show that the
characters $\chi_p$, for $p \in B_n$, transform linearly under a modular
transformation, $\chi_p(X\tau)=\sum_{p'}\,X_{p',p}\,\chi_{p'}(\tau)$ for $X$ in
$\psl$. For $su(3)$ the explicit matrices representing $T$ and $S$ read
\bea
T_{p',p} &=& e^{2i\pi (h_p-c_k/24)}\,\delta_{p,p'} = \xi^{r^2+s^2+rs-n}\,
\delta_{p,p'},\\
S_{p',p} &=& {-i \over n\sqrt{3}} \sum_{w \in W} \varepsilon(w) e^{-2i\pi 
p \cdot w(p')/n} \nonumber\\
&=& {-i \over n\sqrt{3}} \xi^{-(2rr'+2ss'+rs'+r's)} \Big(
1 + \zeta^{tt'-rs'} + \zeta^{tt'-r's} - \zeta^{rr'} - \zeta^{ss'} 
-\zeta^{tt'}\Big), \label{smatrix}
\eea
with $\xi=e^{2i\pi /3n}$ and $\zeta=\xi^3$. The two matrices are symmetric and
unitary, and satisfy $S^2=(ST)^3=C$ with $C$ the charge conjugation, so
that $S$ and $T$ generate a representation of $\sl$ rather than
$\psl$. It has been proved in \cite{kac,bi} that the kernel of this 
representation is of finite index in $\sl$, for any value of $n$, and is even 
contained in some principal congruence subgroup, but a precise description 
of these kernels is still lacking. An obvious relation is $T^{3n}=1$, and it 
is not difficult to see that no smaller power of $T$ equals 1 (except for
$n=3$). One can also show that for  $n \geq 5$, no power $T^a$ for $a<3n$ has
all its eigenvalues equal because this would mean that $a(3-n),a(7-n),a(12-n)$
are equal modulo $3n$, which implies that $3n$ divides $a$. On the other hand,
$T^3$ is central for $n=4$.

Inserting the modular transformations of the characters into the partition
function, one finds that $Z(\tau)$ is modular invariant iff the matrix
$N_{p,p'}$ satisfies $TNT^\dagger = SNS^\dagger = N$, or, by using the unitarity
of $S$ and $T$, iff $N$ is in the commutant of the representation of $\psl$
carried by the characters \footnote{The partition function is given in terms of
the specialized characters, on which the charge conjugation $C$ is trivial.}
\beq
\hbox{$Z(\tau)$ modular invariant} \qquad \Longleftrightarrow \qquad [N,T] =
[N,S] = 0.
\eeq
The commutant of $S$ and $T$, without imposing the positivity condition
$N_{p,p'}\geq 0$, has been worked out in full generality for the affine Lie
algebras of the $su(N)$--series \cite{bi}, but the results extend trivially
to all algebras. It was found that the commutant over $\C$ actually has a basis
of matrices with coefficients in $\Q$, and also that this commutant is rather
big. Its dimension is an arithmetic function, growing roughly like $n^{2N-5}/N!$
for $su(N)$ at level $k=n-N$ \cite{pr}. In view of the fact that so few modular
invariant partition functions satisfy it, it shows that the positivity condition
is really the crucial one, and also the most difficult to handle. Recent
developments have shown that the most efficient way of dealing simultaneously
with the commutation and with the positivity conditions is to use Galois
theory techniques, which beautifully combine the algebraic nature of $S$ with
the rational character of $N_{p,p'}$. Before we review these aspects in the next
section, we mention a last feature of the modular matrices $S$ and $T$.

When $n$ is coprime with 3, $S$ and $T$ have a property which is useful
in actual calculations, namely they can be written as tensor
products. From the  above formulas, one may check that under the cyclic rotation
$\mu(r,s,t)=(t,r,s)$, an automorphism of the extended Dynkin diagram of
$su(3)$, one has, for $\omega=e^{2i\pi /3}$
\beq
T_{\mu(p),\mu(p)} = \omega^{n-r-2s} \, T_{p,p}\,, \qquad 
S_{p,\mu(p')} = \omega^{r+2s} \, S_{p,p'}.
\eeq
The quantity $r+2s$ taken modulo 3 is the triality of $p$. When $n$ and 3 are
coprime, $\mu$ acts on the weights of $B_n$ without fixed points. Thus if
one splits $B_n$ into orbits under the action of $\mu$ by writing $p=\mu^k(r)$
for $k=0,1,2$ and $r \in B_n/ \langle \mu \rangle$ of zero triality, one obtains
\beq
T_{\mu^k(r),\mu^{k'}(r')} = \Big(\omega^{k^2n}\delta_{k,k'}\Big) \, 
T_{r,r'},\qquad
S_{\mu^k(r),\mu^{k'}(r')} = \omega^{nkk'} \, S_{r,r'}.
\eeq
The same property holds in $su(N)$, level $k$, whenever $N$ and $n=N+k$ are
coprime.
 
\subsection{Galois and parity selection rules}

Perhaps the most remarkable property of the modular matrices $S$ and $T$ is
that they are rational combinations of roots of unity, in this case $3n$--roots 
of unity \cite{buffenoir}. Even more remarkable is the fact that this situation
is completely general: any RCFT has matrices $S$ and $T$ that have their
coefficients in cyclotomic extensions of finite degree over $\Q$. For $T$, it
follows from the fact that in a RCFT, the Virasoro central charge $c$ and all
conformal weights are rational numbers \cite{rational}. The corresponding
result for $S$ has been proved in \cite{cg}, whose authors built on results 
from \cite{deBG}.

Let us first fix our notations concerning cyclotomic extensions. For $\zeta_m =
e^{2i\pi /m}$, we will denote by $\Q(\zeta_m)$ the cyclotomic extension of the
rationals by $m$--roots of unity, of degree $\varphi(m)$, the Euler totient
function, over $\Q$. Its Galois group Gal$(\Q(\zeta_m)/\Q)$ consists of the
automorphisms $\sigma_h(\zeta_m) = \zeta_m^h$ for all integers $h$ between 1 and
$m$, coprime with $m$. The Galois group is Abelian, isomorphic to $\Z {m} =
\left(\z/m\z\right)^*$, the group of invertible integers modulo $m$. 

The Galois automorphisms of the algebraic extension where the coefficients of
$S$ lie, have important consequences for the modular problem, which we
now summarize. Each element of the Abelian Galois group of the relevant 
extension induces the unique permutation of the weights of the alc\^ove $\sigma
\;:\; p \rightarrow \sigma(p)$ (we keep the same name for the element of the
Galois group and for the induced permutation), such that
\beq
\sigma (S_{p,p'}) = \e{p} S_{\sigma(p),p'} = \e{p'} S_{p,\sigma(p')},
\eeq
where $\e{p}=\pm 1$ is a cocycle satisfying $\varepsilon_{\sigma\sigma '}(p)=
\e{p} \varepsilon_{\sigma '}(\sigma(p))$. Acting with $\sigma$ on the
commutation relation $[N,S]=0$ and using the fact that the coefficients
$N_{p,p'}$ are rational numbers, one obtains that $N$ must satisfy \cite{cg}
\beq
N_{\sigma(p),\sigma(p')} = \e{p} \e{p'} \, N_{p,p'}, \qquad \hbox{for all
$\sigma$}.
\eeq
This equation is a necessary condition for $N$ to commute with $S$. Its
importance for the modular problem is obvious. Since the entries
of $N$ are to be non--negative integers, it leads to the selection rule:
\beq
N_{p,p'} = 0 \ \hbox{ as soon as there is a $\sigma$ for which $\e{p}
\e{p'}=-1$}.
\label{selection}
\eeq
Its utility is two--fold. First, it turns out to be extremely restrictive,
forcing most of the coefficients to vanish. Even though in actual cases, it may
not be easy to determine which coefficients may or may not vanish, it still
remains much easier than the commutation problem. Second, it facilitates
enormously computer searches, because in the examples we know, which include
all affine algebras, to check the sign of $\e{p} \e{p'}$ is computationally
trivial. The first use of (\ref{selection}) was made in the restricted context
of $su(3)$, $n$ prime \cite{rtw2}, where however neither its generality nor its
Galoisian origin were recognized. They were later generalized to all affine
algebras in \cite{gan2,rtw}, and eventually to all RCFT in \cite{cg} where the
Galoisian nature of the result was transparently brought out.

 From the formula (\ref{smatrix}) for the matrix $S$, in which one may neglect 
the prefactor $-i/ n\sqrt{3}$ since one is interested in commuting $N$ with
$S$, the action of the Galois automorphism $\sigma_h$ amounts formally to
multiply the weight $p$ (or $p'$) by $h$: $\sigma_h(S_{p,p'}) = S_{hp,p'}$. 
This action of $\sigma_h$ is only formal since in general $hp$ is not in the
alc\^ove $B_n$. However one can show that $h$ being invertible modulo $n$
ensures there is a unique affine Weyl transformation which maps $hp$ on some
weight $\sigma_h(p)$ of $B_n$, so we can write 
\beq
\sigma_h(p) = w_{h,p}(hp) + n\alpha, \qquad w \in W,\; \alpha \in \tilde R.
\label{perm}
\eeq
One then obtains from (\ref{smatrix})
\beq
\sigma_h(n\sqrt{-3}S_{p,p'}) = \sum_{w \in W} \varepsilon(w) e^{-2i\pi 
\sigma_h(p) \cdot (w_{h,p}\circ w)(p')/n} = \varepsilon (w_{h,p}) \,
(n\sqrt{-3} S_{\sigma_h(p),p'}).
\eeq
Therefore the permutation of $B_n$ induced by an element of the Galois group is
given in (\ref{perm}), and the cocycle is just the parity of the Weyl
transformation defining the permutation, $\varepsilon_{\sigma_h}(p) = 
\varepsilon (w_{h,p})$, up to the sign $\sigma_h(\sqrt{-3}) \over \sqrt{-3}$,
that only depends on $h$. The same is true of any affine algebra. For that
reason, the cocycles have been termed ``parities'' in the literature. 

We finish this section by showing how the parities can be computed in the case
of $su(3)$. The general algorithm for computing both $\sigma_h(p)$ and
$\varepsilon_{\sigma_h}(p)$ in the $su(N)$ series has been given in
\cite{rtw}. The sign $\varepsilon_{\sigma_h}(p)=\pm 1$ is the signature of the
Weyl transformation which maps the weight $hp$ back in the alc\^ove. By
extension, one can assign all weights a parity $\varepsilon(p)$, which
is just the signature of the Weyl transformation which maps $p$ back in the
alc\^ove. It is well--defined only for those weights which do not lie on the
affine walls, since they would be fixed points of odd Weyl transformations. For
$su(3)$ it means that the parity of $p=(r,s,t)$ is well--defined iff $r,s,t \neq
0 \bmod n$. If $p$ is in a wall, we set $\varepsilon(p)=0$. We have
$\varepsilon(p)=+1$ for all $p$ in $B_n$. A translation by
$n\alpha$, $\alpha$ a co--root, being even, the parity does not change under
such translations, $\varepsilon(p+n\alpha) = \varepsilon(p)$, so that we may
restrict our attention to the six triangles obtained from $B_n$ by the action of
the finite Weyl group. Up to translations by elements of $n\tilde R$, the even
Weyl transformations map $B_n$ onto
\bea
&& B_n = \{(r,s) \;:\; r,s \geq 1, \, r+s \leq n-1\},\\ 
&& w_1w_2(B_n) + n(\alpha_1 + \alpha_2) = \{(n+s,n-r-s)\},\\
&& w_2w_1(B_n) + n(\alpha_1 + \alpha_2) = \{(n-r-s,n+r)\}.
\eea
We note that if $p$ is in $B_n$, then $p+(n,0)$ and $p+(0,n)$ are respectively
in the second and third triangle, so we conclude that the parity of $p=(r,s)$
depends only on the residue $(\rest {r},\rest {s})$ modulo $n$ of $p$. From the
above discussion, $\rest {r},\rest {s}$ and $\rest {r+s}$ are all
different from zero modulo $n$ for any weight which is not in an affine wall.
If we take the residues $\rest {\cdot}$ in $[0,n-1]$, two possibilities remain.
Either $\rest {r}+\rest {s} < n$, in which case the parity $\varepsilon(p)=+1$
since the weights in $B_n$ satisfy this inequality, or else $\rest {r}+\rest
{s} > n$ and $\varepsilon(p)=-1$. In the first case, $\rest {t} = \rest{n-r-s}
= n-\rest{r}-\rest{s}$, while in the second case, $\rest{t}=2n-\rest{r}-
\rest{s}$. Putting all together, one obtains the parity function
\beq
\varepsilon(p) = \varepsilon(r,s,t) = \cases{
0 & if $\rest{r}=0$ or $\rest{s}=0$ or $\rest{t}=0$, \cr
+1 & if $\rest{r}+\rest{s}+\rest{t}=n$, \cr
-1 & if  $\rest{r}+\rest{s}+\rest{t}=2n$. \cr}
\label{parity}
\eeq

Let us summarize the $su(3)$ parity selection rules. With each Galois
automorphism $\sigma_h$ is associated an integer $h$, coprime with $3n$. Given
a weight $p$ in the alc\^ove $B_n$, we compute for each $h$ the parity
$\varepsilon(hp)$ from the formula (\ref{parity}). The parity depends only
on the residue of $hp$ modulo $n$, so we may take $h$ between 1 and $n$. In this
way, we obtain a finite sequence $\{\varepsilon(hp)=\pm 1\}_h$. The parity
selection rules then say that the coefficient $N_{p,p'}$ in the modular
invariant partition function may be non--zero only if the two sequences 
$\{\varepsilon(hp)\}_h$ and $\{\varepsilon(hp')\}_h$ are equal componentwise.
Equivalently, if we collect the $h$'s for which $\varepsilon(hp)=+1$ by defining
\beq 
H_p = H_{r,s,t} =\{h \in \Z {n} \;:\; \rest {hr}+\rest {hs}+\rest {ht}=n \},
\label{defH} 
\eeq
the selection rules imply
\beq
H_p \neq H_{p'} \quad \Longrightarrow \quad N_{p,p'}=0.
\eeq

In this form, the parity condition appears in a completely different context,
namely the study of the complex Fermat curves, of which it governs the
decomposition.

\section{Fermat curves}

\noindent
The parity rule is extremely powerful for the problem of modular
invariance. It is a sufficient condition for $N$, the matrix specifying
a modular invariant, to commute with $S$, and is not concerned at all with the
commutation with $T$. Hence fulfilling the parity rule does not involve the full
complexity of finding the commutant of $S$ (let alone of $S$ and $T$), but at
the same time is constraining enough to encapsulate much of the structure of the
commutant. On the practical side, this makes it a prime tool, as witnessed
by the latest developments \cite{gan3,schel1,schel2}, while conceptually, its
Galoisian origin and its universality \cite{cg} also yielded a renewed viewpoint.

As noted in \cite{rtw}, the parity rule for $\widehat{su(3)}$ is very peculiar
as it has also a key r\^ole in the understanding of the geometry of the Fermat
curves, and more specifically, in the decomposition of the Jacobian of the
Fermat curves into simple factors \cite{kr}. There is no apparent reason for
this, and whether this relationship is deep or accidental was the original
motivation for our investigations. There is a priori no indication whatsoever 
why the Fermat curves should have anything to do with the partition functions
of $\widehat{su(3)}$ CFT's on tori (other curves have, as we shall see in the
fifth section). It is the purpose of this section to merely describe the
connection. We follow the original or standard material available in the
mathematics literature, with a presentation which has no claim to rigor and
directed towards the application at hand. We refer to the original articles for
further (and perhaps more accurate) details.

\subsection{Abelian varieties}

\noindent
A complex Abelian variety of dimension $g$ is a complex
torus ${\Bbb C}^g/L$ equipped with a Riemann form \cite{bost}. $L$
is a lattice (a discrete free Abelian group of rank $2g$ over $\z$), and the
existence of a Riemann form means that there is a positive definite hermitian
form on $\Bbb C^g$, of which the imaginary part takes integral values on $L$. A
prime example, though not generic, of an Abelian variety is the Jacobian of a
Riemann surface. As this example is most relevant to us, we will describe it in
more detail.

If $\S$ is a compact Riemann surface of genus $g$, it is well--known that the
homology group of $\S$ has $2g$ independent cycles $\ga_i$, and that the vector
space of holomorphic 1--forms has dimension equal to $g$. A period of $\S$ is 
the $g$--uple $(\oint_\ga \,\om_1,\ldots,\oint_\ga \,\om_g)$ for some cycle
$\ga$, where the $\om_i$ form a basis for the holomorphic differentials. The
period lattice is the collection of all periods
\beq
L(\S) = \left\{\Big(\oint_\ga \om_1,\oint_\ga \om_2,\ldots,
\oint_\ga \om_g \Big) \;:\; \ga = \sum_i n_i\ga_i \in H_1(\S,\Bbb Z)
\right\} \subset \Bbb C^g.
\eeq
For any fixed point $P_0$ on the surface, it follows that the map (called the
Abel--Jacobi map)
\beq
P \in \S \quad \longmapsto \quad J(P)=\Big(\int_{P_0}^P \,\om_1,\int_{P_0}^P
\,\om_2\ldots,\int_{P_0}^P \,\om_g\Big)
\eeq
is well--defined modulo the periods ({\em i.e.} does not depend on the path
from $P_0$ to $P$), and provides an embedding of the surface into the factor
group $\jac(\S)={\Bbb C}^g/L(\S)$, the Jacobian of $\S$. Clearly, for $g>1$,
the map $J(P)$ is only an embedding, but if we extend the map $J$ to $J_g$ by
setting
\beq
\vec P=(P_1,P_2,\ldots,P_g)  \quad \longmapsto \quad J_g(\vec P)=J(P_1) +
J(P_2) + \ldots + J(P_g) \in \jac(\S),
\eeq
then a fundamental result of Riemann, anticipated and proved in specific cases
by Jacobi, asserts that the map $J_g$ is invertible for ``generic'' points
$\vec P$ in the $g$--th symmetric power of $\S$, ${\rm Sym}_g\,\S=\S^g/S_g$
($S_g$ is the permutation group on $g$ letters). Torelli theorem then shows that
the isomorphism class of the Jacobian in fact determines that of the Riemann
surface. Thus the Jacobian captures the essential features of the surface, and, 
being an affine space, provides a kind of linearization of it. Taking advantage
of that, attention is sometimes focussed on the Jacobians rather than on the
surfaces themselves. For elliptic curves ($g=1$), this is what one is used to, as
the curve is isomorphic to its Jacobian, usually described as a parallelogram
with sides 1 and $\tau$.

An important notion in the study of Abelian varieties is that of isogeny
\cite{swinn,lang2}. A map $\phi \,:\, A \rightarrow B$ is an isogeny if it is a
surjective homomorphism with finite kernel. Isogenies go both ways: if $\phi$ is
an isogeny from $A$ to $B$, there exists another one $\hat\phi$ from $B$ to $A$.
When there are isogenies between them, we say that $A$ and $B$ are isogenous and
write $A \sim B$ (if $A$ and $B$ are Jacobians of algebraic curves, we say by
extension that the two curves are isogenous). Being isogenous is an equivalence
relation. For what follows, it may be convenient to rephrase these properties
in terms of lattices. If we view Abelian varieties as complex tori, say
$A={\Bbb C}^g/L_A$ and $B={\Bbb C}^g/L_B$, an equivalent definition is that $A$
is isogenous to $B$ if and only if there is a complex linear map $\psi$ such
that $\psi(L_A) \subset L_B$ with finite index, say $m$. If this is the case,
one has $m L_B \subset L_A$, which explicitly displays an isogeny from $B$ to
$A$, and shows that isogenies define an equivalence relation (reflexivity and
transitivity are trivial). We then say that the lattices are isogenous,
$L_A \sim L_B$. For instance, in the elliptic case for which the complex lattice
$L$ can be written $L(\tau)=\{a\om_1 + b\om_2 \,:\, a,b \in {\Bbb Z}\}$ with
$\tau=\om_2/\om_1 \not\in \Bbb R$, two lattices $L(\tau) \sim L(\tau')$ are
isogenous if and only if $\tau'={a\tau +b \over c\tau +d}$ for some $\left({a
\atop c}\, {b \atop d}\right)$ in $GL_2(\Bbb Q)$.

The following results show the importance of isogenies.
It may happen that an Abelian variety $A$ contain a non--trivial Abelian
subvariety $A_1$. If $A=\C^g/L_A$, it means that there is a complex vector
space $V_1=\C^h \subset \C^g$ such that $V_1 \cap L_A \equiv L_{A_1}$ is a
lattice in $V_1$ (of rank $2h$). Then the orthocomplement of $V_1$ with respect
to the Riemann form, call it $V_2$, has the same property: $V_2 \cap L_A \equiv
L_{A_2}$ is a lattice in $V_2$ (of rank $2g-2h$). Hence $L_{A_1} \oplus
L_{A_2}$ is of finite index in $L_A$, and $A$ is isogenous to $V_1/L_{A_1}
\times V_2/L_{A_2} \equiv A_1 \times A_2$. Moreover, the Riemann form on $A$
induces by restriction a Riemann form on $A_1$ and $A_2$, so that they are
themselves Abelian varieties. (Note that $A$ being a Jacobian does not
imply that $A_1$ and $A_2$ are Jacobians.) Repeating this decomposition process
as many times as possible, one eventually finds that an Abelian variety is
isogenous to the product of simple Abelian varieties, where simple means that
they contain no proper complex torus. This is the complete reducibility theorem
\cite{swinn,lang2}, due to Poincar\'e. Moreover this decomposition is unique 
up to isogenies.

Decomposing into simple factors the Jacobians of the Fermat curves, defined in
affine coordinates by
\beq
F_n \;:\;\; x^n + y^n = 1, \qquad \hbox{$n$ integer},
\eeq
was precisely the purpose of, first, Koblitz and Rohrlich \cite{kr}, who
partially resolved it, and then of Aoki \cite{aoki}. We are now in position to
detail their work, and the relation to the problem of modular invariance for
$su(3)$.

\subsection{Jacobians}

\noindent
The periods of the Fermat curves have been computed by Rohrlich in \cite{rohr}.
A basis for the holomorphic differentials on $F_n$ is obtained by taking 
$\om_{r,s,t} = \alpha_{r,s,t} \, x^{r-1}y^{s-n}{\rm d}x$, for all admissible
triplets $(r,s,t)$, {\it i.e.} those such that $0<r,s,t<n$ and $r+s+t=n$ (see
also Section 4.2). Its dimension equals the genus of $F_n$, namely $(n-1)(n-2)
\over 2$. This is also the cardinality of the fundamental alc\^ove for
$\widehat{su(3)}$, height $n$. A suitable choice for the normalization constants
$\alpha_{r,s,t}$ yields the following result for the integration of the
differentials along closed curves,
\beq
\oint_{\ga_{i,j}} \; \om_{r,s,t} = \ze_n^{ri+sj}, \qquad \hbox{ $1 \leq i,j
\leq n$,}
\label{period}
\eeq
where $\{\ga_{i,j}\}_{1 \leq i,j \leq n}$ is a generating set of closed loops
\cite{rohr}. Every cycle in $H_1(F_n,{\Bbb Z})$ can be written
$\ga=\sum_{i,j} m_{i,j}\,\ga_{i,j}$, so the period lattice of the $n$--th Fermat
curve is
\beq
L(F_n) = \left\{\Big( \ldots\, , \;\sum_{i,j} m_{i,j} \;\ze_n^{ri+sj}, \,\ldots
\Big)_{0<r,s,t<n} \;:\; m_{i,j} \in {\Bbb Z} \right\}.
\eeq

When all $m_{i,j}$ are varied over $\Bbb Z$, it is clear that the $(r,s,t)$--th
component of the period lattice covers the whole of ${\Bbb Z}(\ze_{n_0})$,
for $n_0$ defined by $\gcd(r,s,t)={n \over n_0}$. If however two
triplets are related by $(r',s',t')=(\rest {hr},\rest {hs},\rest {ht})$ for
some $h \in \Z {n_0}$, the $(r',s',t')$--th component of $L(F_n)$ is just the
Galois transform by $\sigma_h$ of the $(r,s,t)$--th component, so that the two
are not independent. The triplet $(\rest {hr},\rest {hs},\rest {ht})$
is admissible if $h$ is in the set \footnote{If $n_0 < n$, the set defined in
(\ref{defH}) is the trivial extension modulo $n$ of the set defined here.}
\beq
\hrst = \{h \in \Z {n_0} \;:\; \rest {hr} + \rest {hs} + \rest {ht} = n\}.
\label{hrst}
\eeq
We saw in Section 2 that $\hrst$ was crucial for the $su(3)$ parity rule, and
in fact we shall see that it also governs the decomposition of the Jacobian of
the Fermat curves. For the moment we note that $h \in \hrst$ is equivalent to
$-h \not\in \hrst$, so $\hrst$ is a set of representatives of $\Z
{n_0}/\{\pm 1\}$.

If $\{e_{r,s,t} \;:\; {\rm admiss.} \, (r,s,t)\}$ is the canonical basis of
${\Bbb C}^g$, a simple reordering leads to the following writing
\beq
{\Bbb C}^g = \bigoplus_{{\rm admiss.}\, (r,s,t)} {\Bbb C} \, e_{r,s,t} =
\bigoplus_{[r,s,t]} \; \bigoplus_{h \in H_{r,s,t}} {\Bbb C} \,
e_{\rest {hr},\rest {hs},\rest {ht}},
\eeq
where $[r,s,t]$ is the class $\{(\rest {hr},\rest {hs},\rest {ht}) \,:\, h \in
\hrst \}$. Using the same reordering on the period lattice, one easily sees that
\beq
L(F_n) \subset \bigoplus_{[r,s,t]} \; \lrst,
\label{incl}
\eeq
where
\beq
\lrst = \left\{\big(\ldots\,,\sigma_h(z),\,\ldots\big)_{h \in \hrst}
\;:\; z \in \z (\zeta_{n_0}) \right\}.
\eeq
Note that the r.h.s. of (\ref{incl}) is well--defined because $L_{r,s,t} =
L_{r',s',t'}$ if the two triplets belong to the same class $[r,s,t]$ (a
consequence of $H_{\rest {hr},\rest {hs},\rest {ht}}=h^{-1}\hrst$). The
inclusion (\ref{incl}) holds with finite index, since both lattices have same
rank over $\z$,
\beq
(n-1)(n-2) = \sum_{[r,s,t] \in B_n} \varphi(n_0).
\eeq
Consequently the period lattice $L(F_n)$ is isogenous to the direct sum
$\oplus \lrst$, and from this follows the isogeny \cite{rohr}
\beq
\jac(F_n) = {\Bbb C}^g/L(F_n) \sim \prod_{[r,s,t]} \; \Big(
{\Bbb C}^{\varphi(n_0)/2} / \lrst \Big).
\label{jac1}
\eeq
This shows that the (Jacobian of the) curve $F_n$ is far from being simple, but
has a number of factors increasing (at least) linearly with $n$. It is not
difficult to compute the number of factors in (\ref{jac1}). For $n=\prod p^k$,
one finds
\beq
\hbox{\# classes } [r,s,t] = \prod_p \,[\sigma_1(p^k)+\sigma_1(p^{k-1})] -
3\sigma_0(n) + 2,
\eeq
where $\sigma_k(n)$ is the sum of the $k$--th powers of all divisors of $n$
(including 1 and $n$). If $n$ is prime, there are $(n-2)$ classes, which
can be chosen as $[1,s,n-1-s]$ for $s=1,2,\ldots,n-2$. Other particular values
for the number of classes are 1,\,3,\,10,\,12,\,34,\,88 for
$n=3,\,4,\,6,\,8,\,12,\,24$ respectively.

\smallskip
Let us mention that in case $n$ is a prime number, Weil has shown that $\lrst$ 
is in fact the period lattice of the following curve \cite{weil}
\beq
\crst(n) \;:\;\; v^n = u^r(1-u)^s.
\eeq
This is not true in general as the genus of (the irreducible part of)
$\crst(n)$, equal to
\beq
g(\crst(n)) = {n_0-1 \over 2} - {1 \over 2} \big[\gcd(n_0,r) + \gcd(n_0,s)
+ \gcd(n_0,t) - 3 \big],
\eeq
is generically different from ${1 \over 2}\varphi(n_0)$. What is true for
general $n$ is that $\crst(n)$ is the image of $F_n$ under the rational map
$(x,y) \mapsto (u,v)=(x^n,x^r y^s)$, so that the Jacobian of $\crst(n)$ is 
contained in that of $F_n$ \cite{rohr}. There is also a rational map from
$F_n$ to $F_d$ for every divisor $d$ of $n$ (namely the $n/d$--th power map),
--- implying in particular $\jac(F_d) \subset \jac(F_n)$, see {\it e.g.}
(\ref{div24}) below --- so altogether there is a sequence of rational maps
\beq
F_n \longrightarrow \{F_d\} \longrightarrow \{\crst(d)\}, \qquad \hbox{for
any}\, d|n.
\label{maps}
\eeq

The curves $\crst(n)$ have been extensively discussed in \cite{auritz} in the
context of rational billiards. There $C_{r,s,t}(n)$ was associated (through a
Schwarz transformation) with the rational triangle of angles ${r\pi \over
n},{s\pi \over n},{t\pi \over n}$. For that reason, we call them triangular
curves. We will come back to them in Section 4.3, where we will show that 
some of the triangular curves which are rational images of $F_{24}$
are intimately related to the exceptional modular invariants of $su(3)$,
occurring at $n=8,12$ and 24.

\subsection{Complex multiplication}

\noindent
The main result of the previous section was the decomposition (\ref{jac1})
of the Jacobian of $F_n$. The question that remains is whether this
decomposition is complete. 

Let $A=\C^n/L$ be an Abelian variety. The endomorphisms of $A$, denoted by
$\endo (A)$, are the complex endomorphisms of $\C^n$ fixing the lattice $L$, and
have a ring structure. It is clear that $\endo (A)$ contains $\z$, realized as
the multiplication of the elements of $A$ by integers, and that $\z$ is central
in $\endo (A)$. One may broaden the class of transformations and consider
endomorphisms of $A$ up to isogenies, or equivalently endomorphisms of the
isogeny class of $A$, therefore allowing for arbitrary rational factors in the
transformations of $A$ into $A$. This one can do by defining $\endoq (A) = \endo
(A) \otimes \Q$, which is then isogeny invariant. One has now that $\Q$ is
in the center of $\endoq (A)$, but it may happen that the center be larger
than $\Q$. If this is the case, one can show that it is necessarily a number
field $F$, which is either totally real (all its embeddings in $\C$ lie in
$\R$), or else is a totally imaginary quadratic extension of a totally real
number field ($F$ has no embedding in $\R$). In the second case, $F$ is called 
a CM field and $A$ is said to have complex multiplication by $F$
\footnote{Sometimes a more restrictive definition is used, which requires
in addition that the degree of $F$ be twice the dimension of $A$.}.

If $F$ is a CM field of degree $2n$ over $\Q$, let us denote by $\sigma_i$, $1
\leq i \leq 2n$, the distinct embeddings of $F$ in $\C$. (If $F$ is Galois,
the embeddings are related to each other by Galois transformations.) Among the
$\sigma_i$, let us choose a subset $P_+ = \{\sigma_1,\ldots,\sigma_n\}$ such
that no two embeddings in $P_+$ are complex conjugates of each other. Given the
pair $(F,P_+)$, called a CM--type, one may define the lattice
\beq
L(F,P_+) = \Big\{(\sigma_1(z),\ldots,\sigma_n(z)) \;:\; z \in {\cal O}_{F}
\Big\} \subset \C^n,
\eeq
with ${\cal O}_{F}$ the ring of integers of $F$, and then consider the
complex torus $\C^n/L(F,P_+)$. Its special structure allows to put a Riemann
form on it (see \cite{langcm} for the explicit construction), and so to promote
it to an Abelian variety, which, by construction, has complex multiplication by
$F$. Note that $P_+$ and $P_+ \sigma$ lead to the same torus for any embedding
$\sigma$. Conversely, if $A$ is an Abelian variety of dimension $n$, such
that $\endoq (A)$ contains $F$, a CM field of degree $2n$, then A is isogenous
to $\C^n/L(F,P_+)$ for some $P_+$. This shows that the complex multiplication
is a very restrictive property, fixing much of the variety.

Complex multiplication also yields a criterion of simplicity for an Abelian
variety, known as the Shimura--Taniyama (ST) theorem \cite{shita}. Let us assume
that $A=\C^n/L(F,P_+)$ is an Abelian variety of CM--type $(F,P_+)$, and that 
$F$ is Galois over $\Q$ (is a splitting field for any of its defining
polynomials). Set 
\beq
W(P_+) = \big\{ \sigma \in {\rm Gal}\,(F/\Q) \;:\; P_+ \sigma = P_+ \big\}.
\eeq
The ST theorem then states that $A$ is simple if and only if $W(P_+)=\{1\}$
\cite{shita,langcm}. One can moreover prove that if $W(P_+) \neq \{1\}$, $A$ is
isogenous to the product of $|W(P_+)|$ isomorphic simple factors, each one
having complex multiplication by the subfield of $F$ fixed by $W(P_+)$ (see
below). 

All these notions and results have a straight application to the case at hand.
In the decomposition (\ref{jac1}) of Jac$(F_n)$, all factors have complex
multiplication by $\Q(\zeta_{n_0})$, since the lattice $\lrst$ is stabilized
by the multiplication by arbitrary elements of $\z(\zeta_{n_0})$ (note that
the cyclotomic extension $\Q(\zeta_{n_0})$ is the imaginary quadratic extension
of the totally real field $\Q(\zeta_{n_0}+\overline\zeta_{n_0}) = 
\Q(\cos{2\pi \over n_0})$, and is thus a CM field). Observe also that $\lrst$ is
precisely a lattice arising from a CM--type, namely $(\Q(\zeta_{n_0}),\hrst)$.
Indeed $\Q(\zeta_{n_0})$ has an Abelian Galois group over $\Q$, consisting of 
the transformations $\sigma_h \,:\; \zeta \mapsto \zeta^h$ for all $h \in \Z
{n_0}$, and from a previous remark, $\hrst$ is a coset of the Galois group by
$\{\pm 1\}$, and therefore contains no two $h,h'$ such that $\sigma_h = 
\overline \sigma_{h'} = \sigma_{-h'}$. 

Following Koblitz and Rohrlich, one would like to answer two questions:

\smallskip \hskip 1truecm
{\em (i)} are the factors ${\Bbb C}^{\varphi(n_0)/2}/\lrst$ simple ? 

\hskip 1truecm
{\em (ii)} are there isogenies between some of them ? 

\smallskip \noindent
We have just observed that the factors
$\C^{\varphi(n_0)/2}/\lrst$ are Abelian varieties with CM--type
$(\Q(\zeta_{n_0}),\hrst)$, so we can use the ST criterion to solve both
problems. Set
\beq
\wrst = \{ w \in \Bbb Z_{n_0}^* \;:\; w\hrst = \hrst \}.
\label{wrst}
\eeq
 From the above general discussion, two factors related to $(r,s,t)$ and
$(r',s',t')$ will be isogenous if and only if they have the same CM--type up to
a Galois automorphism, {\it i.e.}
\beq
\lrst \sim L_{r',s',t'} \quad \Longleftrightarrow \quad 
\cases{\ \Z {n_0} \sim \Z {n'_0}, & \cr \noalign{\smallskip}
\ \hrst = H_{\rest {xr'},\rest {xs'},\rest {xt'}} \;\;\hbox{for some }x \in
\Z {n}. & }
\label{iso}
\eeq
The first condition is needed to ensure that $\Q(\zeta_{n_0}) =
\Q(\zeta_{n'_0})$, and implies $n_0=n'_0$, or $n_0=2n'_0$ with
$n'_0$ odd, or $n'_0=2n_0$ with $n_0$ odd. That answers problem {\em (ii)}. For
the problem {\em (i)}, one should look at $\wrst$ and see whether it is reduced
to the identity or not. If $\wrst = \{1\}$, then
$\lrst$ is simple, otherwise $\lrst$ splits up into $|\wrst|$ simple factors. 
Since $w\hrst=H_{\rest{w^{-1}r},\rest{w^{-1}s},\rest{w^{-1}t}}$, the
determination of $\wrst$ requires to compare the sets $\hrst$. Therefore
problems {\em (i)} and {\em (ii)}, eventually leading to the complete
decomposition of the Jacobian of $F_n$, boil down to the same question:     
when are two sets $\hrst$ and $H_{r',s',t'}$ equal ? This is precisely      
what the $su(3)$  parity rule requires to know.

This very concrete problem is easily solvable on a case--by--case basis, but
remains difficult to work out for general $n$. Koblitz and Rohrlich solved
it when $n$ is coprime with 6, and when it is a power of 2 or 3.
Recently the decomposition for general $n$ was completed by Aoki, except for 33
values of $n$ between 2 and 180 \footnote{The actual list of excluded integers
is ${\cal E} = \{2, 3, 4, 6, 8, 9, 10, 12, 14, 15, 18, 20, 21, 22, 24, 26,
28, 30,$ $36, 39, 40, 42, 48, 54, 60, 66, 72, 78, 84, 90, 120, 156, 180\}$.}.

We now summarize their results, leaving out the 33 special values of
$n$. We first define an equivalence relation on the admissible triplets: we will
say  that $(r,s,t) \sim (r',s',t')$ iff $(r',s',t') = (\rest {hr},\rest
{hs},\rest {ht})$ up to a permutation, that is, $(r',s',t')$ belongs to the class
$[r,s,t]$, up to a permutation.

Concerning problem {\em (i)}, it has been shown that the only non--simple
$\lrst$ are those with $(r,s,t)$ being equivalent to one of the following
triplets \cite{kr,aoki}
\bea
&&{\textstyle {n \over n_0}}(1,w,n_0-1-w), \quad \hbox{with $w^2=1 \bmod n_0$, $w
\neq \pm 1$, $w \neq {n_0 \over 2}+1$ if $8 | n_0$},\\
&&{\textstyle {n \over n_0}}(1,1,n_0-2), \quad \hbox{if $4 | n_0$},\\
&&{\textstyle {n \over n_0}}(1,w,w^2), \quad \hbox{with $1+w+w^2=0 \bmod n_0$},
\\
&&{\textstyle {n \over n_0}(1,{n_0 \over 2}+1,{n_0 \over 2}-2)}, \quad \hbox{if
$8 | n_0$}.
\eea
Corresponding to these four cases, $\lrst$ factorizes in respectively 2,2,3 and
4 isomorphic simple lattices. For instance, one may check that for $n=7$,
\beq
L_{1,2,4} \sim \big[{\Bbb Z}({\textstyle {1+\sqrt{-7} \over 2}})\big]^3,
\label{l124}
\eeq
so that the factor ${\Bbb C}^3/L_{1,2,4}$ is isogenous to the cube of the
elliptic curve of modulus $\tau={1+\sqrt{-7} \over 2}$.

As to problem {\em (ii)}, obviously we have $\lrst = L_{r',s',t'}$ if
$(r',s',t') \in [r,s,t]$ (as noted after (\ref{incl})), or
if $(r',s',t')$ is a permutation of $(r,s,t)$. These are trivial isogenies (in
fact isomorphisms), and they are the only ones if $n$ is coprime
with 6 \cite{kr}. When 2 or 3 divides $n$, there is a non--trivial isogeny
between $\lrst$ and $L_{r',s',t'}$ if and only if $(r,s,t)$ and $(r',s',t')$ 
are equivalent to elements in one of the following three sets
\cite{aoki}
\bea
&&\{{\textstyle (a,a,n-2a),(a,{n \over 2}-a,{n \over 2}),({n \over 2}-a,
{n \over 2}-a,2a)},\nonumber\\[1mm]
&& \hskip 2truecm {\textstyle ({a \over 2},{n + a \over 2},{n \over
2}-a),({n - 2a \over 4},{3n - 2a \over 4},2a)}\},\\[1mm]
&&\{{\textstyle (a,3a,n-4a),({n \over 2}-a,{n \over 2}-2a,3a)}\},\\[1mm]
&&\{{\textstyle (a,2a,n-3a),({n \over 3}-a,{2n \over 3}-a,2a)}\},
\eea
where the integer $a$ is subjected to two conditions: first, the components of
the triplets should be integers, and second, after $(r,s,t)$ and $(r',s',t')$
have been identified with two triplets in one of the three groups, $n_0$ and
$n'_0$ should be related as in (\ref{iso}), namely $\Z {n_0}$ and $\Z
{n'_0}$ should be isomorphic. On the other hand, $a$ need not be coprime with
$n$. This list, remarkably simple, is  an easy consequence of the corresponding
one for the pairs of triplets satisfying $\hrst=H_{r',s',t'}$, as established in
\cite{aoki}. In the 
$su(3)$ interpretation, it completely solves the parity criterion by giving all 
pairs of affine weights whose characters can be coupled in a modular invariant,
that is, those weights such that the matrix element $N_{p,p'}$ may be non--zero.
This list could of course be used to rederive the classification of the $su(3)$
modular invariant partition functions proved in \cite{gan1} (except at the 33
values of $n$ excluded by Aoki, which can be handled by hand).

It is instructive to compute the decomposition of $\jac(F_n)$ in specific cases.
In order to do this, we first come back to the ST theorem and show how to
compute the splitting of $\lrst$.

Assume $\wrst \neq \{1\}$. We start with two trivial observations:
$\wrst \subset \hrst$ because 1 always belongs to $\hrst$, and $\wrst \subset
\Bbb Z_{n_0}^*$ is a group. Recall the definition of $\lrst$ as
\beq
\lrst = \left\{\big(\ldots\,,\sigma_h(z),\,\ldots\big)_{h \in \hrst}
\;:\; z \in \z(\zeta_{n_0}) \right\}.
\eeq
$\wrst$ acts freely on $\hrst$, so we can write a class decomposition as
$\hrst = A_{r,s,t} \cdot \wrst$, with $|A_{r,s,t}|=|\hrst|/|\wrst|$. Reordering
the entries in $\lrst$ according to this decomposition, we have
\beq
\lrst = \left\{\big(\ldots,(\ldots,\sigma_a\sigma_w(z),\ldots)_{w
\in\wrst},\ldots \big)_{a \in A_{r,s,t}} \;:\; z \in \z(\zeta_{n_0}) \right\}.
\eeq
Thus an element of $\lrst$ is of the form $(\ldots,\sigma_a(\lambda),\ldots)_
{a \in A_{r,s,t}}$ where $\lambda$ is itself a vector of the type $(\ldots,
\sigma_w(z),\ldots)_{w \in \wrst}$.

Let $K$ be the subfield of $\Bbb Q(\ze_{n_0})$ fixed by $\wrst$, and ${\cal
O}_{K}$ be the ring of integers of $K$. Then
$\Bbb Q(\ze_{n_0})$ is an algebraic extension of $K$, with Galois
group Gal($\Bbb Q(\ze_{n_0})/K)=\wrst$. We also let $\theta_i$ be the
elements of a $K$--integral basis of ${\Bbb Z}(\ze_{n_0})$, that is, any
element $z \in {\Bbb Z}(\ze_{n_0})$ can be uniquely written as $z=\sum_i
x_i\theta_i$ with all $x_i$ in ${\cal O}_{K}$. Obviously the number of
$\theta_i$ is equal to $|\wrst|$.

Let $\Lambda = (\ldots,\sigma_w(z),\ldots)_{w \in \wrst}$ for $z$ running over
${\Bbb Z}(\ze_{n_0})$. We define a linear complex map $\psi$ on $\Lambda$ by
\beq
x_i(z) \equiv \big(\psi(\lambda)\big)_i = \sum_{w \in \wrst}\;
\sigma_w(\theta_iz) = \sum_w \;
\sigma_w(\theta_i)\,\sigma_w(z), \qquad 1 \leq i \leq |\wrst|.
\eeq
Clearly all $x_i$ belong to ${\cal O}_{K}$, and we obtain the inclusion
\beq
\psi(\Lambda) \subset [{\cal O}_{K}]^{|\wrst|}, \qquad \hbox{with finite
index.}
\eeq
The index is proved to be finite by noticing that $\psi$ is invertible since
det$\,\psi_{i,w} = {\rm det}\,\sigma_w(\theta_i)$ is the relative discriminant
of ${\Bbb Q}(\ze_{n_0})$ over $K$. Since $\lrst$ is equal to
$(\sigma_a(\Lambda))_{a \in A_{r,s,t}}$, we obtain that it is contained
with finite index in a product of $|\wrst|$ isomorphic factors through the 
linear map $\psi$, from which the isogeny follows
\beq
\lrst \sim \left\{\big(\ldots\,,\sigma(x),\,\ldots\big)_{\sigma \in \hrst/\wrst}
\;:\; x \in {\cal O}_{K} \right\}^{|\wrst|}.
\label{frst}
\eeq
The lattice within the curly brackets has complex multiplication by $K$, has
CM--type $(K,\hrst/\wrst)$, and is simple by the ST theorem. 

For $n=7$, the previous equation implies the decomposition (\ref{l124}). Indeed
one checks that $W_{1,2,4}=H_{1,2,4}=\{1,2,4\}$, and that the subfield of
${\Bbb Q}(\ze_7)$ fixed by $\sigma_2$ and $\sigma_4$, is $K = \Bbb
Q(\sqrt{-7})$, with ${\cal O}_{K} = \Bbb Z({1+\sqrt{-7} \over 2})$.

\subsection{Elliptic curves}

Being one--dimensional, an elliptic curve is the simplest Abelian variety of
all. It is thus a natural question to see if a Fermat curve can decompose in a
maximal way, as a product of elliptic curves. It turns out that this question
has a positive answer, but also that it is far from being generic. We will show
that a necessary condition to have a maximal splitting is that $n$ be a divisor
of 24. Koblitz has solved the more difficult question to list all
lattices $\lrst$ that have a maximal splitting in elliptic curves. Setting
$\gcd (r,s,t)={n \over n_0}$ as above, he finds that no $\lrst$ is isogenous to
a product of elliptic factors unless $n_0$ belongs to the following set
$\{3, 4, 6, 7, 8, 12, 15, 16, 18, 20, 21, 22, 24, 30, 39, 40, 48,
60\}$ \cite{kob}.

The argument is in fact extremely simple. We know that the period lattice of
$F_n$ splits into a product of lattices $\lrst$. If $F_n$ is to be isogenous
to a product of elliptic curves, each $\lrst$ must be isogenous to a product of
1--dimensional lattices. Since $\lrst \subset \Bbb C ^{\varphi(n_0)/2}$, the
Shimura--Taniyama theorem (see previous section) says that this can only happen
if $|\wrst|=|\hrst|={1 \over 2}\varphi(n_0)$.  But $\wrst \subset \hrst$ implies
$\wrst=\hrst$, so that $\hrst$ is a group. Thus $F_n$ is isogenous to a product
of elliptic curves iff $\hrst$ is a group for all admissible triplets $(r,s,t)$.
Note that because of $\hrst = hH_{\rest {hr},\rest {hs}, \rest {ht}}$ for 
$h$ in $\hrst$, the set $\hrst$ in general depends on which representative of 
the class $[r,s,t]$ we choose, except precisely if $\hrst$ is a group. Let
$n=2^mq$, with $q$ odd.

First take $r=s=2^m$. Then $H_{2^m,2^m,n-2^{m+1}} =
\{1,2,\ldots, {q-1 \over 2}\} \cap \Z q$. Since $q$ is odd, 2 and $q-1
\over 2$ belong to $H_{2^m,2^m,n-2^{m+1}}$, but $2 \cdot {q-1 \over 2} = q-1$
does not. Thus $H_{2^m,2^m,n-2^{m+1}}$ is not a group unless $q \leq 3$.

Now take $r=s=q$. Then $H_{q,q,n-2q} = \{1,3,5,\ldots,2^{m-1}-1\} \subset
\Z {2^m}$. But if $2^m \geq 16$, then modulo $2^m$, $(2^{m-2}+1)^2 = 2^{m-1}+1
\not\in H_{q,q,n-2q}$. Thus $H_{q,q,n-2q}$ is not a group if $2^m \geq 16$.

Therefore a necessary condition for $\hrst$ to be a group for all
$(r,s,t)$ is that $n=2^mq$ with $2^m \leq 8$ and $q \leq 3$, or in other words,
that $n$ divides 24. This is a sufficient condition for $n \leq 12$ only. If $n$
divides 24 and is smaller or equal to 12, $\hrst$ is at most of order 2 since
$\varphi(12)=4$. Being a subset of $\Z {24}$, $\hrst$ is automatically a group
because every element of $\Z {24}$ has a square equal to 1 modulo 24 (24 is the
largest integer to have this property). On the other hand, for $n=24$,
$\hrst$ can be of order 4 and it is no longer guaranteed to be a group.
An explicit calculation shows that indeed it is not always a group (see below),
so we conclude that the Fermat curve $F_n$ is isogenous to a product of elliptic
curves if and only if $n \leq 12$ divides 24.

\medskip
It is straightforward to compute the decomposition of $F_n$
for $n\;|\; 24$. For $n=3,4$ and 6, all $\lrst$ are already 1--dimensional,
isogenous to $\Bbb Z(\omega = \exp{2i\pi /3})$ for $n=3$ and 6, and to $\Bbb
Z(i)$ for $n=4$. For $n=8$, $\hrst$ can only be $\{1,3\}$ or $\{1,5\}$ if it
is  of order 2. One finds $K = \Bbb Q(\sqrt{-2})$ if $\hrst=\{1,3\}$ and
$K=\Bbb Q(i)$ if $\hrst = \{1,5\}$. Apart from $(r,s,t)=(2,2,4),(2,4,2)$
and $(4,2,2)$ which have their $\hrst$ equal to $\{1\}$ and their $\lrst$
isogenous to $\Bbb Z(i)$, the complete decomposition of $F_8$ follows by merely
counting how many triplets have a $\hrst$ equal to $\{1,3\}$ or to $\{1,5\}$.
Similarly for $n=12$, a set $\hrst$ of order 2 can only be $\{1,5\}$ or
$\{1,7\}$, yielding respectively $\lrst \sim [\Bbb Z(i)]^2$ or
$\lrst \sim [\Bbb Z(\omega)]^2$. 
Finally for $n=24$, there are 24 triplets $(r,s,t)$ such that their
$\hrst$ is not a group, for instance $H_{1,3,20} = \{1,5,11,17\}$. The
corresponding $\lrst$ are all equal and their product is $[L_{1,3,20}]^{24}$,
with $L_{1,3,20} \subset \Bbb C^4$ simple because $W_{1,3,20}=\{1\}$.
Putting everything together, one obtains
\bea
&& F_3 \sim \Bbb Z(\omega), \quad 
F_4 \sim [\Bbb Z(i)]^3, \quad
F_6 \sim [\Bbb Z(\omega)]^{10}, \\[2mm]
&& F_8 \sim [\Bbb Z(\sqrt{-2})]^{12} \oplus [\Bbb Z(i)]^9, \quad
F_{12} \sim [\Bbb Z(\omega)]^{28} \oplus [\Bbb Z(i)]^{27}, \\[2mm]
&& F_{24} \sim [{\Bbb C}^4/L_{1,3,20}]^{24} \oplus [\hbox{product of 157
elliptic curves}].
\label{div24}
\eea

Let us finally observe that the weights involved in the exceptional
$su(3)$ modular invariants at height $n=8,12$ and 24, correspond to lattices
which have all a maximal decomposition in elliptic curves. Moreover, those
pertaining to a given type I modular invariant have complex multiplication by 
the same CM field, namely $\Q(\sqrt{-2})$ for $n=8$ and 24, and $\Q(i)$ for
$n=12$. This is obvious for triplets that label characters coupled to each 
other since the very fact they can be coupled means they have the same CM--type,
but it is not for characters appearing in different blocks. The Moore--Seiberg
exceptional invariant at $n=12$ has not this property, and involves different
CM--types.

\section{Combinatorial groups for triangulated surfaces}

We gather in this section some constructions that appear naturally
in the context of Fermat curves, triangular billiards and rational
conformal field theories. They lie at the heart of
a deep interplay between combinatorial, complex and arithmetical
structures on closed surfaces. A nice reference about them is
\cite{grot}, and \cite{dessins} contains elementary reviews.
It is a good exercise to read this section and the next in parallel,
with the explicit case of the cubic Fermat curve in mind. 

\subsection{Cartography}

Our starting point is a compact Riemann surface $\S$ together with a holomorphic
map $h$ from $\S$ to the Riemann sphere ramified over three points only, say
$0$, $1$ and $\infty$ \footnote{The compact surfaces $\S$ for which such an $h$
exists have the following characterisation: they are defined over number fields
(Belyi's theorem).}. The Riemann sphere has  a ``standard'' triangulation
consisting of $0$, $1$ and $\infty$ as vertices, the real segments $[\infty,0]$,
$[0,1]$ and $[1,\infty]$ as edges, and the upper and lower half--planes as
faces. This triangulation has the obvious but remarkable property that:

--- the vertices can be assigned labels $0$, $1$ and $\infty$
in such a way that edges do not link vertices with the same label (we
say that vertices are three--coloriable);

--- the faces can be assigned labels black (corresponding to
the lower half--plane) and white (corresponding to the upper--half
plane) in such a way that faces of the same color have no common edge
(faces are two--coloriable). 

Taking the inverse image of this triangulation by $h$, $\S$ can be equipped
with a triangulation which inherits the same colouring properties.

\smallskip
The combinatorial data of the triangulation are conveniently encoded
in the so--called cartographic group. Its definition uses the
orientation of the triangulation (of course the orientation of $\S$ as
a Riemann surface induces an orientation on the triangulation). The
cartographic group permutes the flags of the triangulation. A flag is an 
ordered triple $(v,e,f)$ where $v$ is a vertex, $e$ an edge containing $v$ and
$f$ a face containing $e$ in such a way that with respect to the orientation
of the boundary of $f$, $e$ starts from $v$. The number of flags is
twice the number of edges (or thrice the number of faces for a
triangulation). By orientability, there is a cyclic ordering of the 
flags $(v,.,.)$ (resp. $(.,e,.)$, $(.,.,f)$) that contain a fixed vertex $v$
(resp. an edge $e$, a face $f$). Hence every flag $(v,e,f)$ has a unique vertex
successor $(v,e,f)^\sigma$ (of the form $(v,e',f')$), a unique edge successor
$(v,e,f)^\alpha$ (of the form $(v',e,f')$) and a unique face successor $(v,e,f)
^\varphi$ (of the form $(v',e',f)$). The flag permutations $\sigma$, $\alpha$ 
and $\varphi$ generate the cartographic group $\widehat C$, which encodes all the
combinatorial data of the triangulation. In fact, the cycle decompositions of
$\sigma$, $\alpha$ and $\varphi$ are in one--to--one correspondence with the
vertices, edges and faces of the triangulation. For instance, the cartographic
group of the standard triangulation of the Riemann sphere is isomorphic to
$S_3$ (the permutation group on three letters).

This definition of the cartographic group works for any polygonal decomposition
of an oriented surface without boundary. In general one has $\alpha^2=1$ (an
edge is common to only two faces), but particular to a triangulation is the
relation $\varphi^3=1$. The order of $\sigma$ is equal to lcm$_v(n_v)$, where
$n_v$ is the number of triangles that meet at the vertex $v$. Perhaps less
obvious is the relation $\alpha\sigma\varphi = 1$, valid for general polygonal
decompositions. It can be easily verified with the help of Figure 1.

\begin{figure}[htb]
%\vskip -0.7truecm
\leavevmode
\begin{center}
%\mbox{\epsfscale 1000 \epsfbox{carto.eps}}
\mbox{\epsfbox{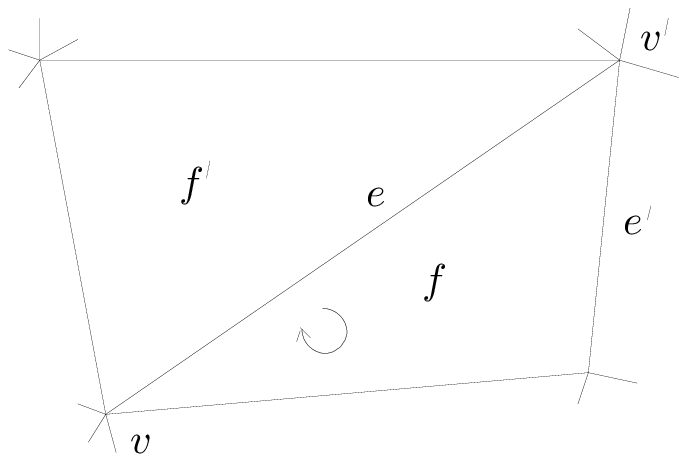}}
\begin{minipage}{14cm}
\bigskip
{\it Figure 1.} The action of the generators of the cartographic group on
the flags is defined in the text. One finds that $(v,e,f)^\varphi =
(v',e',f)$, $(v',e',f)^\sigma = (v',e,f')$ and $(v',e,f')^\alpha = (v,e,f)$,
confirming the relation $\alpha\sigma\varphi = 1$. The arrow indicates the
orientation.
\end{minipage}
\end{center}
\end{figure}

The cartographic group of a triangulation is always a quotient group of the
modular group via the homomorphism $S \rightarrow \alpha$, $T
\rightarrow \sigma$. Indeed as noted above, one has $S^2=1$ and $(ST)^3 =
(\alpha \sigma)^3 = \varphi^{-3} = 1$. This implies that there is a
universal cartographic group, which is the modular group, and a universal
triangulation, of which the flags can be parametrized by the elements of the
modular group. The corresponding triangulation is familiar. The quotient of the
upper--half plane $\h$ by $\G _2$, the principal congruence subgroup of level 2
in $\psl$ is known to be a sphere with three punctures. So there is
a unique holomorphic map from $\h$ to $\C\PP_1-\{0,1,\infty\}$ invariant under
$\G_2$. For $\tau \in \h$, we set $q=e^{2 i\pi \tau}$ and define
\bea
&& U(\tau) =
\left[{q^{1/24}\theta_4(0,\tau) \over
\eta(q)}\right]^4 = \prod_{m=1}^{\infty}(1-q^{m-\frac{1}{2}})^8, \\
&& V(\tau) = -\left[{q^{1/24}\theta_3(0,\tau) \over
\eta(q)}\right]^4 = -\prod_{m=1}^{\infty}(1+q^{m-\frac{1}{2}})^8, \\ 
&& W(\tau) = \left[{q^{1/24}\theta_2(0,\tau) \over \eta(q)}\right]^4 =
16\sqrt{q} \prod_{m=1}^{\infty}(1+q^m)^8.
\eea 
It is a standard identity that $U(\tau)+V(\tau)+W(\tau)=0$. Then the inverse
image of the standard triangulation of the sphere by the map
$\lambda(\tau)=-U(\tau)/W(\tau)$ (invariant under $\G_2$) gives the appropriate
triangulation of $\h$. More precisely, one can check that 
\beq
\lambda({-1 \over \tau}) = {1 \over \lambda(\tau)}, \qquad
\lambda(\tau+1) = 1 - \lambda(\tau),
\eeq 
from which the invariance under $\G_2$ follows, and that
$256\frac{(1-\lambda+\lambda^2)^3}{(\lambda(1-\lambda))^2}$ (a rational function
of degree 6 in $\lambda$) is the standard modular invariant function $j$.
Moreover it is easy to see that $\lambda$ maps 0 to 0, 1 to 1 and $\infty$ to
$\infty$. Thus $\lambda$ defines an homeomorphism of $\overline{\h/\G_2}$ with
the Riemann sphere. From the above product formulas, one checks that
$\lambda(\tau)$ is real negative on the imaginary axis, ranges between 0 and
1 on the big semi--circles going from $\pm 1$ to 0 (they are to be
identified in $\h/\G_2$), and takes all positive values from 1 to $\infty$ on 
the line Re$\,\tau = 1$. Thus the standard triangulation of the 
$\lambda$--sphere consists of the two faces (see Figure 2) 
\bea
&& {\rm B} = \{\tau \in \h \;:\; -1 \leq {\rm Re}\,\tau \leq 0, \; 
|\tau + {\textstyle {1 \over 2}}|^2 \geq {\textstyle {1 \over 4}})\}, \\
&& {\rm W} = \{\tau \in \h \;:\; 0 \leq {\rm Re}\,\tau \leq 1, \; 
|\tau - {\textstyle {1 \over 2}}|^2 \geq {\textstyle {1 \over 4}})\}.
\eea
The action of $\G_2$ on this triangulation yields the universal triangulation 
of $\h \cup \Q \cup \{\infty\}$ shown in Figure 2. For later use, we record the
following remarkable product formula
\beq
{{\rm d}\lambda \over {\rm d}\tau} = {i\pi \over 16\sqrt{q}} \, 
\prod_{m=1}^\infty {(1-q^m)^4 \over (1+q^m)^{16}}.
\label{dlambda}
\eeq

\begin{figure}[htb]
%\vskip -0.7truecm
\leavevmode
\begin{center}
%\mbox{\epsfscale 1200 \epsfbox{tessal.eps}}
\mbox{\epsfxsize=14cm \epsfbox{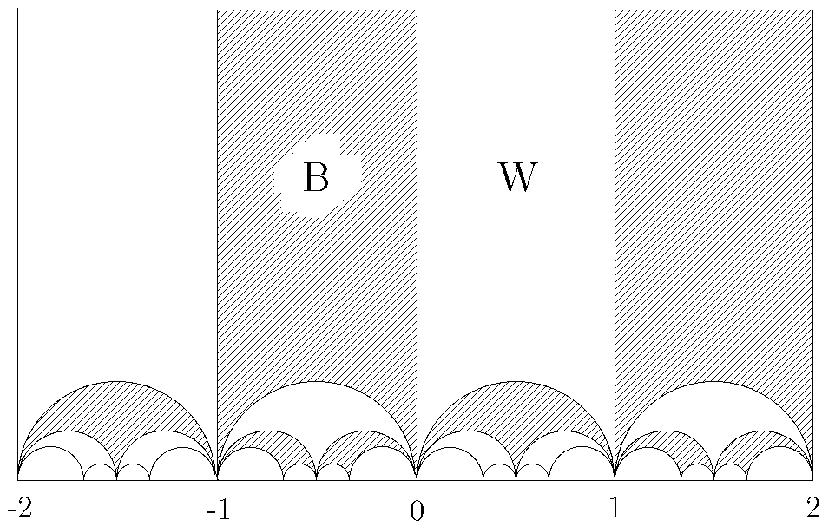}}
\begin{minipage}{14cm}
{\it Figure 2.} Universal triangulation of $\h \cup \Q \cup \{\infty\}$ obtained
by translating the standard triangulation of the $\lambda$--sphere (consisting of
the two faces marked B(lack) and W(hite)) by the principal congruence group
$\G_2$. Any flag is obtained from a fixed one $F^*$ by the action of a unique
element of the modular group. The cartographic group acts by the
standard multiplication: if $c,c'$ are elements of the cartographic group,
$(F^c)^{c'}=g_{c'}g_c(F^*)$ for some $g_c,g_{c'} \in \psl$. For instance the
successors of the flag $F$ containing the vertex 0 and the face W are $F^\sigma
= ST^{-1}S(F)$, $F^\alpha = S(F)$ and
$F^\varphi = TS(F)$ (orientation chosen clockwise).
\end{minipage}
\end{center}
\end{figure}

\smallskip
In this interpretation, the cartographic group of the standard
triangulation of the Riemann sphere, isomorphic to $S_3$, is represented as
$\psl /\G_2$. Let $C$ be the kernel of the homomorphism from $\psl$ to $\widehat
C$. Because of the intertwining property of $h$, an element of the modular group
whose image is trivial in $\widehat C$ must be trivial in $S_3 = \psl /\G_2$. This
means that $C$ is a subgroup of $\G_2$. Let us also define $\widehat D = \G_2/C$. It
is well known that $\G_2$ is the subgroup of $\psl$ generated by
$R_{\infty}=T^2$, $R_0=ST^2S^{-1}$ and $R_1=(TS)T^2 (TS)^{-1}$. They satisfy a 
single relation, namely $R_0R_1R_{\infty} =1$. This implies that $\widehat D$ is
generated by $\mu_{\infty}=\sigma^2$, $\mu_0=\alpha\sigma^2\alpha^{-1}$ and
$\mu_1=(\sigma\alpha)\sigma^2(\sigma \alpha)^{-1}$. The order of $\widehat C=
\psl /C$ is six times that of $\widehat D$.

\medskip
The connectedness of $\S$ implies that $\widehat{C}$ acts transitively on
flags. Hence, the set of flags is a homogeneous space for $\widehat{C}$.
The isotropy subgroup of a flag, well defined up to conjugacy in
$\widehat{C}$, has the following properties :

--- it does not contain any invariant subgroup of $\widehat{C}$ (such an 
invariant subgroup would act trivially on all flags);

--- its image in $S_3$ is trivial (because the cartographic group acts 
compatibly with the map $h$).

In summary, each pair $(\S,h)$ gives rise to a subgroup $B$ of $\G_2$, and an
invariant subgroup $G$ of $\psl$ (the intersection of all the conjugates of $B$
in $\psl$ so $G$ is a subgroup of $B$) such that: {\it (i)} the flags are
parametrized by $B \backslash \psl$ (the set of left cosets $Bg$ for $g \in
\psl$), and, {\it (ii)} the cartographic group is isomorphic to $\psl /G$ acting
on $B \backslash \psl$ on the right.

But there is in fact a reciprocal. If $B$ is any subgroup of $\G_2$,
let $\S '$ be the quotient $\h / B$. It is a Riemann surface
with punctures, and the projection from $\h / B$ to $\h /
\G_2 \cong \C\PP_1 - \{0,1,\infty\}$ is holomorphic and unramified.
This map has a holomorphic extension, say $h'$, from the compactified
surface $\oS'$ (the surface obtained from $\S '$ by ``filling'' the
punctures, see for example \cite{knapp}) to $\C\PP_1$, ramified only above $0$,
$1$ and $\infty$. Now if $B$ comes from a compact Riemann surface $\S$ via a map
$h$ by the above construction, then $\S$ is isomorphic to $\oS'$.
Indeed $\S$ and $\oS'$ can be cut into triangles with the same
combinatorial arrangement. Let us consider pairs of triangles $f$ and $f'$
on $\S$ and $\oS'$, that are mapped by $h$ and $h'$ onto the
same triangle of the standard triangulation of $\C\PP_1$.  Restricted to the
interior of the triangles, these maps are holomorphic, so their composition
defines maps from each triangle of $\S$ to the corresponding triangle
of $\oS'$, which are holomorphic on the interior of the triangles. Along
the edges, they glue so as to yield a continuous one--to--one map from $\S$ to
$\oS'$. To check holomorphicity along the edges, consider a pair of faces
$f_1$ and $f_2$ with a common edge $e$ on $\S$ and the corresponding
pair $f_1'$ and $f_2'$ with common edge $e'$ on $\oS'$. The
maps $h$ and $h'$ send the domains made up of the interiors of the faces
together with the interiors of the common edges holomorphically and one--to--one
to the same open subset of the Riemann sphere.  This ensures
that the composite map is holomorphic, and in particular holomorphic
along the interior of the common edge. So the continuous map we have
constructed from $\S$ to $\oS'$ is holomorphic except maybe
at the marked points.  But then its continuous extension has to be
holomorphic everywhere (a holomorphic map in a pointed disk, bounded
near the puncture has a unique holomorphic extension).

So the cartography of a Riemann surface (defined over $\overline{\Q}$)
eventually leads to its uniformization by $\h /B$ for some Fuchsian
group $B$, in fact a subgroup of $\psl$. This will be extensively
used in the next sections to uniformize the Fermat curves, and other
related curves. 

\smallskip
The symmetry group of a triangulation has a natural definition. It
consists of the relabellings of the flags that do not change
the combinatorial data, i.e.  that commute with the action of the
cartographic group.  As the cartographic group acts transitively on
flags, an element of the symmetry group that fixes one flag has to be
the identity.  For the same reason, the orbits of the symmetry group
all have the same number of elements.  So the order of the symmetry
group divides the number of flags. If this order is the number of
flags, we can choose a flag and then get any other flag by acting with
a unique symmetry. So in that case, the flags can be parametrized by
symmetries. But the cartographic group commutes with symmetries, so an element
of the cartographic group can fix a flag only if it is the identity. In other
words, there is no non--trivial isotropy, and the flags are also parametrized by
elements of the cartographic group ($B=G=C$).

It is almost obvious (and can be checked along lines similar to the
proof above that $\S$ and $\oS'$ are the same Riemann surface)
that the action of the symmetry group induces holomorphic automorphisms of the
associated Riemann surface.

\medskip
As mentioned above, the cartographic group can be used to encode the
combinatorics of a polygonal decomposition of a closed oriented
surface. For the triangulation of a Riemann surface given by the
inverse image of the standard triangulation of the sphere by a map
ramified only over 0, 1 and $\infty$, there is another convenient
combinatorial description. Let us fix such a Riemann surface and such a map. 
This time, we use a group, which we call the triangular group, that permutes the
faces of the triangulation. It is simpler to define than the cartographic group. 
It has three generators of order 2: $\sigma_0$, $\sigma_1$ and
$\sigma_{\infty}$.   If $f$ is a face of the triangulation, the edge of $f$
opposite to the vertex $0$ ({\it i.e.} joining the vertices $1$ and $\infty$) is
common to exactly one other face $f'$, and we set $\sigma_0(f)=f'$. The other
two generators are defined analogously. The triangular group specifies
which triangle is glued to which other triangle and along which edge, so it
gives all is needed to reconstruct the surface. It contains an invariant
subgroup of index 2 with generators 
\beq
\rho_0=\sigma_{\infty}\sigma_1, \qquad
\rho_1=\sigma_0\sigma_{\infty}, \qquad \rho_{\infty}=\sigma_1\sigma_0.
\eeq
Geometrically, $\rho_v$ is a rotation around the vertex $v$, mapping a triangle
touching $v$ to the next one of the same colour. The order of $\rho_v$ is
half the number of triangles meeting at $v$. They clearly satisfy
$\rho_{\infty} \rho_1 \rho_0 =1$, so this subgroup, called the oriented
triangular group for obvious reasons, is a quotient group of $\G_2$. Hence 
there is a unique epimorphism from $\G_2$ to the oriented triangular group
sending $R_0^{-1}$ to $\rho_0$, $R_1^{-1}$ to $\rho_1$ and $R_{\infty}^{-1}$
to $\rho_{\infty}$. The oriented triangular group is important for
two reasons. First it contains the same combinatorial information as
the full triangular group. Indeed, since it maps white faces into white
faces and black faces into black faces, it says how the vertices of
faces of a given color are linked to each other, and thus specifies all the
edge identifications needed to completely reconstruct the surface. Second, it
is easier to compute than the cartographic group, and yet allows to describe
the latter more easily than what was done before. The idea is simple, but 
requires first to find an appropriate parametrization of the flags. With the
flag $(v,e,f)$, we associate a permutation $(jkl)$ of the symbol
$(01\infty)$: the first element is the label of $v$, the second is the label of
the other vertex of $e$ and the third is the last label remaining. Then with
the same flag $(v,e,f)$, we associate the white triangle $t$ that has $e$ in
common with $f$, and we write $(v,e,f) \cong [(jkl),t]$. Let $\varepsilon (jkl)$
be $0$ if $(jkl)$ is an even permutation and $1$ if it is an odd permutation of
$(01\infty)$. It is then easy to check that
\bea
&& \sigma[(jkl),t] = [(jlk),\rho_j^{\varepsilon (jkl)}(t)], \\
&& \alpha[(jkl),t] = [(kjl),t], \\ 
&& \varphi[(jkl),t] = [(klj),\rho_k^{-\varepsilon (klj)}(t)].
\eea 
{}From these, one computes the action of the $\widehat D$ group to be
\bea
&& \mu_0[(jkl),t] = [(jkl),\rho_k t], \\
&& \mu_1[(jkl),t] = [(jkl),\rho_j^{\varepsilon (jlk)}\rho_l\rho_j^{-\varepsilon
(jlk)}t], \\
&& \mu_{\infty}[(jkl),t] = [(jkl),\rho_j t].
\eea 
The equation for $\mu_1$ is a little bit surprising, but is needed to ensure the
relation $\mu_{\infty}\mu_0\mu_1=1$. This shows that although the group 
$\widehat D$ is closely related to the oriented triangular group, they do not
coincide,  the former being in general bigger. The three generators $\mu_0$,
$\mu_1$ and $\mu_{\infty}$ all have the same order $n$, the least common 
multiple of the orders of $\rho_0$, $\rho_1$ and $\rho_{\infty}$.

\medskip
Because it is of particular importance for what follows, we will explore,
for the rest of this section, the case where the oriented triangular group
associated with $(\S,h)$ is Abelian. The above relations simplify to read
\bea
&& \mu_0[(jkl),t] = [(jkl),\rho_k t], \\
&& \mu_1[(jkl),t] = [(jkl),\rho_l t], \\
&& \mu_{\infty}[(jkl),t] = [(jkl),\rho_j t].
\eea 
Thus the group $\widehat D$ is also Abelian, and is isomorphic to a subgroup of
$\z_n \times \z_n$ (or more invariantly of the quotient of $\z_n \times \z_n
\times \z_n$ by the diagonal $\z_n$). 

If $\widehat D \equiv \widehat D_n = \z_n \times \z_n$, one can even give a
presentation of the cartographic group, which we denote by $\widehat C_n$, by
generators and relations, {\it i.e.} one can determine the kernel $C_n$ of the
homomorphism from $\psl$ to $\widehat{C}_n$. As the relation $\mu_0\mu_1\mu_{\infty}
=1$ is automatic in terms of the generators $\alpha$ and $\sigma$, the 
commutation of $\mu_0$, $\mu_1$ and $\mu_{\infty}$ amounts to $\mu_{\infty}\mu_1
\mu_0=1$, or $\sigma^2(\sigma\alpha)\sigma^2(\sigma\alpha)^{-1}\alpha\sigma^2
\alpha^{-1}=1$. Using $\alpha^{-1}=\alpha$ and $\alpha\sigma^{-1}\alpha=\sigma
\alpha\sigma$, this can be simplified to $(\sigma^3\alpha)^3=1$. Moreover,
there is the obvious relation $\sigma^{2n}=1$. The quotient of the
modular group by these two relations is $\widehat{C}_n$ because they ensure the
commutativity and the correct order for $\mu_0$, $\mu_1$ and $\mu_{\infty}$:
\beq
\cases{\widehat D_n = \z_n \times \z_n, & \cr
\widehat C_n = \langle S,T \;|\; S^2,(ST)^3,T^{2n},(T^3S)^3 \rangle, \qquad
|\widehat C_n|=6n^2. & }
\eeq
Though not obvious at this stage, $\widehat C_n$ is isomorphic to the 
semi--direct product $S_3 : (\z_n \times \z_n)$, and possesses an action on
$\C\PP_2$ by
\bea
&& A(x;y;z) = (\zeta x;y;z), \quad B(x;y;z) = (x;\zeta y;z), \qquad (\zeta = 
e^{2i\pi/n}) \\
&& \tau_1(x;y;z) = (y;x;z), \quad \tau_2(x;y;z) = (x;z;y).
\eea
We will prove this isomorphism in the next section, when we consider the
automorphisms of the Fermat curves.

To summarize, there is a natural universal object for pairs $(\S,h)$ with 
Abelian oriented triangular groups. Let $\S_n$ be the quotient $\h /C_n$ with
punctures filled in. It is a compact Riemann surface, and its group of
holomorphic automorphisms permuting the flags is isomorphic to $\widehat{C}_n$.  
The flags of $(\S,h)$ can be parametrized as the quotient of
$\widehat{C}_n$ by the stabilizer subgroup of a given flag, in fact a subgroup 
of $\widehat{D}_n$. This subgroup acts as automorphisms on $\S_n$ and the
quotient is nothing but $\S$. In other words, $\S_n$ is a covering of
$\S$ of degree equal to the common order of the stabilizer subgroups. This gives
a convenient way to relate the geometry of $\S$ to that of $\S_n$. We will use 
it in the following sections.

\subsection{Modular forms associated to Fermat curves}

We have seen in Section 3 that certain important features of the RCFT with an
affine symmetry based on $su(3)$ were governed by quantities that had a very
important r\^ole in the description of the geometry of the Fermat curves. Also
some similarities emerged: for instance the weights in the alc\^ove of $su(3)$
at height $n$ are in one--to--one correspondence with the holomorphic
differentials on $F_n$. In this section, we would like to see whether this
relationship goes beyond the superficial level by making the holomorphic
differentials on the Fermat curves look as much as possible like the characters
of a conformal field theory. Our construction is not canonical in a mathematical
sense, but it is nevertheless quite natural. We use the vocabulary associated
with the combinatorics of triangulations, as presented in the previous section.

Let $F_n$ be the Fermat curve of degree $n$, 
\beq
u^n + v^n + w^n = 0, \quad \hbox{in } \C\PP_2.
\eeq
We shall sometimes use the affine model $x^n+y^n=1$ by setting $x=\xi u/w$ and
$y=\xi v/w$ where $\xi \equiv e^{i\pi /n}$. The map $t=-u/w$ gives an
isomorphism of $F_1$ and $\C\PP_1$. The three base points $u=0$, $v=0$ and
$w=0$ of $F_1$ are mapped to $0$, $1$ and $\infty$. The inverse image of the 
real axis gives a triangulation of $F_1$ with 2 faces, 3 edges, and
the base points as vertices. There is a canonical map of degree
$n^2$ from $F_n$ to $F_1$ given by 
\beq
h_n\;:\;\; (u;v;w) \longrightarrow (u^n;v^n;w^n),
\eeq 
and ramified only over the base points.
Taking the inverse image of the standard triangulation of $F_1$, $F_n$ is
naturally endowed with a triangulation consisting of $2n^2$ faces, $3n^2$ edges
and $3n$ vertices (leading quickly to the genus formula). The $3n$ vertices are 
\bea
&& \hbox{$n$ vertices of type 0:} \quad (0;v_0;w_0) \mbox{ with }
v_0^n + w_0^n = 0, \\
&& \hbox{$n$ vertices of type 1:} \quad (u_1;0;w_1) \mbox{ with }
u_1^n + w_1^n = 0, \\
&& \hbox{$n$ vertices of type $\infty$:} \quad (u_{\infty};v_{\infty};0) 
\mbox{ with } u_{\infty}^n + v_{\infty}^n = 0. 
\eea
For $n \geq 3$, the vertices are nothing but the inflexion points of $F_n$ 
(which are degenerate, the tangent line at a vertex having a contact of order 
$n$ with the curve). Let us parametrize these points more explicitly by setting
$\xi_0= \frac{v_0}{w_0}$, $\xi_1= \frac{w_1}{u_1}$ and $\xi_{\infty}=
\frac{u_{\infty}}{v_{\infty}}$. These numbers are odd powers of $\xi$.

Edges have to joint vertices of different type, so there cannot
be more than $3n \cdot 2n/2$ edges. This is the actual number of edges, so that
there is an edge between any two vertices of different type. It remains to
describe the faces. There is a unique one--to--one holomorphic map
$(u(t);v(t);w(t))$ from the upper--half plane to a (white) triangle on $F_n$ 
such that $-u^n(t)/w^n(t)=t$. If the images of $t=0$ and $t=1$ are given
vertices of type 0 and 1, say $(0;v_0;w_0)$ and $(u_1;0;w_1)$ respectively, then
a straightforward computation shows that the vertex of type $\infty$ in this
triangle is $(u_{\infty};v_{\infty};0)$ such that $\xi_0\xi_1\xi_{\infty}=\xi$.
An analogous computation for black triangles shows that $(0;v_0;w_0)$,
$(u_1;0;w_1)$ and $(u_{\infty};v_{\infty};0)$ are the vertices of a black
triangle if and only if $\xi_0\xi_1\xi_{\infty}=\xi^{-1}$. Note in
particular that any three vertices define the interior of at most one triangle,
unlike the standard triangulation of the sphere.

We are now in position to give the explicit action of the triangular group.
The reflection $\sigma_0$ acts only on $\xi_0$. If the triangle is white (resp.
black) $\sigma_0$ multiplies $\xi_0$ by $\xi^{-2}$ (resp. $\xi^2$). The other 
two generators act analogously (just change the labels). From this we deduce
the action of the oriented triangular group. If $t(\xi_0,\xi_1,\xi_{\infty})$
is the white face with vertices given by $\xi_0$, $\xi_1$, $\xi_{\infty}$
(hence $\xi_{\infty}\xi_1\xi_0=\xi$), we have 
\bea
&& \rho_0 t(\xi_0,\xi_1,\xi_{\infty}) =
t(\xi_0,\xi^{-2}\xi_1,\xi^2\xi_{\infty}),\\
&& \rho_1 t(\xi_0, \xi_1, \xi_{\infty}) = t(\xi^2\xi_0,\xi_1,
\xi^{-2}\xi_{\infty}), \\
&& \rho_{\infty} t(\xi_0,\xi_1,\xi_{\infty}) = t(\xi^{-2}\xi_0,\xi^2\xi_1,
\xi_{\infty}).
\eea
In this form, it is obvious that the oriented triangular group is
commutative, and that $\rho_0 \rho_1 \rho_\infty = \rho_0^n = \rho_1^n =
\rho_{\infty}^n = 1$ are the only relations the generators satisfy. According
to the results of the previous section, it follows that $\widehat D(F_n) = 
\widehat D_n = \z_n \times \z_n$ (equal to the oriented triangular group in 
this case), and that the cartographic group is $\widehat{C}_n$, the group of
order $6n^2$ with presentation $\langle \alpha,\sigma \;|\;
\alpha^2,(\alpha\sigma)^3,\sigma^{2n}, (\sigma^3\alpha)^3 \rangle$. But the 
number of flags on $F_n$ is
$6n^2$, precisely the order of $\widehat{C}_n$, so that $F_n$ is isomorphic, as 
a Riemann surface, to the quotient $\h /C_n$ with punctures filled in. Also the
symmetry group of the triangulation is isomorphic to $\widehat C_n$.

Let us observe that $h_{m}\circ h_{n}=h_{mn}$ and that $h_n$ not only
maps $F_n$ to $F_1$, but also $F_{mn}$ to $F_{m}$ for any $m$. By
construction, as a map from $F_{mn}$ to $F_{m}$, $h_n$ maps
vertices into vertices, edges into edges, faces into faces and
flags into flags, and preserves the colouring properties.
Moreover $h_n$ intertwines the action of $\widehat{C}_{mn}$ and
$\widehat{C}_{m}$, so that $\widehat{C}_{m}$ is a quotient group of 
$\widehat{C}_{mn}$ for any $n$. As a byproduct, the family $\widehat{C}_{n}$ 
indexed by positive integers is a directed projective family of groups, while
the $C_n$ is a directed injective family of groups.

The holomorphic automorphisms of $F_n$ permuting the flags form a
group isomorphic to $\widehat{C}_n$. It is straightforward to get
the corresponding action. In fact, $F_n$ has a number of obvious
automorphisms: permutations of the coordinates, multiplication
of the coordinates by arbitrary $n$--th roots of unity and
combinations thereof. It is clear that this group has order
$6n^2$, and that it must coincide with $\widehat C_n$, the cartographic group we
have just computed. This proves the isomorphism announced in the previous
section,
$\widehat C_n = S_3 :(\z_n \times \z_n)$, which otherwise can be proved
abstractly. For $n > 3$, this turns out to be the full automorphism group
\cite{shi} (see also Appendix A). We can view the automorphism group as a 
quotient of $\psl$ acting on $\h$ if we uniformize $F_n$ by the following
$n$--th roots of the functions uniformizing $\h /\G_2 \cong \C\PP_1$:
\bea
&& u(\tau)=\prod_{m=1}^{\infty}(1-q^{m-\frac{1}{2}})^{8/n}, \\
&& v(\tau)=\xi \,\prod_{m=1}^{\infty}(1+q^{m-\frac{1}{2}})^{8/n}, \\
&& w(\tau)=(16\sqrt{q})^{1/n} \, \prod_{m=1}^{\infty}(1+q^m)^{8/n}. 
\eea 
For definiteness, the above roots are always chosen to be real positive if the
argument is real positive. Then the modular transformations 
\bea
&& S\;: \qquad (u({\textstyle {-1\over\tau}}); v({\textstyle
{-1\over\tau}});w({\textstyle {-1\over\tau}})) = (w(\tau);v(\tau);u(\tau)),
\\ && T\;: \qquad (u(\tau+1);v(\tau+1);w(\tau+1)) =
(\xi^{-2}v(\tau);u(\tau);w(\tau)),
\eea 
generate a group isomorphic to $\widehat{C}_n$. The generators of $\G_2$ act as
\bea
&& R_{0} \;: \qquad (u;v;w) \longrightarrow (\xi^2u;v;w), \\
&& R_{1} \;: \qquad (u;v;w) \longrightarrow (u;\xi^2v;w), \\
&& R_{\infty} \;: \qquad (u;v;w) \longrightarrow (u;v;\xi^2w).
\eea

\smallskip
After these preliminaries, we are ready to associate modular forms (for the
invariant subgroup $C_n$ of $\psl$) to the holomorphic differentials on the
Fermat curves. It is a well--known fact in algebraic geometry that if
the zero set of $P(u,v,w)$, a homogeneous polynomial of degree $n \geq 3$, 
is a smooth curve in $\C\PP_2$, the holomorphic differentials on that
curve take the form 
\beq 
Q(u,v,w)\frac{w^2}{\partial P / \partial v}\; {\rm d}\left(-\frac{u}{w} \right),
\eeq 
where $Q(u,v,w)$ is a homogeneous polynomial of degree $n-3$. More precisely,
this is the expression of a holomorphic differential on the coordinate patch $w
\neq 0$, $\partial P / \partial v \neq 0$, where $-u/w$ is a good local
parameter. Because $\frac{w^2}{\partial P / \partial v}\, {\rm d}(\frac{-u}{w})$
is multiplied, under permutation of the variables, by the signature of the
permutation, the above expression gives the most general everywhere holomorphic
differential. For $F_n$, we thus get a standard basis of holomorphic
differentials $\{\omega_{r,s,t} \;:\; 1 \leq r,s,t \leq n-1,\; r+s+t=n\}$,
where, in the domain $v \neq 0, w \neq 0$,  
\beq
\omega_{r,s,t} = u^{r-1} v^{s-n} w^{t+1} {\rm d}\left(-\frac{u}{w}\right). 
\label{diff}
\eeq 
The differential $\omega_{r,s,t}$ has zeroes of order $r-1$ at the
$n$ vertices of type $0$, of order $s-1$ at the $n$ vertices of type $1$, and
of order $t-1$ at the $n$ vertices of type $\infty$, for a total of
$n(n-3)$ zeroes as expected. Taking $\tau$ as a local parameter, they yield a
basis for the modular forms of degree 2 for $C_n$. Using the relation
(\ref{dlambda}) and the standard identity $\prod_{m \geq 1}\,(1+q^m)(1-q^{2m-1})
=1$, and neglecting a constant factor equal to ${-i \pi \over n}\xi^s 16^{t/n}$,
one eventually arrives at the following expression (we keep the same name for the
differential and for the modular form) 
\beq
\omega_{r,s,t} = q^{t \over 2n} \,
\prod_{m=1}^{\infty}\left(\frac{1-q^\frac{m}{2}} {1+q^\frac{m}{2}} \right)^4
(1+q^m)^{\frac{8(s+2t)}{n}} (1+q^{m-\frac{1}{2}})^{\frac{8(2s+t)}{n}}
\,({\rm d}\tau).
\label{orst}
\eeq 
It would not be difficult to write explicitly the action of the
modular group on this basis of holomorphic forms (see below), and this would
allow to compare the corresponding periods along a fixed cycle, as
was shown using different methods in the third section. We shall
rather examine the similarities and differences between
these modular forms for $C_n$ and the characters of the $su(3)$
affine algebra at level $k=n-3$.

As mentioned earlier, the number of (unrestricted) characters is the same as
the number of $\omega$'s. The restricted characters however are not linearly
independent. Moreover, characters are functions and the differentials on $F_n$
are forms. This is not too serious. One possible remedy is as follows : on $F_3$
there is only one holomorphic differential, and the corresponding modular form is
easily seen to be $\eta^4(\tau)$, whereas the character at height 3 is the
constant function 1. However the denominator of the Weyl--Kac character formula
is $\eta^8(\tau)$, so that the numerator is
naturally a modular form of weight 4, {\it i.e.} a quadratic differential. This
makes plausible the fact that to find analogies, it is perhaps better to
concentrate on the numerators of characters. We shall elaborate a little bit on
that at the end of this section.

Observe that $\omega_{r,s,t}$ does not change if one multiplies
$r$, $s$, $t$ and $n$ by a common factor (this is clearly related to the
map $h_m$ from $F_{mn}$ to $F_{n}$), but this property is not shared by the
characters (though the alc\^ove $B_n$ is properly embedded in $B_{mn}$).
However we have seen in Section 2 that the numerators of characters satisfy 
more involved identities of the same kind having a similar origin.

One might be tempted to see another common point in the fact that both sets 
carry a representation of the modular group for which only a finite quotient
acts, and which is in general highly reducible. For the characters, this is a
well--known fact. For the holomorphic differentials on $F_n$, it is related to
the regularity of the triangulation induced by the map $h_n$. However the two
representations are very different. From the above formulas, one obtains the
modular transformations of (\ref{orst})
\beq
\omega_{r,s,t}(\tau+1) = \xi^t \, \omega_{s,r,t}(\tau), \qquad
\omega_{r,s,t}({\textstyle {-1\over\tau}}) = -\tau ^2 16^{(r-t)/n} \,
\omega_{t,s,r}(\tau).
\eeq
Thus for the holomorphic differentials on $F_n$, the modular group merely
permutes $r$, $s$ and $t$ and multiplies by phases. This is in striking contrast
with the modular transformations of the characters.

One can nevertheless try to push the analogy with the affine $su(3)$ characters
by looking at the modular problem for the differentials on $F_n$. So we set
$\tilde \chi_{r,s,t} = \omega_{r,s,t}/\omega_{1,1,1}$. The $\tilde \chi$ carry
a (non--unitary) representation of $\psl$, and we can look for the modular
sesquilinear forms in the $\tilde \chi$. It turns out to be much easier
than the corresponding affine modular problem.

Let $N$ the matrix specifying a Fermat modular invariant. That it commute with
$T$ and $S$ implies respectively
\bea
&& N_{r,s,t;r',s',t'} = \xi^{t-t'} \, N_{s,r,t;s',r',t'}, \label{tcond} \\
&& N_{r,s,t;r',s',t'} = 16^{(r+r'-t-t')/n} \, N_{t,s,r;t',s',r'}. \label{scond}
\eea
Requiring that the entries of $N$ be positive integers, Equation
(\ref{tcond}) yields $N_{r,s,t;r',s',t'}$ $=0$ if $t \neq t'$. If
$N_{r,s,t;r',s',t'} \neq 0$, then (\ref{scond}) implies $r=r'$ hence $s=s'$, so
that only the diagonal couplings $N_{r,s,t} \equiv N_{r,s,t;r,s,t}$ may be
non--zero. They must satisfy
\beq
N_{r,s,t} = N_{s,r,t}, \quad {\rm and} \quad N_{r,s,t} = 2^{8(r-t)/n}
\, N_{t,s,r}.
\eeq
These conditions mean one can look at the six permutations of $(r,s,t)$
independently of the other triplets, and also that, up to a normalization
factor, the modular invariants involving the six permutations is unique.
The integrality conditions imply $8r=8s=8t \bmod n$, and one finds,
assuming $r \leq s \leq t$, that the unique modular invariants reads
\bea
&& Z_{r,s,t}(F_n) = |\tilde \chi_{r,s,t}|^2 + |\tilde \chi_{s,r,t}|^2 +
2^{8(t-r)/n} \, |\tilde \chi_{t,s,r}|^2 +
2^{8(t-r)/n} \, |\tilde \chi_{s,t,r}|^2 \nonumber\\ \noalign{\smallskip}
&& \hskip 2.5truecm + 2^{8(t-s)/n} \, |\tilde \chi_{t,r,s}|^2 +
2^{8(t-s)/n} \, |\tilde \chi_{r,t,s}|^2.
\eea
Using $r+s+t=n$, the integrality conditions imply $24r=24s=24t=0 \bmod n$. If
$r,s,t$ have a common factor, say $d$, then $\omega_{r,s,t}$ descends to the
differential $\omega_{r/d,s/d,t/d}$ on $F_{n/d}$, whereas if gcd$(r,s,t)=1$, 
then $n$ must divide 24. 

Thus the analogy between the two modular problems is somewhat disappointing,
but there is still a curious fact. The coefficients of the
$q$--expansion of the characters are integers. This is in general not true for
the holomorphic differentials on $F_n$, and in fact happens quite seldom. From
our explicit formula, the $q$--expansion of $\omega_{r,s,t}$ has integer
coefficients (or even bounded denominators, which is a normalization invariant
statement) if and only if $8(s+2t)$ and $8(2s+t)$ are both multiples of $n$. It
is then not difficult to make a catalogue of all triplets satisfying these
conditions. If we assume, without loss of generality, that gcd$(r,s,t)=1$, one
finds the straightforward but puzzling result:

--- $n$ is necessarily a divisor of 24;

--- the $q$--expansion of $\tilde \chi_{r,s,t}$ contains only integer
coefficients if and only if $\tilde \chi_{r,s,t}$ appears in a modular
invariant for $F_n$;

--- the four exceptional $su(3)$ modular invariant partition functions appearing 
at height equal to 8, 12 or 24, involve characters labelled by triplets 
$(r,s,t)$ which all satisfy the above conditions, so that the corresponding 
forms $\omega_{r,s,t}$ have integer coefficients in their $q$--expansion.
They however do not exhaust the list of triplets with this property.

To prove the first two points, one simply notes that $8(s+2t) = 8(2s+t) = 0
\bmod n$ imply $24r=24s=24t=0 \bmod n$.

\subsection{Rational triangular billiards}

A (generalized) billiard is a planar domain with piecewise smooth
boundary. A classical particle moving in such a domain is simply
reflected when it hits the boundary, but moves freely otherwise.
The spectrum of the corresponding quantum mechanical system is
related to the Dirichlet problem for the Laplace operator. The
general case can be very complicated. When the domain is a
Euclidian triangle with rational angles (in units of $\pi$), the
classical phase space has an interesting geometric structure:
it has a foliation by closed topological surfaces. In fact the
leaves have a natural complex structure \cite{berry}. We will
briefly review this construction. On the way we will see that
many quantities we encountered in Section 3 reappear quite naturally. We will
then present yet another intriguing relation with the exceptional modular 
invariants for the $su(3)$ WZNW models.

Na\"\i vely, a point in the classical phase space is a pair $(\vec x,\vec p)$
where $\vec x$ is a position (a point of the triangle) and $\vec p$
a momentum (an arbitrary two--dimensional vector). However, to take
into account reflections when the particle hits the boundary, the
real phase space is a quotient. The points $(\vec x,\vec p)$ and $(\vec
x',\vec p\,')$ are identified if $\vec x=\vec x'$ and if $\vec p$ is obtained
from $\vec p\,'$ by a reflection in the edge containing $\vec x$. Then the phase
space is a union of triangles, labelled by momenta, with some edges identified.
More precisely, let us assume that the angles of the triangle are $\pi r \over
n$, $\pi s \over n$ and $\pi t \over n$, with $r,s,t,n$ four strictly positive
integers satisfying $r+s+t=n$, and gcd$(r,s,t)=1$. Clearly the norm of the
momentum is irrelevant so we can focus on its phase, writing $\vec p = p
e^{i\phi}$. If the triangle lies with its base horizontal, the reflections
through the boundaries change $\phi$ according to 
\bea
\sigma_0 \;&:& \quad \phi \; \longrightarrow \; -\phi - {\textstyle {2\pi s 
\over n}}, \\
\sigma_1 \;&:& \quad \phi \; \longrightarrow \; -\phi + {\textstyle {2\pi r 
\over n}}, \\
\sigma_\infty \;&:& \quad \phi \; \longrightarrow \; -\phi.
\eea
Here $\sigma_v$ denotes the reflection through the boundary opposite to
the vertex $v$, and $0,1$ and $\infty$ correspond respectively to the corners
$r,s,t$. The horizontal base is the edge linking 0 (left corner) to 1 (right
corner). Obviously, the momentum of the particle on its trajectory can take 
$2n$ values (we exclude the momenta leading to singular trajectories in which 
the particle hits the corners). Thus in phase space, the particle moves on a
submanifold consisting of $2n$ copies of the triangle (the billiard), labelled 
by the values of $\phi$ (and a value of $p$), and this gives a foliation of the
phase space. Each such submanifold is a compact combinatorial connected surface
without boundary. It is a surface without boundary because every edge is common
to exactly two triangles, due to the above identification, and it is connected
because one can reach every triangle from any other by a series of reflections.
The surface would not be compact if some angles were irrational, in which case
the leaves of the foliation may well be dense in phase space. From now on we
restrict to the rational case. 

The combinatorial description of the surface made up of the $2n$ triangles is
obviously the same for all initial values of the momentum. The reflection group,
generated by $\sigma_0,\sigma_1$ and $\sigma_\infty$, permutes the triangles
that build  the surface. This action is nothing but the action of the triangular
group as defined in Section 4.1. Our first purpose is to compute the triangular
and cartographic groups, and to find an algebraic model for the surface. Set $r'$
(resp. $s'$, $t'$) for the common factor between $r$ (resp. $s$, $t$) and $n$,
and write $n=r'r''=s's''=t't''$. Let also ${\cal T}_{r, s, t}$ (or simply $\cal
T$ when no confusion is possible) denote the associated combinatorial surface,
which is a generic leaf of the foliation of phase space. 

From the action of $\sigma_v$, the generators $\rho_0$, $\rho_1$, and 
$\rho_{\infty}$ are represented on $\phi$ by a clockwise rotation of angle $2\pi
r \over n$, $2\pi s \over n$ and $2\pi t \over n$ respectively. They commute and 
satisfy $\rho_0^{r''}=\rho_1^{s''}=\rho_\infty^{t''}=1$. They satisfy other
relations as well, as it is clear that the oriented triangular group is
isomorphic to $\z_n$, because, since $r$, $s$ and $t$ are relatively prime, we 
can find integers $a$, $b$ and $c$ such that $ar+bs+ct=1$, so $\rho_0^a\rho_1^b
\rho_{\infty}^c$ is a rotation of angle $\frac{2\pi}{n}$. 
One can also check that the triangular group is the dihedral group of order
$2n$, and that the $\widehat D$ group is
\beq
\widehat D({\cal T}_{r,s,t}) = \cases{
\, \z_{n/3} \times \z_n & \hbox{if } $n=r-s=0 \bmod 3$, \cr
\, \z_n \times \z_n & \hbox{otherwise.}}
\eeq
It is amusing to note that the structure of $\widehat D$ is very reminiscent of 
the complementary series of $su(3)$ modular invariants (also called the
$D$--series). Following Section 4.1, one can use these results to uniformize
${\cal T}_{r,s,t}$. Let us first compute its genus. The triangulation consists
of $2n$ triangles, $3n$ edges, $r'$ vertices of type $0$, $s'$ vertices of type
$1$, and
$t'$ vertices of type ${\infty}$. Indeed for vertices of type 0 for instance,
there are $2r''$, twice the order of $\rho_0$, triangles that meet at each 
vertex of type 0 ($r''$ white and $r''$ black triangles), so that the $2n$ such
vertices get identified by groups of $2r''$, leaving $2n/2r''=r'$ distinct ones.
One obtains the Euler characteristic \cite{auritz}
\beq
2-2g = r' + s' + t' - n. 
\eeq

We already know one way to put a complex structure on $\cal T$ via the
construction of Section 4.1. Let us indicate another equivalent way. We
represent each Euclidean triangle of $\cal T$ in the complex $z$--plane and put
the corresponding complex coordinate in the interior of the triangles back on
$\cal T$. Different choices differ by affinities ($z \rightarrow az+b$), so that
the complex coordinates glue holomorphically along the interior of the edge
common to two triangles. It remains to deal with the vertices. Let us choose
a vertex $v$, of type $\infty$ say. The problem is that at $v$, $2t''$
triangles meet with an incident angle equal to $\pi t/n$, so that the argument 
of $z$ changes by a total amount of $2\pi t \over t'$. If we assume, possibly
after an affinity, that $v$ is at the origin in the complex plane, we can choose
$Z=z^{t'/t}$ as a local parameter in the neighbourhood of $v$ on $\cal T$.
This parameter glues holomorphically with $z$ away from the vertex. Moreover,
the parameters $Z$ of the triangles incident at $v$ glue holomorphically to give
a global coordinate in a small neighbourhood of $v$. The other types of vertices
are treated in an analogous fashion. That this complex structure coincides
with the one given in Section 4.1 is clear. A priori, the Euclidean structure of
the triangle is crucial for the mechanical problem, whereas the complex
structure may look very artificial. However, the differential ${\rm d}z$, which
is crucial for the classical motion (away from the boundary, the equation of
motion says that the velocity $\dot{z}$ is constant), extends holomorphically on
$\cal T$. Close to a vertex, of type $\infty$ say, we have ${\rm d}z 
\propto Z^{\frac{t}{t'}-1}{\rm d}Z$, so that the extension of ${\rm d}z$ has
a zero of order $\frac{t}{t'}-1$. The total number of zeroes is thus 
\beq
r' \left(\frac{r}{r'}-1\right) + s' \left(\frac{s}{s'}-1\right) 
+ t' \left(\frac{t}{t'}-1\right),
\eeq
which is just the opposite of the Euler characteristic, as was to
be expected. We shall see later that the other holomorphic
differentials on $\cal T$ also have a very natural interpretation.

Because the oriented triangular group of $\cal T$ is Abelian, we know from
Section 4.1 that there is a holomorphic map from $F_n$ to the
algebraic curve associated with $\cal T$. Counting the triangles on the two
curves, we see that the map is of degree $n$ and that the algebraic curve
associated to $\cal T$ is the quotient of $F_n$ by a subgroup of
$\widehat{D}_n$ of order $n$, which coincides, if $(n,3)=1$ or $r-s \neq 0 \bmod 3$,
with the isotropy subgroup of a fixed flag. It is not difficult to see that the
group fixing a flag of symbol say $(\infty01)$ on $\cal T$ consists of
the elements $\mu_0^a \mu_1^b\mu_\infty^c$ where the integers $a$, $b$
and $c$ satisfy $ar+bs+ct = 0 \bmod n$. But we know that on the Fermat
curve the corresponding transformation $R_0^{a} R_1^{b} R_{\infty}^{c}$ is
\beq
(u;v;w) \quad \longrightarrow \quad (\xi^{2a}u;\xi^{2b}v;\xi^{2c}w). 
\eeq 
In the affine model the action is $x \rightarrow \xi^{2(a-c)}x$, and $y
\rightarrow \xi^{2(b-c)}y$. The most obvious functions, invariant under these
substitutions, are $X \equiv x^n$, $y^n$ and $Y \equiv x^ry^t$. The first and
third satisfy $Y^n=X^r(1-X)^s$, which is just the equation for $C_{r,s,t}(n)$
defined in the third section. The map $(X,Y) \in C_{r,s,t}(n) \rightarrow X$
has obviously degree $n$, while $(x,y) \in F_n \rightarrow x^n$ has 
degree $n^2$. Thus the intermediate map $(x,y) \in F_n \rightarrow (x^n,
x^ry^t)$ has degree $n$. This implies that the invariants $X$ and $Y$ form a
complete set and that the triangular curve $C_{r,s,t}(n)$ is a (singular) model
for the algebraic curve associated to ${\cal T}_{r,s,t}$.

\smallskip
The holomorphic differentials on $C_{r,s,t}(n)$ are simply the invariant
differentials on $F_n$. Under $R_0^{a}R_1^{b}R_{\infty}^{c}$, the differential
$\omega_{\tilde{r}, \tilde{s}, \tilde{t}}$ given by (\ref{diff}) picks a factor
$\xi^{2(a\tilde{r}+b\tilde{s}+c\tilde{t})}$. For a triplet $(a,b,c)$ satisfying
$ar+bs+ct = 0 \bmod n$, this factor is 1 if and only if there is an integer $h
\in \z_n$ such that $(\tilde{r},\tilde{s},\tilde{t})=(\rest {hr},\rest {hs},
\rest {ht})$ (where as before, $\rest {m}$ is the representative of $m$
modulo $n$ in the interval $[0, n-1]$). The number of such $h$, yielding the
number of holomorphic 1--forms on $C_{r,s,t}(n)$, is equal to the genus. 

Now let $h \in \Z {n}$ be such that $\rest {hr}+\rest {hs}+\rest {ht}=n$. The
triangular group of ${\cal T}_{\rest {hr}, \rest {hs}, \rest {ht}}$ does not
depend on $h$, nor does the description of the complex structure. Hence $C_{r,
s,t}(n)$ and $C_{\rest {hr}, \rest {hs}, \rest {ht}}(n)$ are isomorphic.
Explicitly, if we write $\rest {hr}=hr-\bar{r}n$, $\rest {hs}=hs-\bar{s}n$, we
have the invertible map $(X,Y) \in C_{r,s,t}(n) \longrightarrow (X,
Y^h X^{-\bar{r}}(1-X)^{-\bar{s}}) \in C_{\rest {hr},\rest {hs},\rest {ht}}(n)$.
Moreover, if $h \in \z_n$ has a common factor with $n$, say $d$, the above map 
is still defined, but has degree $d$ and the image is $C_{\rest
{hr}/d,\rest {hs}/d,\rest {ht}/d} (\frac{n}{d})$. Though not
identical, this is quite reminiscent of the third section. We can now come to
another ``coincidence''.

We have seen in Section 2 that the parity rule puts severe restrictions on the
possible couplings among characters in a modular invariant partition function.
This parity rule was expressed in terms of the sets $H_{r,s,t}=\{h \in \Z {n}
\;:\; \rest {hr}+\rest {hs}+\rest {ht}=n \}$ where $(r,s,t)$ were interpreted 
as the affine Dynkin labels of integrable weights. Then the characters
$\chi_{r,s,t}$ and $\chi_{r',s',t'}$ can be coupled only if $H_{r,s,t} = 
H_{r',s',t'}$. We have just seen that $H_{r,s,t}$ also describes the billiards
that are associated to the same triangular curve $C_{r,s,t}(n)$. But there is
a direct and puzzling though incomplete connection between triangular curves
and modular invariants, that in addition involves non--invertible elements of
$\z_n$.

We start with $F_{24}$, the Fermat curve of degree 24. We have seen that
$F_{24}$ is a covering of some triangular curves, which themselves are
coverings of other triangular curves. Let us first consider the triplet
$(1,1,22)$, associated with the character of the identity
operator at height $n=24$, and take all its multiples by elements $h$ of
$\z_{24}$. After reduction modulo 24, we keep only those triplets which have
no zero component and whose sum is equal to 24, obtaining in this way 11
triplets (or triangles). If one classifies them according to the genus of
the associated triangular curve, one finds 4 triangles of genus 11, 2 of genus 
5, 2 of genus 3, 2 of genus 2, and 2 of genus 1, the last two being associated
with two non--isomorphic genus 1 surfaces. Thus there are six different curves
which are involved. They are all isomorphic to $C_{1,1,n_0-2}(n_0)$ for some
$n_0$ dividing 24. The six curves are shown in Figure 3, where the arrows
denote covering maps.

\begin{figure}[htb]
\leavevmode
\begin{center}
%\mbox{\epsfscale 1000 \epsfbox{papillon.eps}}
\mbox{\epsfbox{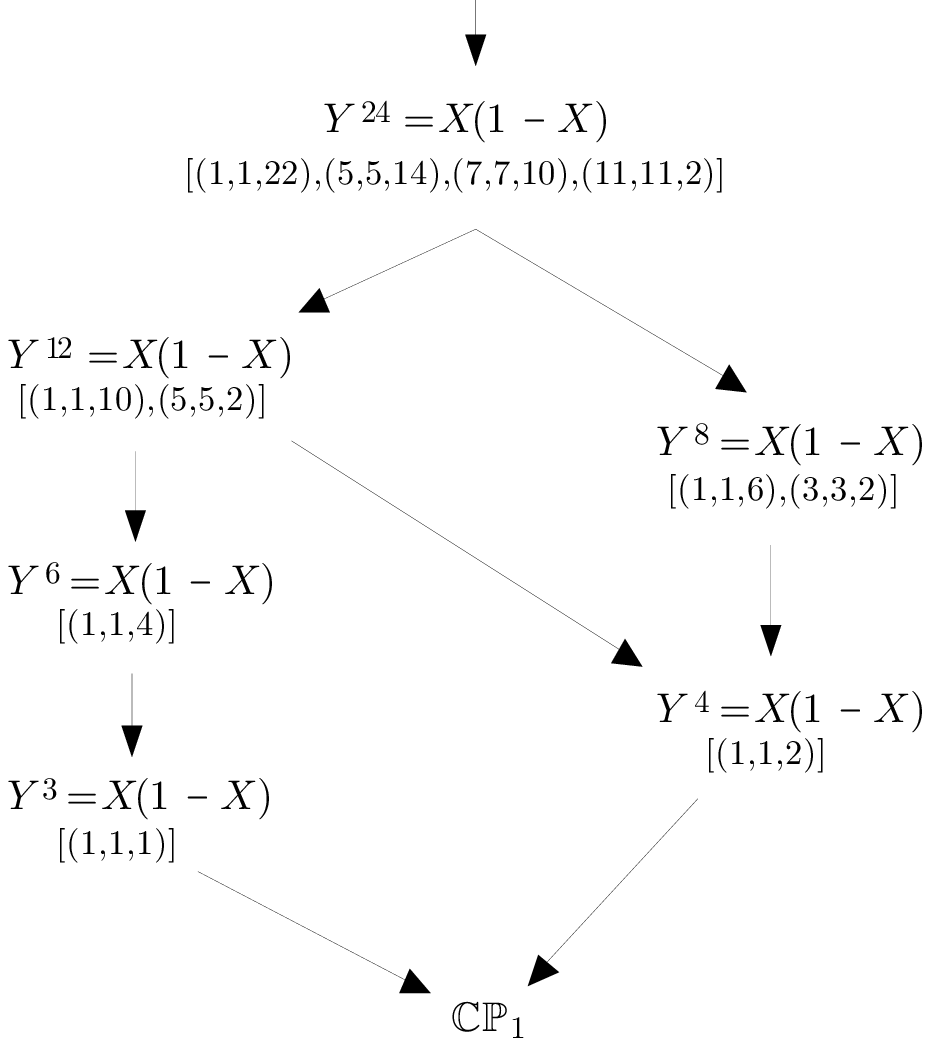}}
\begin{minipage}{14cm}
\bigskip \bigskip \bigskip
{\it Figure 3.} Triangular curves $C_{1,1,n_0-2}(n_0)$ related to $su(3)$ 
modular invariants. The top three specify the identity blocks of three
exceptional invariants, at height $n_0=24, 12$ and 8. The two elliptic curves,
corresponding to $(1,1,2)$ ($n_0=4$) and $(1,1,1)$ ($n_0=3$), are not
isomorphic, having modulus $\tau=i$ and $\tau=e^{2i\pi /3}$ respectively. 
\end{minipage}
\end{center}
\end{figure}

The puzzling observation one can make is the following, and concerns the type I
exceptional $su(3)$ modular invariants (those which can be written as a sum of
squares with only positive coefficients). One observes that the triangles
associated with $C_{1,1,n_0-2}(n_0)$ for $n_0=24,12$ and 8 give precisely the
content of the block of the identity in the exceptional modular invariant at
height $n_0$. The only element which is not encoded in the picture is whether a
character that is labelled by the permutation of a triangle appears or not. More
precisely, one sees that:

--- For $n_0=24$ : the four triangles (of genus 11) are $(1,1,22)$, $(5,5,14)$,
$(7,7,10)$ and $(11,11,2)$. The exceptional invariant at height 24 is 
\beq
E_{24} = |\chi_{(1,1,22)} + \chi_{(5,5,14)} + \chi_{(7,7,10)} + \chi_{(11,11,2)} 
+ \hbox{all perm.}|^2 + \ldots
\eeq 
so all permutations appear.

--- For $n_0=12$ : the two triangles are $(1,1,10)$ and $(5,5,2)$, and the
invariant partition function reads
\beq
E_{12} = |\chi_{(1,1,10)} + \chi_{(5, 5, 2)} + \hbox{all perm.}|^2 + \ldots
\eeq
so again all permutations appear.

--- For $n_0=8$ : there are two triangles, $(1,1,6)$ and $(3,3,2)$. The
partition function reads
\beq
E_8 = |\chi_{(1, 1, 6)} + \chi_{(3, 3, 2)}|^2 + \ldots
\eeq
and the permuted symbols appear in other blocks.

The same pattern persists for the smaller values of $n_0$: for $n_0=6$, the
triangle $(1,1,4)$ specifies the identity block in the diagonal and 
complementary invariants depending on whether permutations are included or not,
whereas for $n_0=4$ and 3, the identity blocks of the diagonal invariants are
reproduced. One is tempted to apply the same idea to the other blocks of the
exceptional invariants, starting for instance with the triangle $(1,7,16)$
appearing in the second block of $E_{24}$. Alas, the outcome is disappointing,
and that is one of the reasons to believe that our observations, however
troublesome, are mere coincidences.

\section{The Riemann surface of a RCFT on the torus}

In this last part, we would like to see to what extent the action
of the modular group on the characters of a general rational
conformal field theory can be related
to its action on algebraic curves, which we might then want to identify. 
In particular, rational conformal field theories like to organize in 
families indexed by integers (for example the height in WZNW models), and
it is therefore a natural question to ask whether these families can be put in
correspondence with families of curves, just like the $su(3)_k$ WZNW models
are related to the Fermat curves in the way detailed in the previous
sections. We show here that a compact Riemann surface can be canonically
associated with any rational conformal field theory. Each such
Riemann surface has an algebraic model, but to compute it explicitly turns
out to be in practice difficult. Nonetheless general features can be
established. We will present the complete details for the
$su(3)$ WZNW models, at level $k=1$ and $k=2$. The most na\"{\i}ve 
hope would have been that the associated algebraic curves are the Fermat curves
of degree 4 and 5 respectively, but as we shall see, this is not the case.
The surprise however is that the curve associated with $su(3)_1$ possesses a
covering by the Fermat curve of degree 12, and is nothing but one of the
triangular curves. 

\subsection{General setting}

We start with some general facts, which are true for the WZNW models,
but that otherwise might well be consequences of the general axioms that a
rational conformal field theory has to fulfill. Since we are not aware of a
complete derivation of them, we content ourselves with listing them as mere
assumptions.

\begin{enumerate} 
\item The theory involves only a finite number $N$ of representations of
the chiral algebra. We denote them by ${\cal R}_p$, $0 \leq p \leq N-1$,
with the convention that ${\cal R}_0$ contains the identity operator (or the
operator of smallest conformal weight in the non--unitary case).

\item Chiral restricted characters are well--defined, that is, 
\beq 
\chi_p(\tau) \equiv tr_{{\cal R}_p} e^{2i\pi\tau(L_0-c/24)} 
\eeq 
is holomorphic in the upper--half plane. Two restricted characters are equal if
they correspond to complex conjugate representations. These are often the
only linear relations among them \footnote{A counterexample is provided
by the affine algebra $\widehat D_4$ at $k=1$, where, because of the triality,
three inequivalent representations have the same restricted character
\cite{kac}.}.

\item There exist unitary matrices $S_{p,p'}$ (symmetric)
and $T_{p,p'}$ (diagonal) such that 
\beq
\chi_p(-1/ \tau)=\sum_{p'} S_{p,p'}\chi_{p'}(\tau), \qquad \chi_p( \tau+1)
= \sum_{p'} T_{p,p'}\chi_{p'}(\tau). 
\eeq 
They both have finite order and their entries are in a finite 
Abelian extension of $\Q$ (a simple consequence of \cite{cg}).
The square of $S$ is the charge conjugation. These matrices 
yield a representation of $\sl$ through the map $\left({0 \atop -1} \; {1 \atop
0}\right) \rightarrow S$, $\left({1 \atop 0} \; {1 \atop 1}\right)
\rightarrow T$. The restriction of this representation to the subspace of
conjugation invariants descends to a representation of $\psl$. This is the
subspace we shall be dealing with in the sequel\footnote{When other linear
relations among the characters exist, we simply pick a maximal set of
linearly independent characters and work with these.}. 

\item The kernel of this representation of $\psl$ (or
equivalently of $\sl$ for the original representation) is very
large \cite{gw}. More precisely, the kernel is an invariant subgroup,
call it $\G$, of finite index in $\psl$. The intuitive reason is
that the characters can be written in terms of theta
functions of a Euclidean lattice, and that $S$ and $T$ are closely
related to the finite Fourier transform on a finite group, namely the
quotient of the lattice by a sublattice of finite index. A proof for
affine algebras is contained in \cite{kac}. A pedestrian approach in the case
of $su(N)_k$ WZNW models, showing that in fact the principal congruence
subgroup $\G _{2N(N+k)}$ is in the kernel, can be found in \cite{bi}.
A general and more conceptual proof, starting from axiomatics of rational
conformal field theory,  would be very interesting. 
\end{enumerate}

These assumptions lead naturally to the following construction. 
The kernel $\G$ is a Fuchsian group, and the quotient of the upper--half
plane $\h$ by $\G$ defines a Riemann surface $\S$ with punctures, which
has a well--defined compactification $\oS$. The surface $\S$ may be described as
the union of $\widehat\g F$ for all $\widehat\g \in \widehat \G \equiv
\psl/\G$, where $F$
is a fundamental domain for $\h/\psl$, for instance
\beq
F=\{\tau \in \h \;:\; |\tau| \geq 1 \hbox{ and } |{\rm Re}\,\tau| \leq 1/2\}.
\eeq
$\S$ has punctures located at the images of $\tau=i\infty$ under $\widehat
\G$. When one compactifies $\S$ by filling these punctures, one gets the
compact surface $\oS$, of which a triangulation is given by $\{\widehat\g F 
\;:\; \widehat\g \in \widehat \G\}$ where those edges and vertices equivalent
under $\G$ are to be identified. 

In the rest of this section, we will consider in more detail the association
RCFT $\longrightarrow$ compact Riemann surface $\oS$ \footnote{Because $\oS$ is
defined from the representation carried by the full set of independent
characters, we could say that it is the surface associated with the diagonal
RCFT. In the same way, one can associate Riemann surfaces to subrepresentations
corresponding to non--diagonal theories.}. More specifically, one would like to
answer three questions:

--- What general features do these Riemann surfaces have ?

--- How explicitly can we describe them, for instance by giving
equations for an embedding in some affine or projective space ?

--- Is there some nice way to characterize the family of
Riemann surfaces that arise in this way from rational conformal field theories ?

\medskip
Although we have not been able to answer the third question, we can
nevertheless make definite statements about these surfaces. 
Useful references for  quotients of $\h$ by Fuchsian groups are
\cite{gunning,josi}.

As $\G$ is an invariant subgroup of $\psl$, we can draw
general conclusions about the quotient $\h/\G$. The projection $\h/\G = \S 
\longrightarrow \h/\psl \cong \C$ ---this last equivalence being via the 
standard modular function $j(\tau)$--- has a holomorphic extension $\oS
\longrightarrow \C \PP _1$, ramified only at $0$, $1728$ and $\infty$
\footnote{At this point, it would be easy to make contact with
the formalism of triangulations briefly presented in Section 4.1.
However, starting from the projection $\oS \rightarrow
\C \PP _1$, the algorithm described in Section 4 would
construct $\S'$, a quotient of $\h$ by a subgroup of $\G _2$
with more punctures than $\S$, but of course such that $\oS =
\oS '$. In particular the cartographic group is closely
related, but not equal, to $\psl /\G$. But other meromorphic functions
ramified only over three points can be used to do cartography. This relationship
with triangulations will be explicited in a specific example, in the next
section.}. The group
$\psl$ has unique invariant subgroups of index 1, $2$ (related to
the fact that $j-12^3=216^2 g_3^2/\eta^{24}\equiv j_{1/2}^2$ is a perfect 
square, the $q$-expansion of $j_{1/2}$ starting as $q^{-1/2}$) and $3$
(related this time to the fact that $j=12^3 g_2^3/\eta^{24} \equiv j_{1/3}^3$ is
a perfect cube \footnote{$j_{1/3}$ is the character of the only representation
of $\widehat E_8$, level $1$.}, the $q$-expansion of $j_{1/3}$ starting as
$q^{-1/3}$). The associated compact Riemann surfaces have
genus 0. If the index $|\widehat{\G}|$ is bigger or equal to 4, the ramification
structure of  the projection map $\oS \rightarrow \C \PP _1$ is fixed : the
ramification index is $2$ above $j=1728$, $3$ above $j=0$ and $n_{\infty}$ (the
order of $T$ in $\widehat\G$) above $j=\infty$. This implies that the Euler 
characteristic of $\oS$ is 
\beq 
2-2g_{\oS}= |\widehat{\G}|\frac{6-n_{\infty}}{6  n_{\infty}}.
\label{genus}
\eeq 
The number of punctures (or cusps) is equal to $|\widehat{\G}|/n_{\infty}$. 
Let us describe in some detail the easiest cases, namely the surfaces of genus
0 and 1. 

If $|\widehat{\G}| \geq 4$ and $g_{\oS}=0$, there are four
possibilities (platonic solids): $n_{\infty}=2$, $|\widehat{\G}|=6$ (the 
dihedral group of order  3, $\G=\G _2$ ); $n_{\infty}=3$, $|\widehat{\G}|=12$
(the symmetry group of the tetrahedron $\G=\G _3$); $n_{\infty}=4$,
$|\widehat{\G}|=24$ (the symmetry group of the octahedron $\G=\G _4$)
and $n_{\infty}=5$, $|\widehat{\G}|=60$ (the symmetry group of the
icosahedron $\G=\G _5$). 

If $n_\infty=6$ the resulting quotient is a torus, and there is now an infinite
sequence of nested invariant subgroups $\Lambda_n$ of $\psl$ containing $T^6$
and of finite index. We denote the corresponding quotients by $\widehat
\Lambda_n$, $n=1,2,3,\ldots$. Let us start with the smallest one, $\widehat
\Lambda_1$, which is the quotient of $\psl$ by its commutator subgroup
$\psl^{comm}$. It has order 6, and is isomorphic to the cyclic group $\z _6$ (a
simple consequence of $S^2=(ST)^3=1$ plus $ST=TS$). Both $j_{1/2}$, the square
root of $j-12^3$, and $j_{1/3}$, the cube root of $j$, carry a one--dimensional
(hence Abelian) representation of $\psl$, and they generate the function field
of the quotient $\h/\psl^{comm}$. This algebraic curve is, as announced, a torus
since $j_{1/2}^2=j_{1/3}^3-12^3$. It  is isomorphic to the cubic Fermat curve,
although its uniformization by $j_{1/2}$ and $j_{1/3}$ is not the one we gave
earlier. The other $\widehat \Lambda_n$ can be constructed as quotient groups 
of their projective limit $\widehat \Lambda_\infty$. 

$\widehat \Lambda_\infty$ is the largest factor group with $n_\infty = 6$,
and by definition is the quotient of $\psl$ by the smallest invariant subgroup
containing $T^6$. It thus has the presentation $\langle S,T \;|\; S^2,(ST)^3,T^6
\rangle$, is of infinite order, and is isomorphic to the cartographic and
symmetry group of the regular triangulation of the plane. To see this
concretely, set $\rho=e^{i\pi/3}$, and let $s$ and $t$ be the Euclidean
transformations of the complex plane given by $s \, : \, z \rightarrow 1-z$ and
$t \, : \, z \rightarrow \rho z$. Then obviously $s^2=t^6=1$, and one checks
that $(st)^3=1$ as well. So the group generated by $s$ and $t$ is a quotient of
$\widehat \Lambda_\infty$. The transformation $a=st^3$ is simply the translation
$z \rightarrow z+1$, and conjugating by $t$, we find that
$b=tst^2$ is the translation $z \rightarrow z+\rho$. Further
conjugations by $t$ give unit translations along the other axes
of the lattice generated by $1$ and $\rho$. So the group generated by $s$ and
$t$ is the semi--direct product of the translations of the lattice and
the rotations generated by $t$, that is, the full symmetry group
of the lattice. On the other hand one can check explicitly that in
$\widehat \Lambda_\infty$, $A=ST^3$ and $B=TST^2$ commute. As a consequence, 
any element of $\widehat \Lambda_\infty$ can be written in a unique way as
$T^jA^pB^q$ with $j$ between $0$ and $5$, and $p$ and $q$ in $\z$.
Indeed we first check that $S=T^3A^{-1}$, $ST=T^4A^{-1}B$, $ST^2=T^5B$,
$ST^3=A$, $ST^4=TAB^{-1}$ and $ST^5=T^2B^{-1}$. Using these, one checks that 
the set of elements of $\widehat \Lambda_\infty$ that have a decomposition
$T^jA^pB^q$ contains $1$, is stable under multiplication on the left and on the
right by  $T$ and $S$, so that this set is $\widehat \Lambda_\infty$. The
decomposition is unique because the corresponding decomposition in the group
generated by $s$ and $t$, a priori a quotient of $\widehat \Lambda_\infty$, is
well--known to be unique. 

Now for finite $n$, $\widehat \Lambda_n$ is the quotient of
$\widehat \Lambda_\infty$ by the further relation $A^n=1$ (or equivalently
$B^n=1$), and has the presentation
\beq
\widehat \Lambda_n = \langle S,T \;|\; S^2,(ST)^3,T^6,(ST^3)^n \rangle.
\eeq 
The order of $\widehat \Lambda_n$ is $6n^2$. The corresponding Riemann surfaces
$\h /\Lambda_n$ are all isomorphic to the cubic Fermat curve $F_3$ (with the
torsion points of order $n$ as punctures). This is because $T$ induces a cyclic
group of automorphisms of order 6 of the associated torus, fixing a point (the
coset of the point at infinity in $\h$). For $n=3$, the uniformization of $F_3$
by $\h /\Lambda_3 = \h /C_3$ is the one we gave earlier in Section 4.2. 

The $\widehat \Lambda_n$, $n \geq 1$, do not exhaust all factor groups of
$\psl$ of finite order with $n_{\infty}=6$, but all of them are factor
groups of the $\widehat \Lambda_n$. For instance, quotienting $\widehat
\Lambda_\infty$ by the relation $AB=1$ ---it implies $A^3=1$---, one obtains a
group of order 18 which is $\widehat \Lambda_3/\langle ST^2ST^{-2} \rangle$.

Another common feature of all the Riemann surfaces arising
from our construction is that they have a rather large group of
automorphisms. This group contains $\widehat{\G}$, but in fact, unless 
$\oS$ is a sphere or a torus (in which case the automorphism group is
infinite), $\widehat{\G}$ is the full automorphism group of $\oS$. That 
$\widehat \G$ is the group of automorphisms of $\S$ is the consequence of a
general result (see f.i. \cite{josi}), and the statement relative to the compact
surface $\oS$ is proved in Appendix A. 

We now come to the question of the explicit and concrete description of $\oS$,
by means of algebraic equations. We do this by looking at the function field of
$\oS$.

The restricted characters are holomorphic on $\S$, and meromorphic on
$\oS$. This is proved by a simple analysis of their behaviour at the punctures.
Characters are meromorphic at the infinite parabolic point because the
eigenvalues of $L_0$ in the representations of the chiral algebra are rational
and bounded below. To conclude for another puncture (necessarily a
rational point on the real axis), one chooses an element $g$ of $\psl$ that maps
it to $\tau=i\infty$ and  use the fact that $\psl$ acts linearly on the
characters. This shows that the singularity of $\chi_p$ at a puncture is
at most as strong as the singularity of $\chi_0$ at $\tau=i\infty$. It is weaker
if the matrix element of $g$ between $\chi_p$ and $\chi_0$ vanishes. This
proof parallels the argument showing that $S_{0,p}$ is real positive for any $p$.

The function field $\cal M$ of $\oS$ certainly contains all rational
functions of the characters $\chi_p$ and of the modular invariant $j$. That it
contains nothing more can be proved in the following way. A classical
theorem states that, given a non--constant meromorphic function $f$
on $\oS$ of degree $d$, the function field ${\cal M}$ is a simple Galois 
extension of $\C(f)$ of degree $d$ \cite{carl}. That is, there exists a function
$g$ satisfying an irreducible polynomial equation of degree $d$ with
coefficients in $\C(f)$, in terms of which ${\cal M} =\C(f)(g)=\C(f,g)$. 
Choosing $f=j(\tau)$, of degree $|\widehat \G|$, shows that ${\cal M}$ is a 
Galois extension of degree $|\widehat \G|$ of $\C(j)$, with Galois group ${\rm
Gal}({\cal M}/\C(j))=\widehat\G$. Now $\widehat\G$ acts linearly on the 
characters, and induces distinct automorphisms of the field ${\cal M}_\chi$
generated by the $\chi$'s and $j$, that all fix $\C(j)$. This implies that
${\cal M}_\chi$ is a subfield of ${\cal M}$ of degree $|\widehat \G|$ over
$\C(j)$, hence equal to ${\cal M}$. 

Thus we have ${\cal M}={\cal M}_\chi=\C(\chi_0,\chi_1,\ldots,\chi_{N-1},j)$, but
since it is also equal to $\C(j,g)$ for some $g$, the field is not freely
generated by the characters and $j$. Let us consider the set $I_{\oS}$ of all
polynomial relations $P(\chi_0,\ldots,\chi_{N-1},j)=0$. It is fixed by 
$\widehat\G$ since $\widehat\G$ acts by automorphisms of
$\C(\chi_0,\ldots,\chi_{N-1},j)$. Moreover, any polynomial
$P(\chi_0,\chi_1,\ldots,\chi_{N-1})$ invariant under the action of $\widehat\G$
on characters is a modular invariant function  of $\tau$, holomorphic in the
upper--half plane without poles at finite distance, so a polynomial in $j$,
say $Q(j)$. Hence
\beq
P(\chi_0,\chi_1,\ldots,\chi_{N-1}) - Q(j) = 0,
\label{seq}
\eeq
and this yields an element of $I_{\oS}$  \footnote{As 
mentioned before, it can be useful to deal with the numerators of
characters rather than with the characters themselves. One way to do it is to
look at the ring of projective invariants (invariants up to a phase for the
action of $\widehat{\G}$ on characters). Those are polynomial in $j_{1/2}$ and
$j_{1/3}$.}.

Every relation of $I_{\oS}$ gives an equation for $\oS$. More precisely, the
locus in $\C^{N+1}$ where all relations are satisfied is (by definition) an
algebraic  variety, and evaluation of $\chi_0, \ldots, \chi_{N-1}$ and $j$ gives
a holomorphic map from $\oS$ into this variety, injective except perhaps at a
finite number of points. But in fact the elements of $I_{\oS}$ corresponding to
invariants under $\widehat\G$ (of the form (\ref{seq})) give a complete set of
equations for $\oS$, because they describe a covering of the $j$--sphere
containing the model for $\oS$ with at most $|\widehat\G|$ leaves. Indeed for 
every value of $j$, we get a value for the invariants $P_i$, and we know from
invariant theory \cite{benson} that there are at most $|\widehat{\G}|$ points in
$\C^N$ corresponding to these values (namely the points in an orbit). Hence this
covering is an affine model for $\oS$.  Let us also observe that this affine
model is well--suited to deal with questions concerning the values characters
can take. For instance, the divisors of the characters (namely the zeroes and
the poles) are nicely encoded.

From the set of Eqs (\ref{seq}) (for all $\widehat\G$--invariant polynomials
$P$), elimination leads to irreducible polynomial relations between every single
character and $j$. One easily sees that, if $\Delta_p$ denotes the stabilizer of
$\chi_p$ in $\widehat\G$, these equations take the form
\beq
\prod_{\widehat \g \in \widehat\G/\Delta_p} \; (X - \widehat\g \chi_p) = 0.
\label{stab}
\eeq
The coefficients of this monic equation are polynomials in $j$, since they
are symmetric functions of the roots, which are themselves linear
combinations of the characters, holomorphic in $\h$. Loosely speaking, this 
means that the characters are algebraic integers over $\Q (j)$.

In fact these arguments also solve the problem of the \lq\lq second generator''
of the function field ${\cal M}$, and provide another model for $\oS$. Since 
the matrices $S$ and $T$ generate a finite group of order $|\widehat\G|$, we 
can  find a linear combination $g=\sum_p c_p\chi_p$ of the characters such  
that its orbit under $\widehat\G$ is of cardinal $|\widehat\G|$. From this
follows that $g$ satisfies the irreducible polynomial equation
\beq
\prod_{\widehat \g \in \widehat\G} \; (X - \widehat\g g) = 0.
\label{model}
\eeq
As before the coefficients of $X^k$ are polynomials in $j$. The simple
extension of $\C(j)$ obtained by adding $g$ and its powers is of degree
$|\widehat\G|$, and is therefore equal to the whole of the function field
${\cal M}=\C(j,g)$. The equation (\ref{model}) is a plane curve that is a 
(highly) singular model for $\oS$. 

\smallskip
Finally let us comment about some automorphisms of $\widehat\G$. 
Set $M_{\rm restr}=\Q(S^{\rm restr}_{p,p'},$ $T_{p,p'})$, the algebraic 
extension generated by the matrix elements of $T$ and $S$, {\it acting on the
independent restricted characters}. We want to show that the Galois group
Gal($M_{\rm restr}/\Q$) acts as automorphisms of $\widehat\G$. This is obvious
for $T$, because for every Galois transformation, $\sigma(T)=T^h$ for
an integer $h$ coprime with $n_\infty$, the order of $T$. That $\sigma(S)$ is
also a word in $T$ and $S$ is less trivial, but can be seen as follows. 

First we show that the ideal $I_{\oS}$ can be generated by polynomials with
integral coefficients. In the language of algebraic geometry, this says that
$\oS$ is defined over $\Q$, a property shared with the Fermat curves. The point
is that the insertion of the Puiseux series for $\chi_p$ and $j$ ---they all 
have integral coefficients--- in a polynomial $P(\chi_0,\ldots,\chi_{N-1},j)$
shows that the condition for $P$ to belong to $I_{\oS}$ is expressed by a
linear system with integral entries, the unknowns being the coefficients
of $P$. We can therefore choose a basis of solutions with integral or
rational coefficients. We call it an integral basis.

For $\sigma \in {\rm Gal}(\overline {\Q}/\Q)$, we extend the action of $\sigma$
on polynomials by acting trivially on the characters and $j$. Now let $P$
be a polynomial in $I_{\oS}$ with integral coefficients and $X$ be an
element of $\widehat\G$ (for instance $S$). We make the following observations.
First $P(\sigma(X) \cdot \chi,j) = \sigma \big(P(X \cdot \chi,j)\big)$
because $P$ has integral coefficients. But $P(X \cdot \chi,j)=P(X \cdot
\chi,X \cdot j)$  because $j$ is invariant under $\widehat\G$. Next 
$P(X \cdot \chi,X \cdot j)(\tau)= P(\chi,j)(X\tau)$ which is identically $0$
because $P(\chi,j)$ belongs to $I_{\oS}$. Hence $P(X \cdot \chi,j)$ belongs
to $I_{\oS}$, and can be expressed as a linear combination (with complex
coefficients) of elements of an integral basis for $I_{\oS}$. The Galois
transformation $\sigma$ acts trivially on the integral basis, so that 
$P(\sigma(X) \cdot \chi,j)=\sigma \big(P(X \cdot \chi,j)\big)$ is in
$I_{\oS}$ as well. This means that $\sigma(X)$, a linear transformation of the
characters, induces an automorphism of $\S$ fixing $j$. The proof is finished
since $\widehat\G$ is the set of all such automorphisms. This
proves at the same time that the extension $M_{\rm restr}$ of $\Q$ is a Galois
extension, as already known from \cite{cg}.

When $S$ and $T$ correspond to the modular transformations of affine
characters, one can be more explicit. As mentioned above, there is a principal
congruence subgroup, say $\G_N$, in the kernel of the representation generated
by $S$ and $T$, so that they form a representation of $PSL_2(\z_N)$, for some
$N$. In this case, it is conjectured (and shown for large families of examples)
that the cyclotomic Galois transformation $\sigma_h$ acts on $S$ by multiplication
by the matrix representing the group element $\left({h \atop 0} \; {0 \atop
h^{-1}}\right)$ of $PSL_2(\z_N)$ (thus $h^{-1}$ is the inverse of $h$ modulo $N$)
\cite{bauer}. This result implies the following action of $\sigma_h$ on $S$
\beq
\sigma_h(S) = ST^{h^{-1}}ST^hST^{h^{-1}}S.
\label{galois}
\eeq
Illustrations  of this formula are given in the next section and in Appendix B.

\subsection{The Riemann surface of su(3) level 1}

We could illustrate the machinery of the previous section on various rational
conformal field theories, but as $su(3)$ was central in our previous
investigations, we shall give here the complete treatment of $su(3)$ at level 1,
relegating to an appendix the case of $su(3)$, level 2, already much more
complex. It will soon become clear that explicit computations of the Riemann
surface of a rational conformal field theory tend to be painful. To compute
even the genus of the surface is quite a challenge, since most of the
time very little is known about the finite group $S$ and $T$ generate.

\medskip
The affine Lie algebra $\widehat{su(3)}_1$ has three integrable representations,
corresponding to the shifted weights $(1,1)$, $(1,2)$ and $(2,1)$.
To simplify the notations, we denote the three independent characters as
$\chi_0 \equiv \chi_{(1,1)}$, $\chi_1 \equiv \chi_{(1,2)}$, $\chi_2 \equiv
\chi_{(2,1)}$. Setting $\xi= e^{2i\pi /12}$ and $\om=e^{2i\pi /3}$, the
expressions for $S$ and $T$ in the basis $(\chi_0,\chi_1,\chi_2)$ are 
\beq
S = {1 \over \sqrt{3}} \pmatrix{1&1&1 \cr 1&\om&\om^2 \cr 1&\om^2&\om \cr},
\qquad
T = \pmatrix{\xi^{-1}&0&0 \cr 0&\xi^3&0 \cr 0&0&\xi^3 \cr}.
\eeq
The restricted characters $\chi_1$ and $\chi_2$ being equal, we are left
with a two--dimensional representation of the modular group, given in the basis
$(\chi_0,{\chi_1 + \chi_2 \over 2})$ by
\beq
S = {1 \over \sqrt{3}} \pmatrix{1&2 \cr 1&-1 \cr}, \qquad
T = \pmatrix{\xi^{-1}&0 \cr 0&\xi^3 \cr}.
\label{repres}
\eeq
The extension defined in the previous section is clearly $M_{\rm
restr}=\Q(\xi)$, with Galois group over $\Q$ isomorphic to
$\z^*_{12}=\{1,5,7,11\}$. One easily obtains its action on $\widehat\G$: 
$\sigma_5 (S,T) = (ST^6,T^5)$, $\sigma_7(S,T) = (S,T^7)$ and  
$\sigma_{11}(S,T) = (ST^6,T^{11})$, in agreement with the general formula 
(\ref{galois}).

Finally the Weyl--Kac formula gives the following Puiseux series for the
restricted characters 
\bea
&& \chi_0 = q^{-1/12} \, [1 + 8q + 17q^2 + 46q^3 + 98q^4 + \ldots], \\
&& \chi_1 = q^{1/4} \, [3 + 9q + 27q^2  + 57q^3  + 126q^4 + \ldots]. 
\eea

\smallskip
As before, let $\G$ be the kernel (in $\psl$) of the representation
(\ref{repres}) and $\widehat{\G}$ be the corresponding quotient. Let us recall
that our main interest is the study of the Riemann surface with punctures $\S =
\h/\G$ and the corresponding compact surface $\oS$. Our first task is to
compute the order of $\widehat\G$. 

It is quite clear that $T$ has order 12 in $\widehat{\G}$, but one can also 
observe that $T^3$ is central, namely it commutes with $S$. Consequently
$\widehat\G$ is a quotient group of $\tilde{\G} = \langle S,T \;|\;
S^2,(ST)^3,T^{12}, ST^3ST^{-3} \rangle$. The tetrahedron group $\G_3 = \langle
S,T \;|\; S^2,(ST)^3,T^3 \rangle$, of order 12, is itself a quotient group of
$\tilde{\G}$ by the cyclic subgroup generated by $T^3$. Since $T^3$ is of order
4 in $\tilde{\G}$, it follows that $|\tilde{\G}|=48$. On the other hand, the
elements $T^i$ and $ST^i$ for $0 \leq i \leq 11$, are all distinct in 
$\widehat\G$ because $S$ is not a power of $T$ (its matrix representation is not
diagonal). Moreover $STS=T^{-1}ST^{-1}$ gives still a distinct element, because 
$T^{-1}ST^{-1}$ is not a power of $T$, and $STS$ is not equal to $S$ times a
power of $T$. Thus $|\widehat\G| > 24$, which implies $\widehat\G = 
\tilde{\G}$ (since the order of $\widehat\G$ must divide that of $\tilde{\G}$).
The genus of $\oS$ is then given by (\ref{genus}), which yields $g=3$ (same
genus as the Fermat curve $F_4$). The number of punctures of $\S$ is
$|\widehat{\G}|/n_{\infty}={48 \over 12}=4$.

As we have seen in the previous section, invariant theory for the 
two--dimensional representation of $\widehat{\G}$ on characters will give us a
description of $\oS$. Invariants for finite groups can be computed in a
systematic way, but the procedure is generally cumbersome, so that it may
be more efficient to find shortcuts based on geometrical insights. We used both
to obtain the following results (as well as those of Appendix B).

The generating function for the number of invariants of degree $n$, call it
$d_n$, under the representation (\ref{repres}) of $\widehat\G$ is given by the
Molien series \cite{benson}
\beq
F(t) = \sum_{n=0}^\infty d_nt^n = {1 \over |\widehat\G|} \sum_{\widehat\gamma 
\in \widehat\G} {1 \over {\rm det}\, (1-t\widehat\gamma)} = {1 \over
(1-t^4)(1-t^{12})}.
\eeq
The ring of invariants is freely generated \footnote{An $n$--dimensional 
representation of a finite group gives rise to a ring of invariants which is 
freely generated by algebraically independent invariants iff the group is
generated by pseudo--reflections in an $n$--dimensional complex space
\cite{pseudoref}. This last paper establishes the classification of such groups. 
The list, containing 37 entries, can also be found in \cite{benson}.} by two
algebraically independent invariants, of degree 4 and 12, which one can choose
as 
\bea
&& P_4(\chi_0,\chi_1) = \chi_0^3\chi_1 - \chi_1^4 = 3, \label{p4} \\
&& P_{12}(\chi_0,\chi_1) = [\chi_0^4 + 8 \chi_0\chi_1^3]^3 = j. \label{p12}
\eea
As mentioned before, these invariants are modular invariant functions of $\tau$,
holomorphic in $\h$, that are determined from the Puiseux series of $\chi_0$ 
and $\chi_1$ by looking at the singular terms in $q$. 

The first invariant (\ref{p4}) alone yields a plane curve which is a 
non--singular model for the Riemann surface associated with the affine
model $su(3)$ at level 1:
\beq
x^3y - y^4 = 3z^4. 
\label{proj}
\eeq
Being smooth of degree 4, its genus is equal to 3, as expected. The
second invariant (\ref{p12}) gives $\oS$ as a covering of the
$j$--sphere of degree 48, and allows to compute the divisors of the two
restricted characters. In terms of the projective coordinates $(x;y;z)$, 
$\chi_0$ has four simple zeroes located at $(0;y_0;1)$ with $y_0^4=-3$ (where
$j=0$), and four simple poles at $(1;0;0)$ and $(\om^k;1;0)$ for $k=0,1,2$.
Similarly $\chi_1$ has a triple zero at $(1;0;0)$ and three simple poles at 
$(\om^k;1;0)$ for $k=0,1,2$. In particular, $\chi_0$ and $\chi_1$ are of degree
4 and 3 respectively. 

It is not difficult to obtain the polynomial equations relating the characters
to $j$, as in (\ref{stab}). The stabilizer of $\chi_1$ in $\widehat\G$
is of order 3 ($T^4 \chi_1 = \chi_1$), so that the irreducible equation
relating $\chi_1$ to $j$ is of degree 16, but easily seen to be of degree 4 in
$\chi_1^4$ because $\chi_1,\xi^3\chi_1,\xi^6\chi_1$ and $\xi^9\chi_1$ all solve
the same equation. The coefficients of the polynomial are easily determined
from the Puiseux series of $\chi_1$, and one finds
\beq
(1+3X^4)^3(3+X^4) - {j \over 27}X^4 = 0.
\eeq
This polynomial defines an algebraic extension of $\C(j)$ of degree
48, isomorphic to the function field of $\oS$. The other character $\chi_0$ has
a trivial stabilizer, and therefore generates, along with $j$, the function
field of $\oS$. It satisfies a polynomial equation of degree 48, this time of
degree 4 in $\chi_0^{12}$ (same reason as above), but more complicated:
\bea
&&\hskip -2.5truecm X^{48} + {\textstyle{4(27\,648-61j) \over 243}}X^{36} +
{\textstyle{2(6\,115\,295\,232+16\,809\,984j+ 365j^2) \over 177\,147}}X^{24}
\nonumber\\ 
&&\hskip -1.7truecm + {\textstyle
{4(338\,151\,365\,148\,672-256\,842\,399\,744j+42\,633\,216j^2-547j^3) \over
387\,420\,489}}X^{12} + {\textstyle{j^4 \over 387\,420\,489}} = 0.
\eea
This extension, again of degree 48 over $\C(j)$, defines the function field as
the simple extension $\C(j,\chi_0)$, and provides another projective model for
$\oS$, this time very singular.

It remains to show that $\oS$ is distinct from the Fermat curve $F_4$. This
one can do by decomposing the Jacobian of $\oS$, and by proving that it splits
into two elliptic curves with modular invariant $j=1728$, and one elliptic curve
with invariant $j=0$. This will definitely establish the non--isomorphism of
$\oS$ with $F_4$ since the latter is isogenous to the cube of the elliptic 
curve with invariant $j=1728$.

There are many ways to show the decomposition of the Jacobian of $\oS$, but 
an instructive one is to resort to yet another algebraic model of the surface,
which is in itself interesting since it turns out to be one of the curves we
discussed in Section 4.3, in relation with rational billiards. The idea is
precisely to use the cartographic machinery. Our general discussion showed that
$\oS$ is defined over $\Q$ (as is confirmed by (\ref{proj})), so that it can be
realized as a covering of the Riemann sphere, ramified over three points. A
possible choice is the following. Corresponding to $y=0$, there is one ``point 
at infinity'', namely $(x;y;z)=(1;0;0)$, while away from $y=0$, one sets
$x^3=ty^3$ and $3z^4=(t-1)y^4$, so that altogether
\beq
(x;y;z) = \cases{(\ze_3^k \, \sqrt[3]{t}\,;\,1\,;\,\ze_4^\ell \, \sqrt[4]
{\textstyle {t-1 \over 3}}) & for $t \neq \infty$, $1 \leq k \leq 3$, 
$1 \leq \ell \leq 4$, \cr
(1;0;0) & fot $t=\infty$.}
\eeq
It yields a covering of degree 12, ramified over 0,1 and $\infty$, where the
ramification indices are respectively 3,4 and 12. Thus the corresponding
triangulation obtained by lifting the standard triangulation of $\C\PP_1$
consists of 24 faces, 36 edges and 8 vertices, from which one cross--checks that
the genus is 3. If one labels the four points above 0 by the numbers 1,3,5,7
(from top to bottom), and the three points above 1 by the numbers 2,4,6, one
obtains the triangulation depicted in Figure 4, where the center represents the
point at infinity.

\begin{figure}[htb]
\leavevmode
\begin{center}
%\mbox{\epsfscale 1000 \epsfbox{chariot.eps}}
\mbox{\epsfbox{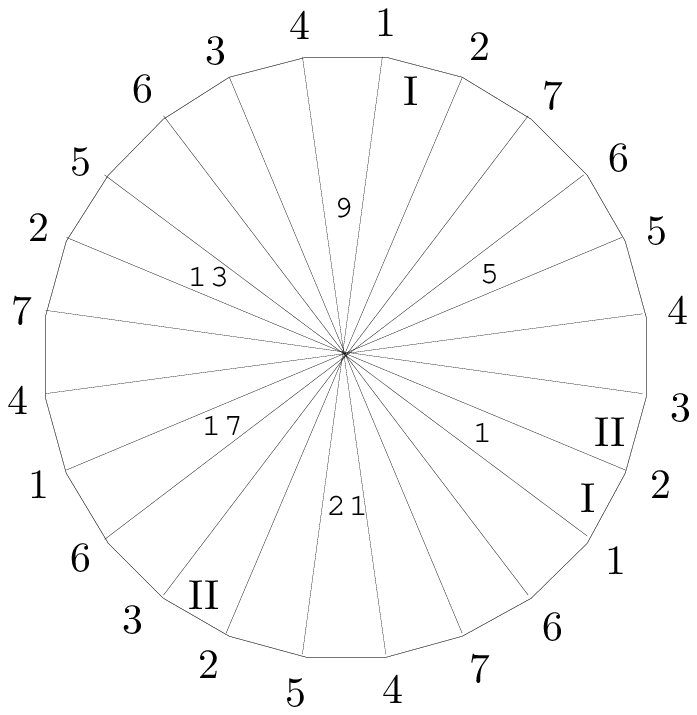}}
\begin{minipage}{14cm}
\bigskip \bigskip \bigskip
{\it Figure 4.} Triangulation of $\oS$ as a covering of degree 12 of the
sphere. The points above 0 are labeled 1,3,5,7, while those above 1 are
numbered 2,4,6. The central point is the point at infinity. There is no
identification among the radial edges, but perimetric ones are to be identified
by pairs, as exemplified by the edges marked I and II. This leaves 36
distinct edges. The small numbers close to the center label the faces from 1 to
24. 
\end{minipage}
\end{center}
\end{figure}

It is not difficult to compute the triangular and cartographic groups, and we
only quote the results. To be consistent with the way the vertices have been 
numbered, we say that the vertices of type 0, lying above 0, are those
labeled by an odd number. Numbering the triangles as in Figure 4, from 1 up to
24, one obtains that the action of the generators of the triangular group on the
$i$--th triangle is
\beq
\sigma_0(i) = i - (-1)^i \bmod 24, \quad
\sigma_1(i) = i + (-1)^i \bmod 24, \quad
\sigma_\infty(i) = i - 7(-1)^i \bmod 24,
\eeq
where the representatives modulo 24 are taken between 1 and 24. From this, one
easily computes the action of the oriented triangular group
\beq
\rho_0(i) = i + 8(-1)^i \bmod 24, \quad
\rho_1(i) = i - 6(-1)^i \bmod 24, \quad
\rho_\infty(i) = i - 2(-1)^i \bmod 24.
\eeq
The generators satisfy $\rho_0 = \rho_\infty^{-4}$, $\rho_1 = \rho_\infty^3$
and $\rho_\infty^{12} = 1$, so that the oriented triangular group is isomorphic
to $\z_{12}$, from which it follows that the Fermat curve $F_{12}$ covers $\oS$.
Because of the above relations between $\rho_0,\rho_1$ and $\rho_\infty$, an
element $\mu_0^a \mu_1^b \mu_\infty^c$ of $\widehat  D(\oS)$ fixes a flag of
symbol $(\infty 0 1)$ if and only if $-4a + 3b + c = 0 \bmod 12$. The quotient
surface was discussed in Section 4.3, and is nothing but the triangular curve 
$C_{8,3,1}(12)$. This yields a third algebraic model for $\oS$. 

The projection $F_{12} \rightarrow C_{8,3,1}(12)$ allows to compute the period
lattice. (The periods of all triangular curves $\crst (n)$ can be found in
\cite{lang2}.) From the results of Section 4.3, the holomorphic differentials on
$C_{8,3,1}(12)$ are the $\omega_{r,s,t}$ with $(r,s,t) = (\rest {8h}, \rest
{3h}, \rest h) \bmod 12$, that is, $\omega_{8,3,1}, \omega_{4,3,5}$, and
$\omega_{4,6,2}$. From (\ref{period}), the periods of these three differentials
along the homology cycles in $F_{12}$ equal
\beq
(\int_{\gamma_{i,j}} \omega_{8,3,1}\; ;\; \int_{\gamma_{i,j}} \omega_{4,3,5} \;
; \; \int_{\gamma_{i,j}} \omega_{4,6,2}) =
(\xi^{8i+3j};\xi^{4i+3j};\xi^{4i+6j}).
\label{per}
\eeq
Since cycles in $F_{12}$ descend to cycles in $C_{8,3,1}(12)$, the period
lattice of the latter contains the lattice in $\C^3$ formed by all integer
combinations of the vectors (\ref{per}). Noticing that the second component of
(\ref{per}) is obtained from the first component by the Galois automorphism
$\sigma_5(\xi)=\xi^5$, one sees that this lattice is equal to
$\{(z,\sigma_5(z),w) \;:\; z \in \z(\xi), \, w \in \z(\omega)\}$. It is
of rank 6 over $\z$, hence of finite index in the full period lattice, so that 
the two are isogenous:
\beq
L(\oS) \sim \{(z,\sigma_5(z)) \;:\; z \in \z(\xi)\} \oplus 
\z(\omega).
\eeq
The first factor, which we may call $L_{8,3,1}$, has been analyzed in Section
3.3, where it was found to be isogenous to the square of $\z(i)$.
Altogether we obtain
\beq
L(\oS) \sim [\z(i)]^2 \oplus \z(\omega).
\eeq

\section{Conclusions}

In this paper, we have tried to give some substance to a suggestion that had 
been made recently, concerning a possible connection between the 
modular invariant partition functions of WZNW models based on the affine 
algebra $\widehat {su(3)}$ and the geometry of the complex Fermat curves. There
are many technical similarities between the two problems. In particular, we
have shown that the decomposition of the Jacobian of the degree $n$ Fermat curve
$F_n$ into simple Abelian varieties is essentially equivalent to the modular
problem for the affine $su(3)$, at level $k=n-3$. The relation was seen
at the technical level through the $su(3)$ parity selection rules. 

Besides this technical observation, which was at the origin of the suggestion,
we have pointed out some intriguing coincidences with a third problem, namely
that of the rational triangular billiards and the related algebraic curves. We
have described at length the three circles of ideas, and found that many of the
concepts in one of them have counterparts in the other two, like for instance
holomorphic differentials against affine characters, parity selection
rules against complex multiplication, ... Despite these fine
mathematical relationships, we have not been able to find a clear and 
definitive way to relate them to the list of modular invariants for $su(3)$,
nor even to give an indication as to why the list of invariants is what it is.

In an attempt to take the relationhip between the modular problem and
algebraic curves in a broader sense, we have shown that a Riemann surface can
be canonically associated with any rational conformal field theory. This could
be done as a consequence of the fact that the matrices $S$ and $T$, describing
the modular transformations of the characters, generate a representation of
$\psl$ which has a finite index subgroup in its kernel. The actual Riemann
surfaces can be computed using invariant theory, as was illustrated in the
cases of $su(3)$, levels 1 and 2. Characterizing all the Riemann surfaces that
arise from conformal field theories in the way described in the text may be 
hard, but the few explicit examples we have analyzed so far suggest the
following questions. 

We have seen throughout this paper several infinite families of algebraic
curves, and in particular the Fermat curves $F_n$. Is it true that every $F_n$
is the Riemann surface associated with some RCFT, and if so, which one(s)
correspond to a given $F_n$ ? What about the triangular curves ? Is there a
more intrinsic way to see if two RCFT's have the same Riemann surface ? 
Is it true that the conformal theories containing dual affine Lie algebras, in
the sense of the rank--level duality, have related Riemann surfaces ?

According to the discussion in Section 3.4, the complete decomposition into
elliptic curves is something rather rare in the context of Fermat curves, and
happens for very special values of $n$ only. Is it true that the Riemann
surfaces coming from conformal theories have a generically large number of
elliptic curves ? 

There are some indications that the Riemann surfaces arising from RCFT are
somewhat special, since for example they have a large group of automorphisms. 
In general the center of the automorphism group is trivial, but there is no
reason to believe that it implies that the Jacobian has no complex
multiplication. We saw for instance in $su(3)$, level 1, that the field of
complex multiplication was larger than what should have been expected on the
sole consideration of the automorphism group. Moreover this field was clearly
related to the eigenvalues of the $T$ matrix. Is this more generally so ? Do
the surfaces coming from RCFTs have always complex multiplication ? And if so,
is it related to $T$, and in which precise way since $T$ is not central ?

We have no general answers to these questions, but looking at the algebras
$su(2)$, $su(3)$ (see the text), $so(8)$, $E_6$ and $E_7$, all at level 1, we
found the following encouraging results: all have Riemann surfaces isomorphic
to triangular curves ($\widehat{so(8)}_1$ has the Fermat curve $F_3 \sim
C_{1,1,1}(3)$), which all have complete decomposition in elliptic curves, and
which all have complex multiplication. The other two algebras with two
independent restricted characters, namely $F_4$ and $G_2$, also at level 1,
are more complicated, but they have the same Riemann surface.

\vskip 2truecm \noindent
{\bf Acknowledgements:} During the course of this work, some of us and 
sometimes all of us have benefited of illuminating discussions with
Arnaud Beauville, Paula Cohen, Pascal Degiovanni, Terry Gannon, Luc Haine, 
Victor Kac and J\"urgen Wolfart. They also made us aware of some important
references. It is a pleasure to thank them warmly. A. C. thanks Jean Lascoux,
Joseph Oesterl\'e and Jean-Bernard Zuber for kind hospitality.

\newpage
\appendix
\section{On automorphisms of quotient surfaces}

Let $\G$ be an invariant subgroup of $\psl$ with finite index. Let $\S$ be the
Riemann surface with punctures $\h/ \G$ and let $\oS$ be the associated
compact Riemann surface. It is known that $\widehat \G \equiv \psl / \G$ is the 
group of automorphisms of $\S$  (see for instance Th. 5.9.4 in \cite{josi}), and
thus is a subgroup of Aut$\,\oS$, the automorphism group of $\oS$. Our goal in
this appendix is to show that if the genus $g$ of $\oS$ is bigger than 1, then
$\widehat{\G}= {\rm Aut}\,\oS$ is the full automorphism group. 

\smallskip
Let us assume that $\widehat \G$ is of index $I$ in Aut$\,\oS$ (a finite group 
since the genus of $\oS$ is greater than 1). It is well--known that the quotient
of $\oS$ by a subgroup $G$ of Aut$\,\oS$ has a natural structure of Riemann
surface, the Euler characteristics, hence the genuses, of the two surfaces being
related by the Riemann--Hurwitz formula. The number of pre--images of a point
$P \in \oS /G$ by the projection map $\oS \rightarrow \oS/ G$ is $|G|/m_P$ where
$m_P$ is the common order of the stability groups of the pre--images of $P$.
Thus $P$ is a ramification point of order $m_P$ and multiplicity $|G|/m_P$. A
straightforward application of the Riemann--Hurwitz formula then gives
\beq
\chi(\oS) = 2-2g = |G| \left(\chi(\oS/G) -
\sum_P \left(1-\frac{1}{m_P}\right)\right).
\label{rh}
\eeq 
The sum over $P$ is actually finite because only a finite number of points 
have $m_P \neq 1$. 

The projection map $\S \rightarrow \S/\widehat\G$ has a holomorphic extension 
$\oS \rightarrow \oS/\widehat{\G} \cong \overline{\h/\psl}$ $\cong \C\PP_1$,
ramified only over 0, 1728 and $\infty$, where the ramification order is
respectively 3, 2 and $n_\infty$ (the order of $T$ in $\widehat\G$). By the above
formula, we have (as mentioned in the text)
\beq
\chi(\oS) = -|\widehat{\G}| \left(\frac{1}{6}-\frac{1}{n_{\infty}}\right). 
\label{first}
\eeq

On the other hand, the other projection, $\oS \rightarrow \oS/{\rm Aut}\,\oS$,
can be decomposed as $\oS \rightarrow \oS/\widehat\G \cong \C\PP_1 \rightarrow \oS/
{\rm Aut}\,\oS$. Yet another application of the Riemann--Hurwitz formula ensures
that a holomorphic map from the Riemann sphere to a compact Riemann surface can
exist only if the latter is also a Riemann sphere. Therefore $\oS/{\rm
Aut}\,\oS \cong \C\PP_1$ and the last map, from $\C\PP_1$ to $\C\PP_1$, can
be normalized in such a way that it fixes the point at infinity, implying
$m_\infty > 1$. Putting $G={\rm Aut}\,\oS$ in (\ref{rh}), the genus of
$\oS$ can be computed from this second projection, and comparison with
(\ref{first}) yields
\beq
{1 \over 6} - {1 \over n_\infty } = I \left(-2 + \left(1-{1 \over
m_\infty}\right) + \sum_{P \neq \infty} \left(1-\frac{1}{m_P}\right)\right).
\label{index}
\eeq
It follows from (\ref{first}) that $g>1$ is equivalent to $n_\infty > 6$, so
that the l.h.s. is positive. This implies that the sum over $P$ has at least
two terms, since $1-{1 \over m_P} < 1$ for $m_P > 1$. Also because $1-{1 \over
m_P} \geq {1 \over 2}$, the value of the sum over $P$ is at least $3 \over 2$
if it contains three terms or more, whereas if it contains two terms, its
minimal value is $7 \over 6$, corresponding to $m_{P_1}=2$, $m_{P_2}=3$. (For
$m_{P_1}=m_{P_2}=2$, the sum is smaller, being equal to 1, but it renders the
r.h.s. of (\ref{index}) negative, and must be excluded.) Thus we obtain the
inequality
\beq
\frac{1}{6}-\frac{1}{n_{\infty}} \geq I \left(\frac{1}{6}-
\frac{1}{m_{\infty}} \right).
\eeq
Finally one can observe that $m_\infty$ is a non--zero multiple of
$n_\infty$. This is because the elements in $\widehat\G$ which fix a point on
$\oS$ form a subgroup of those in Aut$\,\oS$ which fix that point. In
particular, $m_\infty \geq  n_\infty > 6$ and this forces $I=1$. Therefore
$\widehat\G$ is the full automorphism group of $\oS$, as announced. 

Using the results of Sections 4 and 5 on the Fermat curves $F_n$, this gives
another proof that the full automorphism group of $F_n$, $n > 3$, is
$\widehat{C}_n$, of cardinal $6n^2$.

\section{su(3) level 2}

The $\widehat{su(3)}_2$ WZNW model has six chiral integrable representations,
labelled by the six dominant weights $(1,1),\, (1,2),\, (2,1),\, (1,3),\,
(2,2)$ and $(3,1)$, with corresponding characters $\chi_p$. As explained in
Section 2, they split into two orbits under the automorphisms. The $S$ and $T$
matrices accordingly factorize into a two--by--two piece acting on the orbit
space, and a three--by--three Fourier kernel acting within each orbit. Following
our general  philosophy, we are interested in the independent restricted
characters, and the representation of the modular group they carry,
four--dimensional in this case. This amounts going to the subspace of 
conjugation invariant characters. One may check that, if one puts the
restricted characters in a matrix as follows,
\beq
\underline{\chi} \equiv \pmatrix{\chi_0 & \chi_1 \cr \chi_2 & \chi_3} = 
\pmatrix{\chi_{(1,1)} & \chi_{(2,2)} \cr
{1 \over 2}[\chi_{(1,3)} + \chi_{(3,1)}] & 
{1 \over 2}[\chi_{(1,2)} + \chi_{(2,1)}] \cr},
\eeq
the action of the modular group can be written as
\beq
S \;:\; \underline{\chi}\;\longrightarrow \; S_l \,\underline{\chi}\, S_r^{-1},
\qquad 
T \;:\; \underline{\chi} \;\longrightarrow \; T_l \,\underline{\chi}\, T_r^{-1},
\eeq
where ($\om=e^{2i\pi /3}$)
\beq
S_l = {i \over \sqrt{3}} \pmatrix{1&1 \cr 2&-1}, \qquad
T_l = \pmatrix{\om^2 & 0 \cr 0 & \om},
\eeq
and ($\zeta=e^{2i\pi /5}$)
\beq
S_r = {2i \over \sqrt{5}} \pmatrix{\sin{\pi \over 5} & \sin{2\pi \over 5} \cr
\sin{2\pi \over 5} & -\sin{\pi \over 5}}, \qquad
T_r = \pmatrix{\zeta^4 & 0 \cr 0 & \zeta}.
\eeq

The elements of all these matrices belong to $M_{\rm restr} = \Q(\zeta_{15})$.
The corresponding Galois group, of order 8, consists of $\sigma_h$, for $h \in
\z_{15}^*$. A general formula for the action of the Galois group on $S$ has been
given in Section 5.1, Eq. (\ref{galois}). In the present case, $S$ and $T$
generate a representation of $PSL_2(\z_{15})$, so that the formula yields
($2^{-1}=8 \bmod 15$ and $7^{-1}=13 \bmod 15$)
\beq
\sigma_2(S) = S T^8 S T^2 S T^8 S, \qquad 
\sigma_7(S) = S T^{13} S T^7 S T^{13} S.
\eeq
These two elements generate the full Galois group. Its action on $T$ is just
$\sigma_h(T)=T^h$.

Useful for what follows are the Puiseux series of the restricted characters:
\bea
&& \chi_{(1,1)} = q^{-2/15} \, [1 + 8q + 44q^2 + + 128q^3 + 376q^4 + \ldots], \\
&& \chi_{(2,2)} = q^{7/15} \, [8 + 37q + 136q^2 + 404q^3 + 1\,072q^4 + \ldots],
\\ 
&& \chi_{(1,3)} = q^{8/15} \, [6 + 24q + 93q^2 + 264q^3 + 708q^4 + \ldots], \\
&& \chi_{(1,2)} = q^{2/15} \, [3 + 24q + 90q^2 + 288q^3 + 777q^4 + \ldots].
\eea

\medskip
One easily checks that $S_l^2 = (S_lT_l)^3 = -1$ and $T_l^3 = 1$, and
similarly $S_r^2 = (S_rT_r)^3 = -1$ and $T_r^5 = 1$. Of course the 
normalizations in the left and right factors are arbitrary since only their
product matters, but our choice makes all four matrices have determinant 1. Then
$S_l$ and $T_l$ on the one hand, $S_r$ and $T_r$ on the other hand, generate
subgroups of $SL_2(\C)$. If we consider the quotients of these subgroups by $-1$,
and keep the classification of finite subgroups of the special linear group in
two complex dimensions in mind, we see that the left group
$\widehat\G_l$ is the double cover of the tetrahedron group, hence of order $2
\times 12$, and that the right group $\widehat\G_r$ is the double cover of the
icosahedron group, of order $2 \times 60$.

Let us denote the matrix that acts on the characters as $R_l \,\underline\chi\,
R_r^{-1}$ by $R_l \times R_r$. Then $T^6$ acts as $1
\times T_r$, and $T^{10}$ acts as $T_l \times 1$. Similarly 
$(ST^6)^9$ acts as $S_l \times 1$, while $(ST^{10})^9$ acts as $1 \times
S_r$. This is enough to show that there is a surjection from $\widehat\G_l
\times \widehat\G_r$ onto $\widehat\G$, the group generated by the matrices
$S$ and $T$ on the restricted characters, and that $\widehat\G \cong 
\widehat\G_l \times \widehat\G_r$ modulo the kernel of this map, which is the
diagonal $\z_2=\{1 \times 1,\, -1\times -1\}$ of order 2. Therefore the order of
$\widehat\G$ is ${(2\cdot 12)(2\cdot 60) \over 2}=1440$. On the other hand the
order of $T$ is manifestly 15, so the general formula (\ref{genus}) implies that
the genus of the Riemann surface associated with $su(3)_2$ is equal to 73. The
non--compactified surface $\S$ has $1440/15=96$ punctures. 

This number looks frightening, and to compute an algebraic model for it seems
hopeless. Let us recall that according to the general discussion of Section
5.1, what we have to do is to find a basis of polynomials in the characters
which are invariant for the action of $\widehat\G$. There is at least one
polynomial invariant which is easy to obtain: since all left and right matrices
have determinant 1, the determinant of $\underline \chi$ is invariant under
$\widehat\G$. A look at the Puiseux series shows that it is regular at $q=0$, and
that the zero--th order coefficient is equal to 3, so that it is exactly
equal to 3:
\beq
P_2(\chi_i) = \chi_0 \chi_3 - \chi_1 \chi_2 = 3.
\label{determ}
\eeq

The ring of polynomial invariants for $\widehat\G$ is more complicated than in 
the level 1 case. The Molien series for the number of invariants is 
\beq
F(t) = {1+t^{12}+t^{20}+t^{24}+2t^{30}+t^{36}+t^{40}+t^{48}+t^{60} \over
(1-t^2)(1-t^{12})(1-t^{20})(1-t^{30})}.
\label{ffull}
\eeq
It shows that the ring is not freely generated by four invariants,
but that there are more generators with relations among them. In this case,
the ring of invariants is a free module over $\C(P_2,P_{12},P_{20},P_{30})$ with
basis $\{1,R_{12},R_{20},R_{24},R_{30},R_{30}',R_{36},R_{40},R_{48}, 
R_{60}\}$, where the $P_i$'s are ``fundamental'' invariants of degree
$i$, and the $R_j$'s are ``auxiliary'' invariants. Every $R_j$ can be
expressed algebraically in terms of the $P_i$'s, but not polynomially. What we
want to do is to compute enough independent irreducible invariants, namely
three or four depending on whether they are equal to trivial functions of $j$
or not. We have already found one, namely $P_2$ given in (\ref{determ}), which
clearly accounts for the factor $(1-t^2)^{-1}$ in $F(t)$.

\medskip
In order to compute invariants for $\widehat\G$, one can take advantage of the
quasi--factorized form of $\widehat\G = \widehat\G_l \times \widehat\G_r /\z_2$.
In fact we deal with a true (even if not faithful) representation
$\widehat\G_l \times \widehat\G_r$ so the $\z_2$ factor is automatically taken
into account. Given four indeterminates $w,x,y,z$
(they will soon become $\chi_0,\ldots,\chi_3$), an element $\widehat\g_l
\times \widehat\g_r$ acts on them by left and right multiplication $\widehat\g_l 
\big({w \atop y} \; {x \atop z}\big) \widehat\g_r^{-1}$. If we write the
variables $w,x,y,z$ in terms of four others $u_1,u_2,v_1,v_2$ through 
\beq
\pmatrix{w & x \cr y & z} = \pmatrix{u_1 \cr u_2} \times \pmatrix{v_1 & v_2},
\label{tensor}
\eeq
then obviously $\widehat\g_l$ acts on the $u$'s whereas $\widehat\g_r$ acts on the
$v$'s. It is clear that upon this substitution, a polynomial invariant
$P(w,x,y,z)$ for $\widehat\G$ becomes a (perhaps identically vanishing) linear
combination of product polynomials $P^l(u_1,u_2)P^r(v_1,v_2)$ with $P^l$
invariant under  $\widehat\G_l$ and $P^r$ invariant under $\widehat\G_r$. 

We would like to go backwards as well, {\it i.e.} start with polynomial
invariants $P^l(u_1,u_2)$ for $\widehat\G_l$ and $P^r(v_1,v_2)$ for
$\widehat\G_r$ and construct a polynomial invariant $P(w,x,y,z)$ for $\widehat\G$.
This involves a nice analogy with Wick contractions. 
We define formal 2--point functions $\langle u_1v_1 \rangle = w$, $\langle 
u_1v_2 \rangle = x$, $\langle u_2v_1 \rangle = y$, $\langle u_2v_2 \rangle = z$,
and all others zero. The expectation value of a monomial in $u_1$, $u_2$, $v_1$
and $v_2$ is the polynomial in $w$, $x$, $y$ and $z$ obtained by doing all
Wick contractions (obviously, this gives $0$ unless the degrees in the $u$
and $v$ variables are equal), and we extend to all polynomials by linearity. We
also include a normalization factor and shall use the explicit general formula
\beq
\langle u_1^\alpha u_2^{n-\alpha} v_1^\beta v_2^{n-\beta} \rangle = 
{n \choose \alpha}^{-1} \; \sum_{a=\max(0,\alpha+\beta-n)}^{\min(\alpha,\beta)}
{\beta \choose a}\,{n-\beta \choose \alpha -a} \; w^a x^{\alpha -a}  y^{\beta
-a} z^{n-\alpha-\beta-a}.
\label{wick}
\eeq
As is clear from the properties of correlation functions in field theory, for
consistency  Wick contractions must transform covariantly under linear
transformations of the fields. A formal argument is easy to build. 
In particular if one does a linear transformation of the fields that leaves
the polynomial in the fields invariant,
the correlator is invariant. Thus, if  $P^l(u_1,u_2)$ is an invariant for
$\widehat\G_l$ and $P^r(v_1,v_2)$ an invariant for $\widehat\G_r$,
$P(w,x,y,z)=\langle P^l(u_1,u_2)P^r(v_1,v_2)\rangle$ is an invariant for
$\widehat\G$.

Moreover, with the chosen normalization, substitution of the $u$ and $v$
variables in $P(w,x,y,z)=\langle P^l(u_1,u_2)P^r(v_1,v_2)\rangle$ gives back
$P^l(u_1,u_2)P^r(v_1,v_2)$.

So if we start from an arbitrary polynomial invariant $P(w,x,y,z)$, substitute
the $u$ and $v$ variables according to Eq. (\ref{tensor}), and take the
expectation value, we get a new polynomial $Q(w,x,y,z)$ such that $P-Q$ gives 
$0$ upon the substitution of the $u$ and $v$ variables. Now we use a trivial 
fact from algebra, related to the simplest Pl\"ucker embedding of algebraic
geometry: the ideal of polynomials in $w$, $x$, $y$ and $z$ vanishing
identically upon substituting the $u$ and $v$ variables is principal and
generated by the determinant $wz-xy$. Hence $P-Q$ has to be a multiple of
$wz-xy$ and the quotient is of course invariant because $wz-xy$ is. We can
repeat the construction . At every step the degree decreases so the procedure
must stop. This shows that all invariants are polynomials in the
determinant $wz-xy$ with coefficients being the expectation value invariants. 

If $d_n^{(l)}$ and $d_n^{(r)}$ denote the number of degree $n$
invariants for $\widehat\G_l$ and $\widehat\G_r$, the above construction yield
$d_n^{(l)} d_n^{(r)}$ invariants for $\widehat\G$ of degree $n$. From 
\bea
&& F_l(t) = \sum_{n=0}^\infty d_n^{(l)} t^n = {1 + t^{12} \over
(1-t^6)(1-t^{8})}, \label{fleft} \\
&& F_r(t) = \sum_{n=0}^\infty d_n^{(r)} t^n = {1 + t^{30} \over
(1-t^{12})(1-t^{20})} \label{fright},
\eea
one finds the generating function for the number of invariants for $\widehat\G$
induced from those of $\widehat\G_l$ and $\widehat\G_r$:
\beq
\sum_{n=0}^\infty d_n^{(l)}d_n^{(r)} t^n = {1+t^{12}+t^{20}+t^{24}+
2t^{30}+t^{36}+t^{40}+t^{48}+t^{60} \over (1-t^{12})(1-t^{20})(1-t^{30})}.
\eeq
As announced, the missing factor is due to the contribution of
$wz-xy$. We now proceed to give the invariants of lowest degree explicitly.

\smallskip
{}From (\ref{fleft}), all polynomial invariants for $\widehat\G_l$ can be
expressed in terms of only three invariants, of degree 6,8 and 12:
\bea
&& P^l_6 = 8 \, u_1^6 - 20 \, u_1^3 u_2^3 - u_2^6, \\
&& P^l_8 = 8 \, u_1^7 u_2 + 7 \, u_1^4 u_2^4 - u_1 u_2^7, \\
&& R^l_{12} = 64 \, u_1^{12} + 704 \, u_1^9 u_2^3 + 88 \, u_1^3 u_2^9 - u_2^{12}.
\eea
There are two left invariants of degree 12, $[P^l_6]^2$ and $R^l_{12}$, and two
of degree 20, namely $[P^l_6]^2 P^l_8$ and $R^l_{12} P^l_8$. 

Likewise for the right factor, all invariants can be written in terms of three
invariants of degree 12, 20 and 30:
\bea
&& P^r_{12} = v_1^{11} v_2 + 11 \, v_1^6 v_2^6 - v_1 v_2^{11}, \\
&& P^r_{20} = v_1^{20} - 228 \, v_1^{15} v_2^5 + 494 \, v_1^{10} v_2^{10} 
+ 228 \, v_1^5 v_2^{15} + v_2^{20}, \\
&& R^r_{30} = v_1^{30} + 522 \, v_1^{25} v_2^5 - 10\,005 \, v_1^{20} v_2^{10} 
- 10\,005 \, v_1^{10} v_2^{20} - 522 \, v_1^5 v_2^{25} + v_2^{30}. \qquad
\eea
There is one right invariant of degree 12, and one of degree 20.

\smallskip
{}From the generating function (\ref{ffull}), the degrees of the five lowest
irreducible invariants for $\widehat\G$ are 2, 12, 12, 20 and 20. In terms of
left and right invariants, they are given by the following Wick products:
\bea
&& P_2 = wz - xy \quad (= 3), \\
&& P_{12} = \langle \,[P^l_6]^2 \, P^r_{12} \,\rangle \quad (= \textstyle{512
\over q} + {294\,816 \over 7} + \ldots), \\
&& P'_{12} = \langle \, R^l_{12} \, P^r_{12} \,\rangle \quad (= \textstyle{512
\over q} + 66\,432 + \ldots), \\
&& P_{20} = \langle \, [P^l_6]^2 \, P^l_8 \, P^r_{20} \,\rangle \quad (= 
\textstyle{3\,072 \over q^2} + {13\,621\,248 \over 17q} -
{22\,400\,979\,840 \over 187} + \ldots), \qquad \\
&&P'_{20} = \langle \, R^l_{12} \, P^l_8 \, P^r_{20} \,\rangle \quad (= 
\textstyle{3\,072 \over q^2} - {7\,612\,416 \over 17q} -
{15\,570\,897\,408 \over 17} + \ldots). 
\eea
We have indicated in parenthesis the first terms of the Puiseux series of the
invariants when one substitutes $\chi_0,\chi_1,\chi_2,\chi_3$ for $w,x,y,z$.
One sees, upon the same substitution, that $P_2$, $P_{12}-P_{12}'$ and $
P_{20}-P_{20}'-{512 \over 17}P_2^4\,P'_{12}$ are modular
invariant functions, holomorphic in the upper half--plane, hence equal to pure
constants, given respectively by 3, $-{170\,208 \over 7}$ and
${118\,573\,144\,704 \over 187}$. They form a complete set of algebraic
equations that describe the Riemann surface associated to $su(3)$ level 2. Their
explicit form is just a matter of computing Wick contractions using
(\ref{wick}). For completeness, we quote the final results:
\beq
\chi_0 \chi_3 - \chi_1 \chi_2 = 3, 
\eeq

\bea 
&& 28\,\big[\chi_1\,\chi_2 + \chi_0\,\chi_3\big]\,
\big[24\,(\chi_0^5\,\chi_2^5 - \chi_1^5\,\chi_3^5)  \nonumber \\
&& \qquad - (64\,\chi_0^3\,\chi_1^3 + 3\,\chi_2^3 \chi_3^3)\,
(2\,\chi_0^2\,\chi_3^2 + 7\,\chi_0\,\chi_1\,\chi_2\,\chi_3 +
2\,\chi_1^2\,\chi_2^2)\big] \nonumber \\ 
&& - 14\,\big[\chi_0\,\chi_3 + 3\,\chi_1\,\chi_2\big]\,
\big[64\,\chi_0^8\,\chi_2^2 - 3\,\chi_1^2\,\chi_3^8\big] 
- 14\,\big[3\,\chi_0\,\chi_3 + \chi_1\,\chi_2\big]\,
\big[3\,\chi_0^2\,\chi_2^8 - 64\,\chi_1^8\,\chi_3^2\big]
\nonumber \\
&& + 16\,\big[\chi_1^6\,\chi_2^6 + 36\,\chi_0\,\chi_1^5\,\chi_2^5\,\chi_3 +
225\,\chi_0^2\,\chi_1^4\,\chi_2^4\,\chi_3^2 + 
400\,\chi_0^3\,\chi_1^3\,\chi_2^3\,\chi_3^3 \nonumber \\
&& \qquad + 225\,\chi_0^4\,\chi_1^2\,\chi_2^2\,\chi_3^4 +
36\,\chi_0^5\,\chi_1\,\chi_2\,\chi_3^5 +
\chi_0^6\,\chi_3^6\big] \nonumber \\
&& + 7\,\chi_2^{11}\,\chi_3 + 77\,\chi_2^6\,\chi_3^6 - 7\,\chi_2\,\chi_3^{11} +
85\,104 = 0,
\qquad
\eea

\bea
&& \Big\{ 247\,\chi_2^9\,\chi_3^9\,\chi_0\,\chi_3  
+ \chi_0\,\chi_2^4\,\big(8\,\chi_0^3 - \chi_2^3\big)^4\,\big(\chi_0^3
+ \chi_2^3\big) 
- 57\,\chi_2^{14}\,\chi_3^4\,\big(3\,\chi_0\,\chi_3 + \chi_1\,\chi_2\big)
\nonumber \\
&&\textstyle + {26 \over 17}\,\big[4\,096\,\chi_0^6\,\chi_1^6 -
31\,\chi_2^6\,\chi_3^6\big] \times \big[14\,{\chi_1^4}\,{\chi_2^4} +
80\,\chi_0\,{\chi_1^3}\,{\chi_2^3}\,\chi_3 
+ {135 \over 2}\,{\chi_0^2}\,{\chi_1^2}\,{\chi_2^2}\,{\chi_3^2}\big]
\nonumber \\
&&\textstyle + {4\,992 \over 17}\,\big[\chi_0^5\,\chi_2^5 -
\chi_1^5\,\chi_3^5\big] \times 
\big[21\,{\chi_1^5}\,{\chi_2^5} + 175\,\chi_0\,{\chi_1^4}\,{\chi_2^4}\,\chi_3
+ 450\,{\chi_0^2}\,{\chi_1^3}\,{\chi_2^3}\,{\chi_3^2}\big] \nonumber \\
&& \textstyle + {208 \over 17}\,\big[64\,\chi_0^3\,\chi_1^3 +
11\,\chi_2^3\,\chi_3^3\big]  \times \nonumber \\ 
&& \qquad \big[2\,{\chi_1^7}\,{\chi_2^7} 
+ 35\,\chi_0\,{\chi_1^6}\,{\chi_2^6}\,\chi_3 
+ 189\,{\chi_0^2}\,{\chi_1^5}\,{\chi_2^5}\,{\chi_3^2}  
+ 420\,{\chi_0^3}\,{\chi_1^4}\,{\chi_2^4}\,{\chi_3^3}\big]  \nonumber \\
&&\textstyle - {832 \over 187}\,\big[{\chi_1^{10}}\,{\chi_2^{10}} 
+ 100\,\chi_0\,{\chi_1^9}\,{\chi_2^9}\,\chi_3 
+ 2\,025\,{\chi_0^2}\,{\chi_1^8}\,{\chi_2^8}\,{\chi_3^2} 
+ 14\,400\,{\chi_0^3}\,{\chi_1^7}\,{\chi_2^7}\,{\chi_3^3}
\nonumber \\
&& \qquad \qquad  + 44\,100\,{\chi_0^4}\,{\chi_1^6}\,{\chi_2^6}\,{\chi_3^4} 
+ 31\,752\,{\chi_0^5}\,{\chi_1^5}\,{\chi_2^5}\,{\chi_3^5} \big]
\nonumber \\
&& \textstyle - {4 \over 17}\,\big[4\,096\,\chi_0^{11}\,\chi_1 +
31\,\chi_2\,\chi_3^{11}\big] \times \big[273\,{\chi_1^4}\,{\chi_2^4}
\nonumber \\ 
&& \qquad \qquad  + 455\,\chi_0\,{\chi_1^3}\,{\chi_2^3}\,\chi_3 +
210\,{\chi_0^2}\,{\chi_1^2}\,{\chi_2^2}\,{\chi_3^2} +
30\,{\chi_0^3}\,\chi_1\,\chi_2\,{\chi_3^3} + {\chi_0^4}\,{\chi_3^4}\big]
\nonumber \\ 
&& \textstyle - {8 \over 17}\,\big[64\,\chi_0^8\,\chi_2^2 -
11\,\chi_1^2\,\chi_3^8\big] \times \big[1287\,{\chi_1^5}\,{\chi_2^5} +
5005\,\chi_0\,{\chi_1^4}\,{\chi_2^4}\,\chi_3 
\nonumber \\ 
&& \qquad \qquad + 6006\,{\chi_0^2}\,{\chi_1^3}\,{\chi_2^3}\,{\chi_3^2} 
+ 2730\,{\chi_0^3}\,{\chi_1^2}\,{\chi_2^2}\,{\chi_3^3} +
455\,{\chi_0^4}\,\chi_1\,\chi_2\,{\chi_3^4} +
21\,{\chi_0^5}\,{\chi_3^5}\big] \nonumber \\
&&\textstyle + {256 \over 17} \Big[\chi_0\,\chi_3 - \chi_1\,\chi_2\Big]^4
\times \Big[64\,{\chi_0^{11}}\,\chi_1 + 352\,{\chi_0^6}\,{\chi_1^6} 
- \chi_2^{11}\,\chi_3 - {11 \over 2}\,\chi_2^6\,\chi_3^6 \nonumber \\
&& \qquad \qquad + 22\,\big(8\,{\chi_0^8}\,{\chi_2^2} - \chi_1^2\,\chi_3^8\big)\,
\big(\chi_0\,\chi_3 + 3\,\chi_1\,\chi_2 \big) 
\nonumber \\
&& \qquad \qquad + 44\,\big(8\,\chi_0^3\,\chi_1^3 + \chi_2^3\,\chi_3^3\big)\,
\big(2\,{\chi_1^3}\,{\chi_2^3} +
9\,\chi_0\,{\chi_1^2}\,{\chi_2^2}\,\chi_3\big) \Big] \nonumber \\
&&\textstyle  + (\chi_0,\chi_1,\chi_2,\chi_3) \longrightarrow
(\chi_1,-\chi_0,\chi_3,-\chi_2) \Big\} 
+ {59\,286\,572\,352 \over 187} = 0. 
\eea


\newpage

\begin{thebibliography}{xx}
         
\bibitem{ciz} A.  Cappelli, C. Itzykson and J.-B. Zuber, {\em The A-D-E 
classification of minimal and $A_1^{(1)}$ conformal invariant theories}, 
Commun. Math. Phys. 113 (1987) 1--26.\\  
A. Kato, {\em Classification of modular invariant partition functions in two
dimensions}, Mod. Phys. Lett. A2 (1987) 585--600.  

\bibitem{gan1} T. Gannon, {\em The Classification of affine su(3) modular
invariant partition functions}, Commun. Math. Phys. 161 (1994) 233--264.

\bibitem{gan2} T. Gannon, {\em WZW commutants, lattices, and level--one
partition functions}, Nucl. Phys. B396 (1993) 708--736.

\bibitem{rtw} P. Ruelle, E. Thiran and J. Weyers, {\em Implications of an
arithmetical symmetry of the commutant for modular invariants}, Nucl. Phys.
B402 (1993) 693--708.

\bibitem{cg} A. Coste and T. Gannon, {\em Remarks on Galois symmetry in
rational conformal field theories}, Phys. Lett. B323 (1994) 316--321.

\bibitem{kr} N. Koblitz and D. Rohrlich, {\em Simple factors in the Jacobian of
a Fermat curve}, Can. J. Math. XXX (1978) 1183--1205.

\bibitem{auritz} E. Aurell and C. Itzykson, {\em Rational billiards and
algebraic curves}, J. Geom. and Phys. 5 (1988) 191--208.

\bibitem{dessins} Contributions by P. Cohen, J. Wolfart, M. Bauer and  C.
Itzykson in {\em The Grothendieck theory of dessins d'enfants}, L. Schneps ed., 
LMSLNS 200, Cambridge Univ. Press.
                 
\bibitem{kac} V.G. Kac, {\em Infinite dimensional Lie algebras}, 3rd edition,
Cambridge University Press, Cambridge 1990.

\bibitem{drouffeitz}  C. Itzykson and J.-M. Drouffe, {\em Th\'eorie
statistique des champs}, \'editions du CNRS, Paris 1989.  

\bibitem{kawa} V.G. Kac and M. Wakimoto, {\em Modular and conformal invariance
constraints in representation theory of affine algebras}, Adv. Math. 70 (1988)
156--236.

\bibitem{witten} E. Witten, {\em Non--Abelian bosonization in two dimensions},
Commun. Math. Phys. 92 (1984) 455--472.

\bibitem{godol} P. Goddard and D. Olive, {\em Kac--Moody and Virasoro algebras
in relation to quantum physics}, Int. J. Mod. Phys. A1 (1983) 303--414.

\bibitem{moosei} G. Moore and N. Seiberg, {\em Naturality in conformal field
theory}, Nucl. Phys. B313 (1989) 16--40.

\bibitem{cardy} J. Cardy, {\em The operator content of two--dimensional
conformally invariant theories}, Nucl. Phys. B270 (1986) 186--204. 

\bibitem{nahm} W. Nahm, {\em Lie group exponents and $SU(2)$ current algebras},
Commun. Math. Phys. 118 (1988) 171--176.

\bibitem{dizu} P. Di Francesco and J.-B. Zuber, {\em SU(N) lattice integrable
models associated with graphs}, Nucl. Phys. B338 (1990) 602---646.

\bibitem{gan3} T. Gannon, {\em The level two and three modular invariants of
SU(n)}, preprint (November 1995).

\bibitem{gan4} T. Gannon, {\em Towards a classification of su(2) $\oplus \ldots
\oplus$ su(2) modular invariant partition functions}, J. Math. Phys. 36 (1995)
675--706.

\bibitem{grw} T. Gannon, P. Ruelle and M. Walton, {\em Automorphism modular
invariants of current algebras}, to appear in Commun. Math. Phys. (hep-th
9503141)

\bibitem{gan5} T. Gannon, {\em Kac--Peterson, Perron--Frobenius, and the
classification of conformal field theories}, preprint (q-alg 9510026).

\bibitem{bi} M. Bauer and C. Itzykson, {\em Modular transformations of
$SU(N)$ affine characters and their commutant}, Commun. Math. Phys. 127 (1990)
617--636.

\bibitem{pr} P. Ruelle, {\em Dimension of the commutant for the $SU(N)$ affine
algebras}, Commun. Math. Phys. 133 (1990) 181--196.

\bibitem{buffenoir}  E. Buffenoir, A. Coste, J. Lascoux, P. Degiovanni and A.
Buhot, {\em Precise study of some number fields and  Galois actions
occurring in conformal field theory}, Ann. Inst. Poincar\'e, Theor. Phys. 63
(1995) 41--79.  
 
\bibitem{rational} C. Vafa, {\em Toward classification of conformal theories},
Phys. Lett. 206B (1988) 421--426.\\
G. Anderson and G. Moore, {\em Rationality in conformal field
theory}, Commun. Math. Phys. 117 (1988) 441--450.

\bibitem{deBG} J. de Boer and J. Goeree, {\em Markov traces and $II_1$
factors in conformal field theory}, Commun. Math. Phys. 139 (1991) 267--304.

\bibitem{rtw2} P. Ruelle, E. Thiran and J. Weyers, {\em Modular invariants for
affine $\widehat{SU}(3)$ theories at prime heights}, Commun. Math. Phys. 133
(1990) 305--322.

\bibitem{schel1} J. Fuchs, B. Gato-Rivera, B. Schellekens and C. Schweigert,
{\em Modular invariants and fusion rule automorphisms from Galois theory},
Phys. Lett. B334 (1994) 113--120.

\bibitem{schel2} J. Fuchs, B. Schellekens and C. Schweigert, {\em Galois
modular invariants of WZW models}, Nucl. Phys. B437 (1995) 667--694.

\bibitem{bost} J.-B. Bost, Les Houches lectures ``Introduction to compact
Riemann surfaces, Jacobians and Abelian varieties'', in {\em From Number
Theory to Physics}, edited by M. Waldschmidt, P. Moussa, J.--M. Luck and C.
Itzykson, Springer 1990.

\bibitem{swinn} H.P.F. Swinnerton--Dyer, {\em Analytic theory of Abelian
varieties}, Cambridge University Press, Cambridge 1974.

\bibitem{lang2} S. Lang, {\em Introduction to algebraic and Abelian functions},
GTM 89, Springer 1982.

\bibitem{aoki} N. Aoki, {\em Simple factors of the Jacobian of a Fermat curve 
and the Picard number of a product of Fermat curves}, Amer. J. Math. 113 (1991)
779--833.

\bibitem{rohr} D. Rohrlich, appendix to B. Gross, {\em On the periods of
Abelian integrals and a formula of Chowla and Seiberg}, Invent. Math. 45 (1978)
193--211.

\bibitem{weil} A. Weil, {\em Sur les p\'eriodes d'int\'egrales ab\'eliennes},
Commun. Pure and Appl. Math. XXIX (1976) 813--819.

\bibitem{langcm} S. Lang, {\em Complex multiplication}, Springer 1983.

\bibitem{shita} G. Shimura and Y. Taniyama, {\em Complex multiplication of
Abelian varieties and its applications to number theory}, Publ. Math. Soc.
Jap., no. 6, 1961.

\bibitem{kob} N. Koblitz, {\em Gamma function identities and elliptic
differentials on Fermat curves}, Duke Math. J. 45 (1978) 87--99.
 
\bibitem{grot} A. Grothendick, {\em Esquisse d'un programme.}

\bibitem{knapp} A.W. Knapp, {\em Elliptic curves}, Princeton Univ. Press,
Princeton 1992.

\bibitem{shi} {\em Arithmetic and Geometry of Fermat Curves} Proceedings of the
Algebraic Geometry Seminar, Singapore (1987).

\bibitem{berry} P.J. Richens and M.V. Berry, {\em Pseudo--integrable systems in
classical and quantum mechanics}, Physica 2D (1981) 495--512.

\bibitem{gw} D. Gepner, E. Witten, {\em String theory on group manifolds}, Nucl.
Phys. B278 (1986) 493--549.  

\bibitem{gunning} R.C. Gunning, {\em Lectures on modular forms} Annals of Mathematics Studies 48, Princeton Univ. Press, Princeton 1992.

\bibitem{josi} G.A. Jones and D. Singerman, {\em Complex functions}, Cambridge
University Press, Cambridge 1987.

\bibitem{carl} C.L. Siegel, {\em Topics in complex function theory}, Vol. II,
Wiley \& Sons 1971.

\bibitem{bauer} M. Bauer, unpublished.

\bibitem{benson} D.J. Benson, {\em Polynomial invariants of finite groups},
LMSLNS 190, Cambridge Univ. Press, Cambridge 1993.

\bibitem{pseudoref} G.C. Shephard and J.A. Todd, {\em Finite unitary reflection
groups}, Can. J. Math. VI (1954) 274--304.

\end{thebibliography}
\end{document}